\documentclass[preprint,3p]{elsarticle}      

\usepackage{amsmath}
\usepackage{natbib}
\usepackage{amssymb}
\usepackage{graphicx}
\usepackage{dcolumn}
\usepackage{bm}
\usepackage{mathrsfs}
\usepackage{color}
\usepackage{psfrag}
\usepackage{url}
\usepackage{soul}
\usepackage{float}  

\usepackage{cancel}  

\usepackage{array}
\usepackage{caption}

\newcolumntype{P}[1]{>{\centering\arraybackslash}p{#1}}
\usepackage[normalem]{ulem}
\usepackage{color}

\usepackage{booktabs}

\usepackage[pdftitle={The PDF title},
  pdffitwindow=true,
  breaklinks=true,
  colorlinks=true,
  urlcolor=blue,
  linkcolor=black,
  citecolor=black,
  anchorcolor=red]{hyperref}

\newcommand{\et}{\textit{et al.}}
\newcommand{\eg}{\textit{e.g.}}
\newcommand{\ie}{\textit{i.e.}}
\newcommand{\fig}{Figure}

\newcommand{\eq}{Eq.}
\newcommand{\eqs}{Eqs.}
\newcommand{\sect}{Section}

\newcommand{\rf}{Ref.}
\newcommand{\rfs}{Refs.}
\newcommand{\tab}{Table}

\newcommand{\sta}{{\rm stall}}

\newcommand{\cov}{\text{cov}}
\newcommand{\BE}{\mathbb{E}}
\newcommand{\BS}{\mathbb{S}}
\newcommand{\BR}{\mathbb{R}}

\newcommand{\BV}{\mathbb{V}}
\newcommand{\BX}{\mathbb{X}}
\newcommand{\BT}{\mathbb{T}}

\newcommand{\cU}{\mathcal{U}}
\newcommand{\cN}{\mathcal{N}}

\newcommand{\dd}{\mbox{d}}

\newcommand{\fzero}{\mathbf{0}}
\newcommand{\fx}{\mathbf{x}}

\newcommand{\hf}{\hat{f}}

\newcommand{\ft}{\mathbf{t}}

\newcommand{\fthet}{\bm{\theta}}
\newcommand{\betas}{\kappa}
\newcommand{\bd}{\bar{d}}
\newcommand{\rey}{\mbox{Re}}
\newcommand{\reyt}{\mbox{Re}_\tau}
\newcommand{\reyb}{\mbox{Re}_b}

\newcommand{\dxp}{{\Delta x^+}}

\newcommand{\dzp}{{\Delta z^+}}

\newcommand{\lut}{\langle u_\tau \rangle}

\newcommand{\lu}{\langle u \rangle}

\begin{document}

\title{Efficient prediction of turbulent flow quantities using a Bayesian hierarchical multifidelity model}

\author[focal3,focal1,focal2]{S. Rezaeiravesh\corref{cor1}}
\ead{saleh.rezaeiravesh@manchester.ac.uk}
\author[focal1,focal2]{T. Mukha}
\ead{tmu@mech.kth.se}
\author[focal1,focal2]{P. Schlatter}
\ead{pschlatt@mech.kth.se}
\cortext[cor1]{Principal Corresponding Author}
\cortext[cor2]{Corresponding Author}
\address[focal3]{Department of Mechanical, Aerospace and Civil Engineering, The University of Manchester, UK}
\address[focal1]{SimEx/FLOW, Engineering Mechanics, KTH Royal Institute of Technology, SE-100 44 Stockholm, Sweden}
\address[focal2]{Swedish e-Science Research Centre (SeRC), Stockholm, Sweden}

\date{\today}


\begin{abstract}
High-fidelity scale-resolving simulations of turbulent flows quickly become prohibitively expensive, especially at high Reynolds numbers. 
As a remedy, we may use multifidelity models (MFM) to construct predictive models for flow quantities of interest (QoIs), with the purpose of  uncertainty quantification, data fusion and optimization.
For numerical simulation of turbulence, there is a hierarchy of methodologies ranked by accuracy and cost, which include several numerical/modeling parameters that control the predictive accuracy and robustness of the resulting outputs.
Compatible with these specifications, the present hierarchical MFM strategy allows for simultaneous calibration of the fidelity-specific parameters in a Bayesian framework as developed by Goh~\et~\cite{goh:13}. The purpose of the multifidelity model is to provide an improved prediction by combining lower and higher fidelity data in an optimal way for any number of fidelity levels; even providing confidence intervals for the resulting QoI.
The capabilities of our multifidelity model are first demonstrated on an illustrative toy problem, and it is then applied to three realistic cases relevant to engineering  turbulent flows.  
The latter include the prediction of friction at different Reynolds numbers in turbulent channel flow, the prediction of aerodynamic coefficients for a range of angles of attack of a standard airfoil, and the uncertainty propagation and sensitivity analysis of the separation bubble in the turbulent flow over periodic hills subject to the geometrical uncertainties. 
In all cases, based on only a few high-fidelity data samples (typically direct numerical simulations, DNS), the multifidelity model leads to accurate predictions of the QoIs accompanied with an estimate of confidence. 
The result of the UQ and sensitivity analyses are also found to be accurate compared to the ground truth in each case. 
Supported by the present results, the hierarchical multifidelity model can be proposed as a viable solution to the outer-loop problems in a range of computational fluid dynamics (CFD) involving turbulent flows, especially in situations when there is no direct correlation between the data of low- and high-fidelities; occurring for instance in wall modeling for turbulent flows. In addition, the methodology is applicable in other problems in computational physics, as long as there can be a hierarchy of simulation fidelities defined, including climate modeling and biomedical applications.
\end{abstract}

\maketitle

\section{Introduction}\label{sec:intro}
In science and engineering, different computational models can be derived to make realizations of the quantities of interest (QoIs) of a process or an event happening in reality.
High-fidelity (HF) models can result in highly accurate and robust realizations, but running them is typically computationally expensive. 
In contrast, different low-fidelity (LF) models with lower computational cost can be developed for the same process which, however, potentially lead to lower accuracy QoIs due to partial or completely-missing physics captured by the model.
On the other hand, in different applications arising in uncertainty quantification (UQ), data fusion, and optimization numerous realizations of the QoIs are required associated to the samples taken from the space of the inputs/parameters in order to make reliable estimations (these non-intrusive problems are referred to as the outer-loop problems).
In this regard, multifidelity models (MFM) can be constructed by combining realizations of the HF and LF models such that a balance between the overall computational cost and predictive accuracy is achieved. The goal is to provide, by combining HF and LF models, an estimate of the QoI that is better than any of the models alone.

In the recent years, different types of MFMs have been applied to a wide range of problems, see \eg~the recent review by Peherstorfer \et~\cite{pehersrofer:18}.
The use of the MFMs in studies of turbulent flows can be greatly advantageous, considering the wide range of engineering applications relying on these flows and also the high cost generally involved in the HF computations (such as scale-resolving simulations) and experiments of the turbulent flows.
There is a distinguishable hierarchy in the fidelity of the computational models utilized for simulation of turbulence, see \eg~\rf~\cite{sagaut:13}.
Let us consider the wall-bounded turbulent flows where a turbulent boundary layer forms at the wall boundaries.
Direct numerical simulation (DNS) can provide the highest-fidelity results for a given turbulent flow, however it can become prohibitively expensive at high Reynolds numbers which are relevant to practical applications.
The computational cost can be reduced by employing large eddy simulation (LES) which aims at directly resolving the scales larger than a defined size and modeling the unresolved effects.
At the lowest cost and fidelity level, Reynolds-averaged Navier-Stokes (RANS) simulations can be performed which avoid directly resolving any flow fluctuations by resorting to a statistical description of turbulence.
Between RANS and wall-resolving LES, other approaches such as hybrid RANS-LES and wall-modeled LES can be considered, see \cite{sagaut:13,larsson:16}.
Although this clear hierarchy is extremely beneficial when constructing MFMs as will be thoroughly discussed and demonstrated in the present paper, there is a challenge to be dealt with: the realizations of different turbulence simulation approaches are, in general, sensitive to various modeling and numerical parameters as well as inputs.
At lower fidelities like RANS, modeling effects are dominant while as moving towards LES and DNS, numerical factors become more relevant, including grid resolution and discretization properties.
Hereafter, these fidelity-specific controlling parameters are referred to as tuning or calibration parameters.

Combining training data from different turbulence-simulation approaches, MFMs are constructed over the space of design and uncertain parameters/inputs.
An appropriate approach to construct MFMs for turbulent flow problems should systematically allow for simultaneous calibration of the tuning parameters.
An appropriate methodology which is employed in the present study is the hierarchical multifidelity predictive model developed in~\cite{higdon:04,goh:13} in which the calibration parameters of the involved fidelities are estimated using the data of the higher fidelity models.
This MFM which is hereafter referred to as~HC-MFM, can also incorporate the observational uncertainties.
The HC-MFM can be seen as an extension of the model by Higdon~\et~\cite{higdon:04} which was employed to combine experimental (field) and simulation data.
A fundamental component of this class of MFMs is the Bayesian calibration of the computer models as described in the landmark paper by Kennedy and O'Hagan~\cite{koh:01}.
At each level of the MFM, the Gaussian process regression (GPR)~\cite{rasmussen:05} is employed to construct surrogates for the simulators.

The application of the HC-MFM in the field of computational fluid dynamics (CFD) and turbulent flows is novel, and we make specific adaptations suitable for turbulence simulations. In this regard, the present paper aims at assessing the useful potential of the HC-MFM by applying it to three examples relevant to engineering wall-bounded turbulent flows. 
To highlight the contributions of the present work, the existing studies in the literature devoted to the development and application of the multifidelity models to CFD and turbulent flows are briefly reviewed here by classifying them according to their underlying MFM strategy. 
1. A model was originally introduced by Kennedy and O'Hagan~\cite{koh:00} where a QoI at each fidelity is expressed as a first-order autoregressive model of the same QoI at the immediately lower fidelity.
Co-Kriging using Gaussian process regression to construct surrogates are classified in this category, see \eg~\rf~\cite{gomez:18,voet:21} for applications in turbulence simulations.
To enhance the computational efficiency of the co-Kriging for several fidelity levels, recursive algorithms have been proposed and applied to CFD problems, see~\cite{leGratiet:14, perdikaris:15}.
2. A class of MFMs has been developed based on non-intrusive polynomial chaos expansion (PCE) and stochastic collocation methods, see \cite{ng:12,palar:16}.
Here, an additive or a multiplicative term is considered to correct the LF model's predictions against the HF model. 
3. The multi-level multifidelity Monte Carlo (MLMF-MC) models~\cite{fairbanks:17,geraci:17} are appropriate for the UQ forward problems.   
These models are developed by combining multilevel~\cite{giles:08} and control-variate~\cite{pasupathy:12} MC methods to improve the rate of convergence of the stochastic moments of the QoIs compared to the the standard MC method.  
Jofre~\et~\cite{jofre:18} applied MLMF-MC models to an irradiated particle-laden turbulent flow.
The HF model was considered to be DNS and the two LF models were based on a surrogate particle approach and lower resolutions for flow and particles.
4. Other models including the hierarchical Kriging model based on GPR where the predictions of a LF model are taken as the trend in the HF Kriging, see Han and G{\"o}rtz~\cite{han:12}.

Recently, Voet \et~\cite{voet:21} compared inverse weighted distance-, PCE- and co-Kriging-based MFMs using the data of RANS and DNS for the turbulent flow over a periodic hill, and concluded that the co-Kriging model outperforms the the others in terms of accuracy.
This is the first (and to our knowledge only) study where multifidelity models have been applied to engineering-relevant RANS and DNS data for the purpose of uncertainty propagation. 
Voet~\et~\cite{voet:21} also found that the performance of the co-Kriging can deteriorate when there is no significant correlation between the RANS and DNS data and at the same time there is a significant deviation between them. 
Motivated by this deficiency, we adapt and use the HC-MFM where the discrepancy between the data (and not their correlation) over the space of design parameters is learned using independent Gaussian processes. 
The model is absolutely generative and can be extended to an arbitrary number of fidelity levels. 
Besides the systematic calibration of the fidelity-specific parameters during its training stage, the HC-MFM is also capable of handling uncertain data, as for instance happens when QoIs are turbulence statistics computed over a finite time-averaging interval (recently, a framework is proposed to combine these observational uncertainties with parametric ones, see \rf~\cite{uqFrame:22}). 
Relying on these characteristics, the HC-MFM is suitable to be applied to the data of various turbulence simulation methodologies to address different types of the outer-loop problems. 
In contrast to all the previous studies (at least in CFD), we adopt a Bayesian inference to construct the HC-MFM, a feature which results in more accurate models as well as estimating confidence intervals for the predictions.

The rest of the paper is organized as follows.
In \sect~\ref{sec:method}, various elements of the HC-MFM approach are introduced and explained. 
\sect~\ref{sec:results} is devoted to the application of the HC-MFM to an illustrative example, turbulent channel flow, polars for an airfoil, and analysis of the geometrical uncertainties in the turbulent flow over a periodic hill.
The summary of the paper along with the conclusions is presented in \sect~\ref{sec:conclusion}.

\section{Method}\label{sec:method}
In this section, the hierarchical MFM with calibration (HC-MFM) developed by Goh~\et~\cite{goh:13}, which forms the basis for the present study is reviewed.
We will proceed by sequentially going through the aspects of Gaussian process regression (GPR), model calibration, and eventually the HC-MFM formulation.

\subsection{Gaussian Process Regression}
In general, the Gaussian processes (GPs) provide a way to systematically build a representation of the QoI as a function of the various inputs to the model. Eventually, regression can be performed by evaluating the Gaussian process at new inputs not seen by the model before.
Let $\fx\in \BX\subset \BR^{p_\fx}$ represent the controllable inputs and parameters, adopting the notation from~\rf~\cite{koh:01}.
As a convention, all boldface letters are hereafter considered to be a vector or a matrix. 
The design and uncertain parameters appearing in optimization and UQ analyses, respectively, can also be classified as~$\fx$.
A Gaussian process~$\hf(\cdot)$, see~\eg~\rf~\cite{rasmussen:05}, can be employed to map the inputs~$\fx$ to the QoIs or outputs~$y\subset \BR$ of the computer codes (simulators) or field data.
For a finite set of training samples $\{\fx_1, \fx_2,\ldots,\fx_n\}$ with corresponding observations $\{y_1,y_2,\ldots,y_n\}$, the collection of $\{\hf(\fx_1),\hf(\fx_2),\ldots,\hf(\fx_n)\}$ will have a joint Gaussian (multivariate normal) distribution,~\cite{rasmussen:05}. The GP $\hf(\fx)$ is written as,
\begin{equation}\label{eq:GPdef}
\hat{f}(\fx)\sim \mathcal{GP}\left(m(\fx),k(\fx,\fx')\right) \,,
\end{equation}
which is fully described by its mean $m(\fx)$ and covariance function $k(\fx,\fx')$ defined as,
\begin{eqnarray}
  m(\fx) &=& \BE[\hat{f}(\fx)] \,, \label{eq:GPmean}\\
  k(\fx,\fx') &=& \BE[(\hat{f}(\fx)-m(\fx)) (\hat{f}(\fx')-m(\fx'))] \label{eq:GPcov} \,.
\end{eqnarray}
In general, the GPs can be used in the case of having observation noise $\bm{\varepsilon}$ in the~$y$~data.
Using an additive error model, we have,
\begin{equation}
  y(\fx)=\hat{f}(\fx)+\bm{\varepsilon} \,,
\end{equation}
where the noises are assumed to be independent and have Gaussian distribution~$\bm{\varepsilon}\sim \cN(0,\sigma^2)$. 

In the Gaussian process regression (GPR), given a set of training data $\mathcal{D}=\{\fx_i,y_i\}_{i=1}^n$ the posterior and posterior predictive distributions of~$\hf(\cdot)$ and~$y$, respectively, at test inputs $\fx^* \in \BX$ can be inferred in a Bayesian framework, see \eg~\rf~\cite{rasmussen:05}.
To this end, first a prior distribution for $\hf(\fx)$, see \eq~(\ref{eq:GPdef}), is assumed through specifying functions for the mean and covariance in \eqs~(\ref{eq:GPmean}) and (\ref{eq:GPcov}), where there are unknown hyperparameters~$\bm{\beta}$ in the functions. 
Using the training data, the posterior distribution of~$\bm{\beta}$ is learned. 
As a main advantage of the GPR, the predictions at test samples will be accompanied by an estimate of uncertainty.

\subsection{Model Calibration}
As pointed out in \sect~\ref{sec:intro}, the outputs of computational models (simulators) at a given~$\fx$ may depend on different tuning or calibration parameters, $\ft\in \BT\subset \BR^{p_\ft}$.
Given a set of observations, these parameters can be calibrated through conducting a UQ inverse problem  which can be expressed in a Bayesian framework, see~Kennedy and O'Hagan~\cite{koh:01}.
The calibrated model can then not only be employed for prediction, but also for fusion of the field and simulation data, see Higdon \et~\cite{higdon:04}.
Consider~$n_1$ data samples $\{(\fx_i,y_i)\}_{i=1}^{n_1}$ are observed for a physical process~$\zeta(\fx)$.
To statistically model the observations, a simulator~$\hf(\fx,\fthet)$ can be employed in which the~$\fthet$ are the true or optimal values of~$\ft$ and are to be estimated from the training data.
However, in general, it is possible that even the calibrated simulator~$\hf(\fx,\fthet)$ produces observations which  systematically deviate from reality.
To remove such a bias, a model-discrepancy term~$\hat{\delta}(\fx)$ can be added to the simulator, \cite{koh:01,higdon:04}.
In many applications, particularly in CFD and turbulent flow simulations, the flow solver can be run only a limited number of times.
In any realization, the adopted values for the tuning parameters~$\ft$ are not necessarily optimal and hence potentially lead to the outputs which are systematically different from the QoIs in reality.
For the described calibration problem, the~Kennedy~and~O'Hagan~model~\cite{koh:01}~reads~as,
\begin{equation}\label{eq:hegdonModel}
  \begin{cases}
  y_i = \hat{f}(\fx_i,\fthet) + \hat{\delta}(\fx_i) +\varepsilon_i \,&,\quad i=1,2,\cdots,n_1 \\
  y_{i} = \hat{f}(\fx_i,\ft_i)  \,&,\quad i=1+n_1,2+n_1,\cdots,n_2+n_1
  \end{cases} \,,
\end{equation}
where, $\hat{\cdot}$ specifies a GP and~$n_2$ is the number of simulated data.
Note that the samples $\{\fx_i\}_{i=1+n_1}^{n_2+n_1}$ are not necessarily the same as $\{\fx_i\}_{i=1}^{n_1}$ at which the observations are made.
The index~$i$ should be seen as a global index which implies that a different model is used for each of the two subranges of~$i$.  
Given the~$n_1+n_2$ data, the posterior distribution for the calibration parameters~$\fthet$ along with that of the hyperparameters in the GPs,~$\bm{\beta}$ is estimated.
Further details are provided in the section below.

\subsection{The Hierarchical Multifidelity Model with Automatic Calibration (HC-MFM)}
Goh \et~\cite{goh:13} extended the model~(\ref{eq:hegdonModel}) to an arbitrary number of fidelity levels which together form a modeling hierarchy for a physical process.
As a main feature of the resulting MFM, each fidelity can, in general, have its own calibration parameters and also share some calibration parameters with other fidelities.
The basics of the MFM comprising three fidelity levels are explained below, noting that adapting the formulation to any number of fidelities with different combinations of parameters is straightforward. 
We assume that the fidelity of the models decreases from~$M_1$ to~$M_3$, and in practice due to the budget limitations, the number of training data decreases with increasing the model fidelity.
The HC-MFM for three fidelities reads as~\cite{goh:13},
\begin{eqnarray}\label{eq:mfModel}
  \begin{cases}
   y_{M_1}(\fx_i) \, = {\hat{f}(\fx_i,\fthet_3,\fthet_s)} +
   {\hat{g}(\fx_i,\fthet_{2},\fthet_{s})}+
   {\hat{\delta}(\fx_i)} + \varepsilon_{1_i} \,&, i=1,2,\cdots,n_1 \\
   y_{M_2}(\fx_{i}) \, = {\hat{f}(\fx_i,\fthet_{3},\ft_{s_i})} +
   {\hat{g}(x_i,\ft_{2_i},\ft_{s_i})}+ \varepsilon_{2_i}  \,&, { i=1+n_1,2+n_1, \cdots , n_2+n_1} \\
   y_{M_3}(\fx_{i}) = {\hat{f}(\fx_i,\ft_{3_i},\ft_{s_i})}+ \varepsilon_{3_i}  \,&, i=1+n_1+n_2,2+n_1+n_2,  \cdots , n_3+n_1+n_2
  \end{cases} \,,
\end{eqnarray}
where subscript~$s$ denotes the parameters which are shared between the models, whereas~$\ft_2$ and~$\ft_3$ are the calibration parameters specific to fidelities~$M_2$ and~$M_3$, respectively.
The noises are assumed to have Gaussian distributions with zero mean.
At each fidelity level, the associated simulator is created by adding a model discrepancy term to the simulator describing the immediately lower fidelity.
Concatenating all training data, an augmented vector~$\mathbf{Y}$ of size $n_1+n_2+n_3$ is obtained, for which the covariance matrix can be written in terms of the covariances of $\hf(\cdot)$, $\hat{g}(\cdot)$, $\hat{\delta}(\cdot)$ and the observational noise,
\begin{eqnarray}\label{eq:mfCovs}
\bm{\Sigma} = {\bm{\Sigma}_{f}}
   +
   \begin{bmatrix}
   {\bm{\Sigma}_{g}} & \fzero_{(n_1+n_2)\times n_3} \\
   \fzero_{n_3\times (n_1+n_2)} & \fzero_{n_3\times n_3} \\
   \end{bmatrix}
   +
   \begin{bmatrix}
   {\bm{\Sigma}_\delta} & \fzero_{n_1\times (n_2+n_3)} \\
   \fzero_{(n2+n3)\times n_1} & \fzero_{(n_2+n_3)\times (n_2+n_3)}  \\
   \end{bmatrix}
   +
   \begin{bmatrix}
   \bm{\Sigma}_{\varepsilon_1} & \fzero_{n_1\times n_2} & \fzero_{n_1\times n_3} \\
   \fzero_{n_2\times n_1} & \bm{\Sigma}_{\varepsilon_2} & \fzero_{n_2\times n_3} \\
   \fzero_{n_3\times n_1} & \fzero_{n_3\times n_2} & \bm{\Sigma}_{\varepsilon_3}\\
   \end{bmatrix} \,.
\end{eqnarray}
Appropriate kernel functions should be chosen to express the structure of the covariances.
Using samples~$i$ and~$j$ of the inputs and parameters, the associated element in the covariance matrix~$\bm{\Sigma}_f$ will be obtained~from,
\begin{equation}\label{eq:Kf_kernel}
\bm{\Sigma}_{f_{ij}}=\cov(\fx_i,\ft_{3_i},\ft_{s_i},\fx_j,\ft_{3_j},\ft_{s_j})
=\lambda_f^2 \, k(\bd_{f_{ij}})     \,,
\end{equation}
where~$\lambda_f$ is a hyperparameter and~$\bd_{f_{ij}}$ is the scaled Euclidean distance between two samples~$i$ and~$j$ over the space of~$(\fx,\ft_3,\ft_s)$:
\begin{eqnarray}
\bd^2_{f_{ij}} &=& \bd^2(\fx_i,\fx_j) + \bd^2(\ft_{3_i},\ft_{3_j})+ \bd^2(\ft_{s_i},\ft_{s_j}) \label{eq:euclid1}\\
&=&
\sum_{l=1}^{p_\fx} \frac{(x_{l_i}-x_{l_j})^2}{\ell^2_{f_{x_l}}} + 
\sum_{l=1}^{p_{\ft_3}} \frac{(t_{3_{l_i}}-t_{3_{l_j}})^2}{\ell^2_{f_{{t_3}_l}}} +
\sum_{l=1}^{p_{\ft_s}} \frac{(t_{s_{l_i}}-t_{s_{l_j}})^2}{\ell^2_{f_{{t_s}_l}}} \label{eq:euclid2} \,.
\end{eqnarray}
Here,~$p_\fx$,~$p_{\ft_3}$, and~$p_{\ft_s}$ specify the dimension of~$\fx$,~$\ft_3$, and~$\ft_s$, respectively.  
Correspondingly, the length-scale over the $l$-th dimension of each of these spaces is represented by~$\ell_{f_{x_l}}$,~$\ell_{f_{{t_3}_l}}$, and~$\ell_{f_{{t_s}_l}}$, respectively.
These length-scales are among the hyperparameters~$\bm{\beta}$ to be learned when constructing the HC-MFM.  
There are various options for modeling the covariance kernel function~$k(\cdot)$, see \eg~\rfs~\cite{rasmussen:05,gramacy:20}, among which the exponentiated quadratic and {Matern-5/2}~\cite{matern:86} functions are used in the examples in \sect~\ref{sec:results}. 
These two functions respectively read as, 
\begin{equation}\label{eq:exponKernel}
    k(\bd_{f_{ij}}) = \exp(-0.5 \,\bd^2_{f_{ij}}) \,,
\end{equation}
and, 
\begin{equation}\label{eq:mat52Kernel}
    k(\bd_{f_{ij}}) = \left[ 1+\sqrt{5}\,\bd_{f_{ij}} +\frac{5}{3}\bd^2_{f_{ij}} \right] \exp(-\sqrt{5}\, \bd_{f_{ij}}) \,.
\end{equation}
Similar expressions can be derived for $\bm{\Sigma}_{g_{ij}} = k_{g}(\fx_i,\ft_{2_i},\ft_{s_i},\fx_j,\ft_{2_j},\ft_{s_j}) $ and ${\bm{\Sigma}_{\delta_{ij}} =k_\delta(\fx_i,\fx_j)}$ appearing in \eq~(\ref{eq:mfCovs}).
This leads to introducing new hyperparameters associated to the GPs.
Note that, given how the training vector~$\mathbf{Y}$ and associated inputs are assembled, correct combinations of training data for the inputs and parameters will be used in the kernels.
The unknown parameters to estimate include calibration parameters in different models,~$\fthet$, and hyperparameters~$\bm{\beta}$ appearing in the GPs.
Following the Bayes rule, the posterior distribution of these parameters given the training data~$\mathbf{Y}$ can be inferred from~\cite{goh:13,koh:01,higdon:04},
\begin{equation}\label{eq:bayes}
    \pi(\fthet,\bm{\beta}|\mathbf{Y}) \propto \pi(\mathbf{Y}|\fthet,\bm{\beta}) \pi_0(\fthet) \pi_0(\bm{\beta}) \,,
\end{equation}
where $\pi(\mathbf{Y}|\fthet,\bm{\beta})$ specifies the likelihood function and $\pi_0(\cdot)$ represents a prior distribution.
Note that all priors are assumed to be independent. 
For all GPs in \sect~\ref{sec:results}, the prior distribution for~$\lambda$ appearing in the covariance matrices such as \eq~(\ref{eq:Kf_kernel}) is taken to be half-Cauchy whereas the length-scales~$\ell$ in \eqs~(\ref{eq:exponKernel}) and~(\ref{eq:mat52Kernel}) are assumed to have Gamma distributions. 
The exact definition of the priors will be provided later for each case in \sect~\ref{sec:results}, and \tab~\ref{tab:pdfs} summarizes the formulation of the standard distributions used as priors.
The standard-deviation of the noises are assumed to be the same for which a half-Cauchy prior is adopted. 
For the calibration parameters~$\fthet$, Gaussian or uniform priors are considered.
In some cases, we may consider a constant mean function for the GPs, where a Gaussian distribution is used as the prior.
Due to this, the predictions of a trained MFM when it is used to extrapolate in~$\fx$ (outside of the range of training samples) should be used with caution.
To avoid potential inaccuracies, in general, more elaborate mean functions can be used when constructing the HC-MFMs (this, however, is not the subject of the present study).

Given the training data~$\mathbf{Y}$, a Markov Chain Monte Carlo (MCMC) technique can be used to draw samples from the posterior distributions of~$\fthet$ and~$\bm{\beta}$, and hence construct a HC-MFM.
In the present study, the described HC-MFM~(\ref{eq:mfModel}) has been implemented in \texttt{Python} using the \texttt{PyMC3}~\cite{pymc3} package with the NUTS MCMC sampling approach~\cite{nuts}.
As it will be shown in \sect~\ref{sec:MCMC_MAP}, the MCMC sampling method may lead to more accurate results compared to the point estimators. 

After being constructed, an HC-MFM can be used for predicting the QoI~$y$ for any new sample taken from the space of inputs~$\fx$.
The accuracy of the predicted QoIs will be assessed by measuring their deviation from the validation data of the highest fidelity~${M_1}$.
As detailed in \rf~\cite{goh:13}, the joint distribution of the training~$\mathbf{Y}$ and new~$y^*$ (associated to a test sample~$\fx^*$) conditioned on $\fthet,\bm{\beta}$ will have a multivariate normal distribution with a covariance matrix of the same structure as~$\bm{\Sigma}$ in \eq~(\ref{eq:mfCovs}).
For any joint sample drawn from the posterior distribution of $\pi(\fthet,\bm{\beta}|\mathbf{Y})$, a sample prediction for~$y^*$ is made.
Repeating this procedure for a large number of times, valid estimations for the posterior of the~predictions~$y^*$~can~be~achieved.
Therefore, estimating the confidence in the predictions is straightforward. 
Note that at this stage, various UQ analyses can be performed using the HC-MFM as a surrogate of the physical process over~$\fx$.

\section{Results and Discussion}\label{sec:results}
Four examples are considered to which the HC-MFM described in the previous section is applied.
The first example in \sect~\ref{sec:ex1} is used to validate the implementation of the MFM, and the next three examples are relevant to fundamental and engineering analysis of wall-bounded turbulent flows.

\subsection{An Illustrative Example}\label{sec:ex1}
Consider the following analytical model taken from Forrester~\et~\cite{forrester:07} to generate high- and low-fidelity samples of the QoI~$y$ for input~$x\in[0,1]$,
\begin{equation}\label{eq:ex1}
\begin{cases}
y_{H}(x)=(\theta x-2)^2 \sin (2\theta x-4)   \\
y_{L}(x)=y_{H}(x) + B(x-0.5)-C  \\
\end{cases} \,.
\end{equation}
In \rf~\cite{forrester:07},~$\theta$ is taken to be fixed and equal to~$6$, but here it is treated as an uncertain calibration parameter that is to be estimated during the construction of the MFM.
Note that the notations of the general model~(\ref{eq:hegdonModel}) can be adopted for \eq~(\ref{eq:ex1}). For simulator~$\hat{f}(x,t)$ and~model discrepancy~$\hat{\delta}(x)$ the covariance matrix in~\eq~(\ref{eq:Kf_kernel}) is used with the exponentiated quadratic kernel~(\ref{eq:exponKernel}). 
The following prior distributions are considered: $\lambda_f, \lambda_\delta \sim\mathcal{HC}(\alpha=5)$, $\ell_{f_x}, \ell_{f_t}, \ell_{\delta_x} \sim \Gamma(\alpha=1,\beta=5)$, $\varepsilon\sim \cN(0,\sigma)$ with $\sigma~\sim \mathcal{HC}(\alpha=5)$, and $\theta\sim\cU[5.8,6.2]$. 
Here,~$\mathcal{HC}$,~$\Gamma$,~$\cN$ and~$\cU$ denote the half-Cauchy, Gamma, Gaussian, and uniform distributions, respectively, see \tab~\ref{tab:pdfs}. 
The HF training samples are taken at~$x=\{0,0.4,0.6,1\}$, therefore,~$n_H=4$ is fixed.
To investigate the effect of~$n_L$, three sets of LF samples of size~$10$,~$15$, and~$20$ are considered which are generated by Latin hypercube sampling from the admissible space $[0,1]\times[5.8,6.2]$ corresponding to~$x$ and~$t$ (uncalibrated instance of~$\theta$), respectively.

\begin{figure}[t]
    \centering
    \begin{tabular}{ccc}
         \includegraphics[scale=0.25]{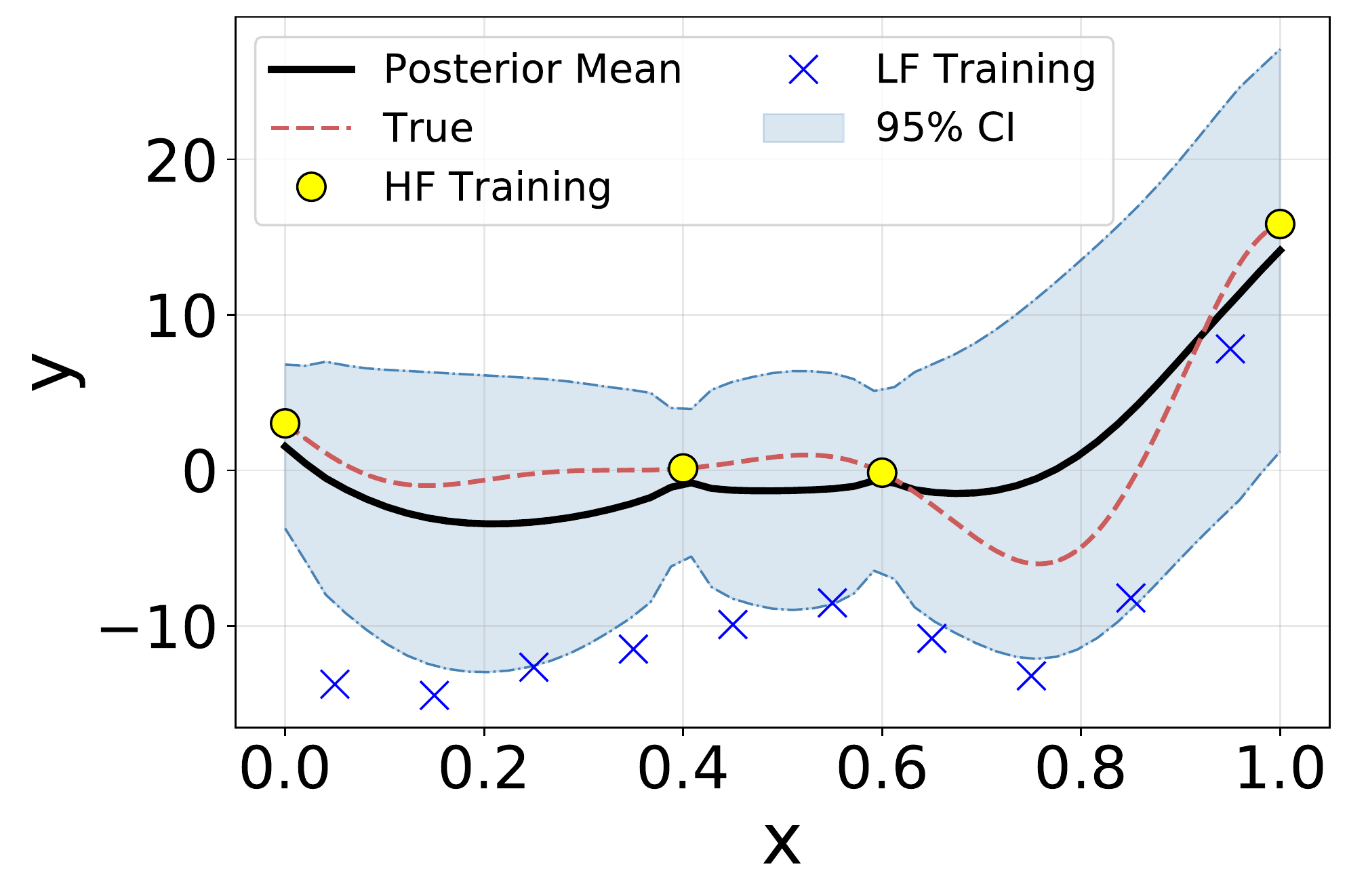} & \hspace{-0.3cm}
         \includegraphics[scale=0.25]{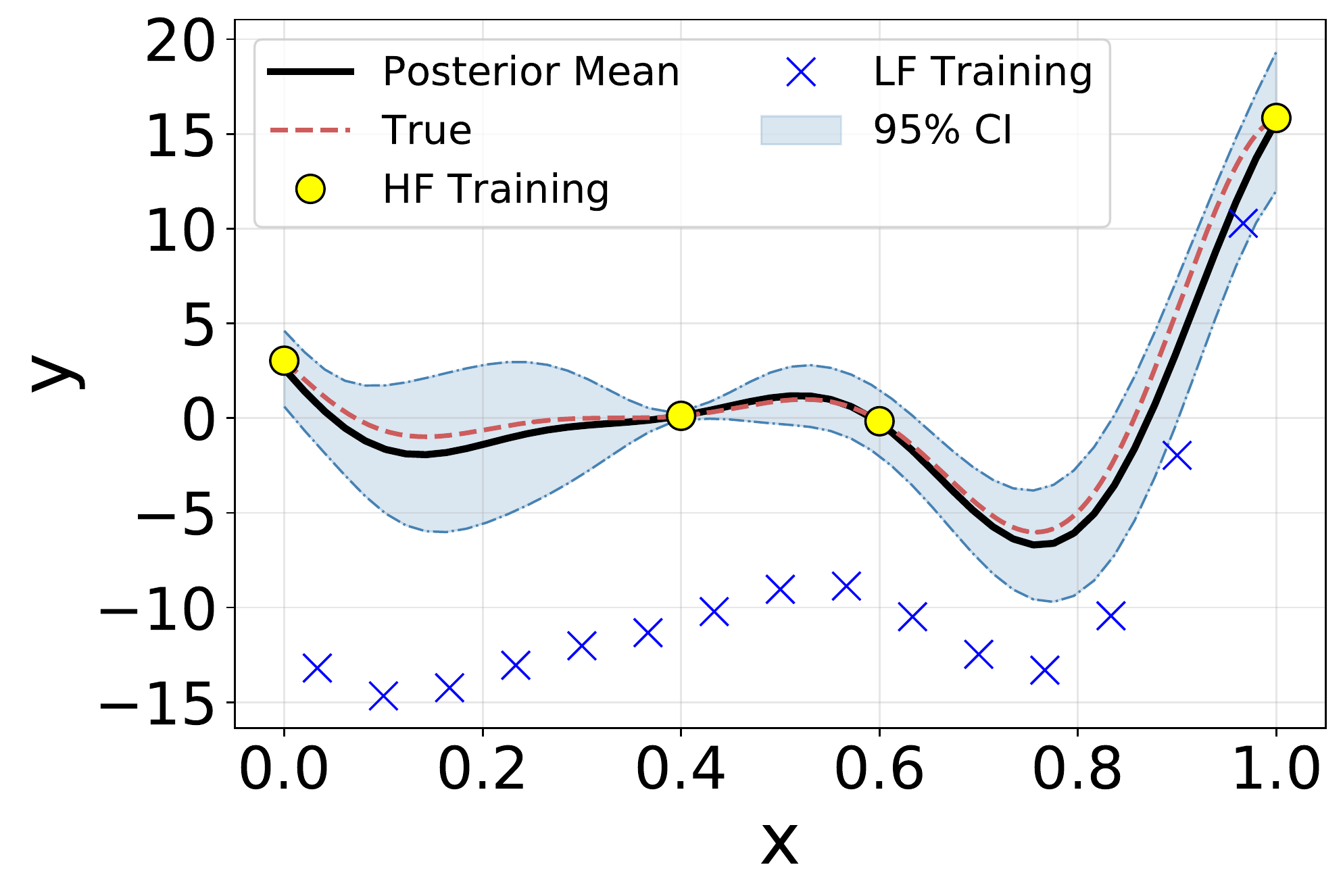} & \hspace{-0.3cm}
         \includegraphics[scale=0.25]{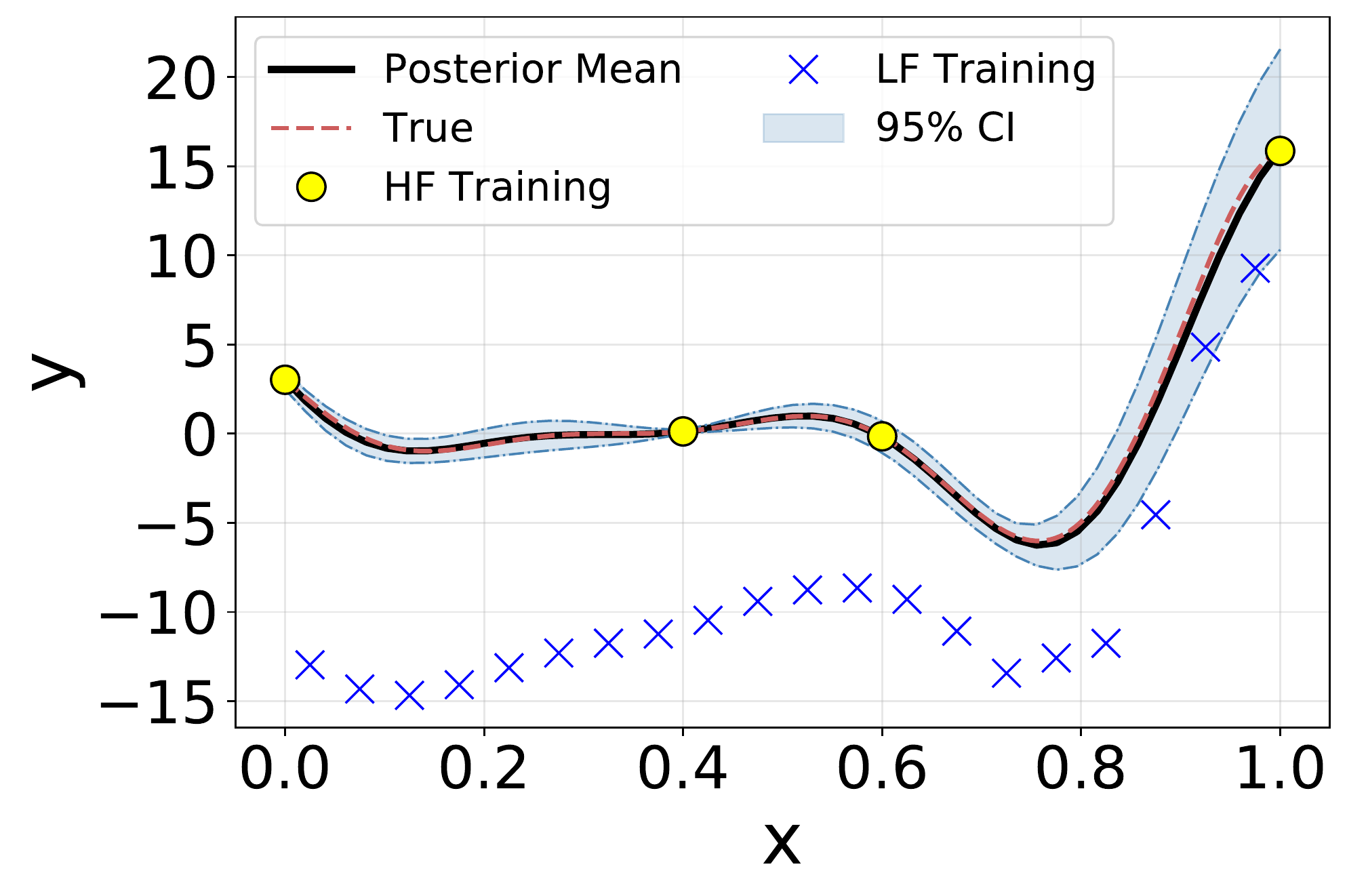} \\
         \includegraphics[scale=0.22]{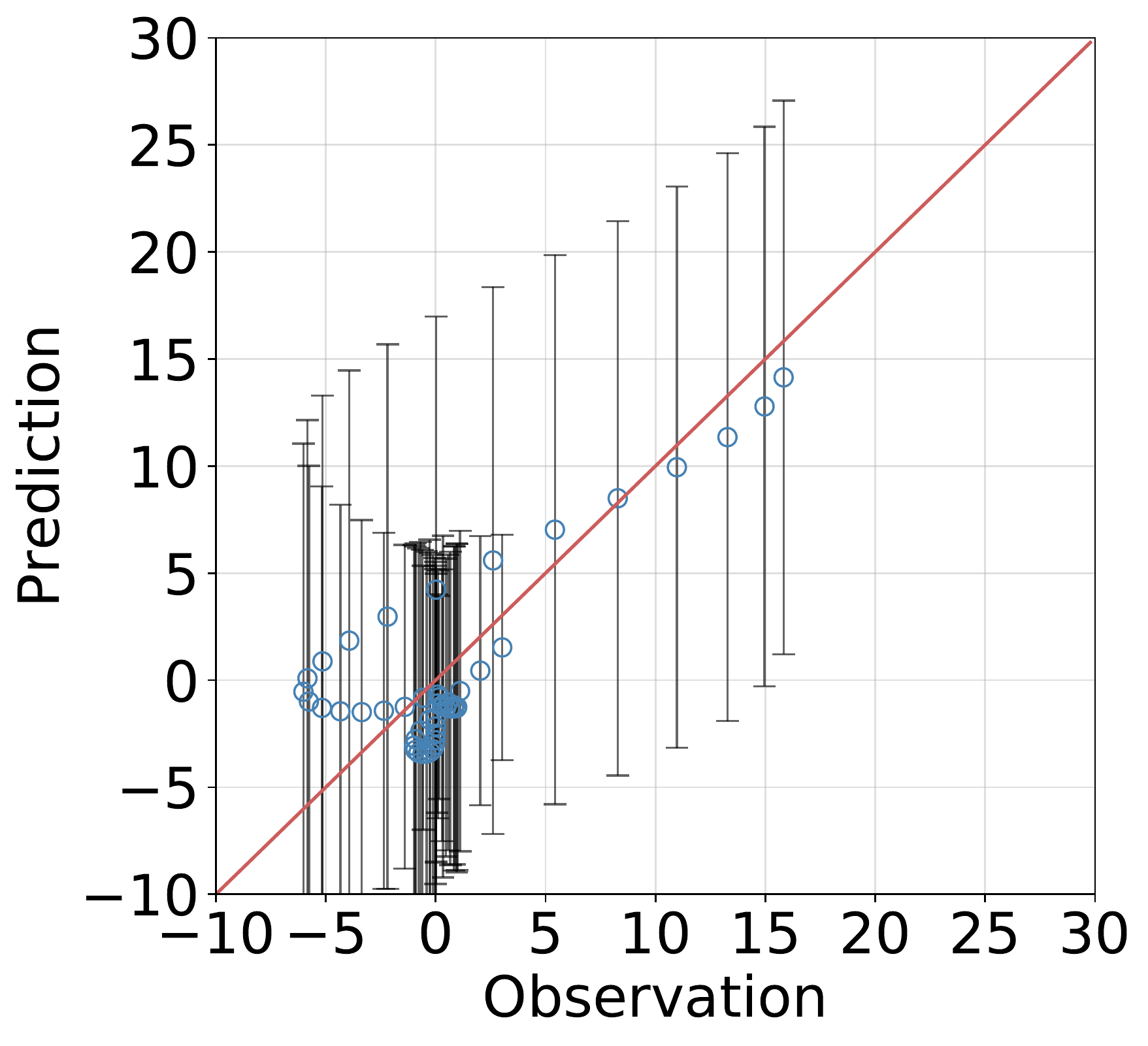} & \hspace{-0.5cm}
         \includegraphics[scale=0.22]{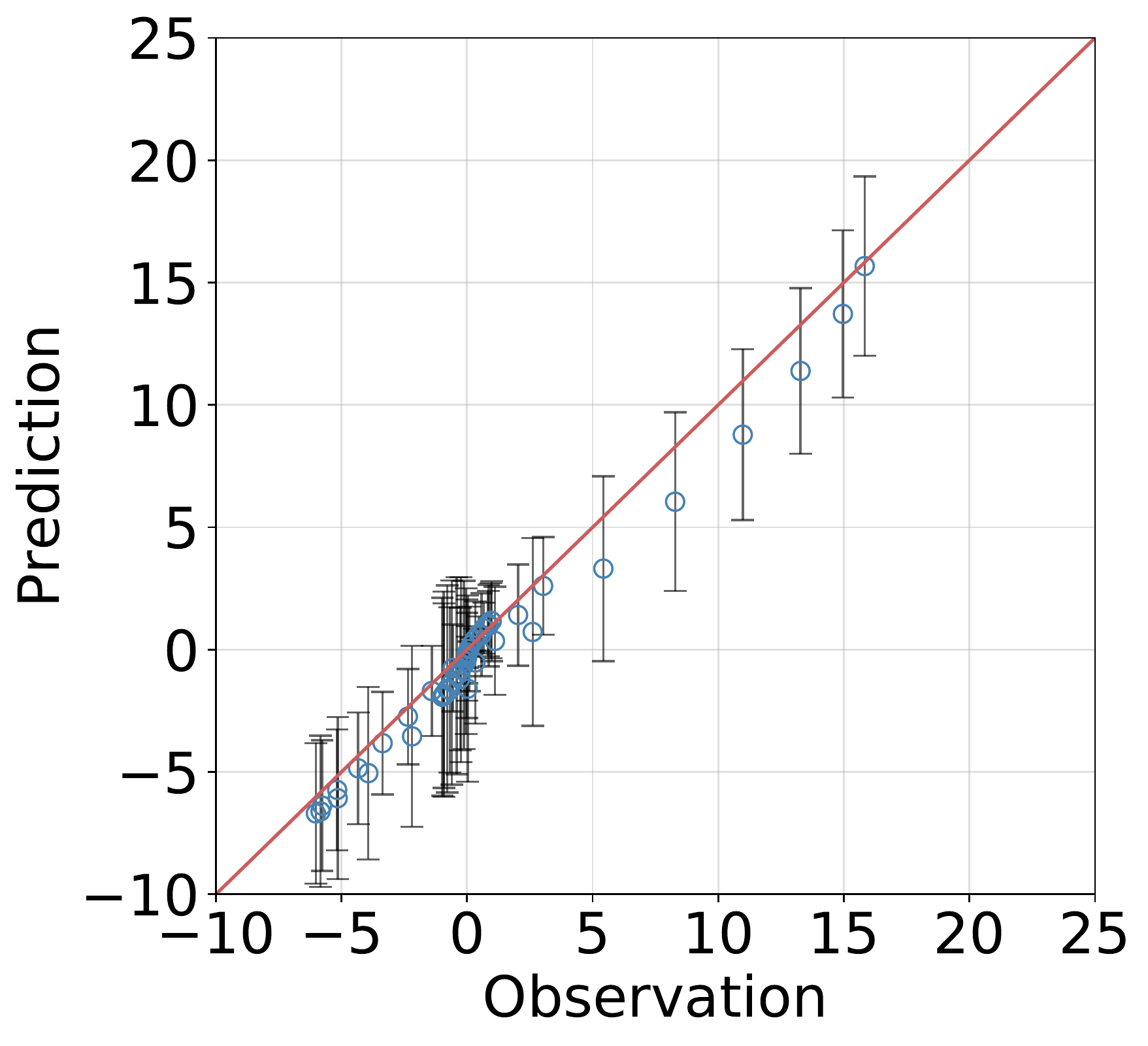} & \hspace{-0.5cm}
         \includegraphics[scale=0.22]{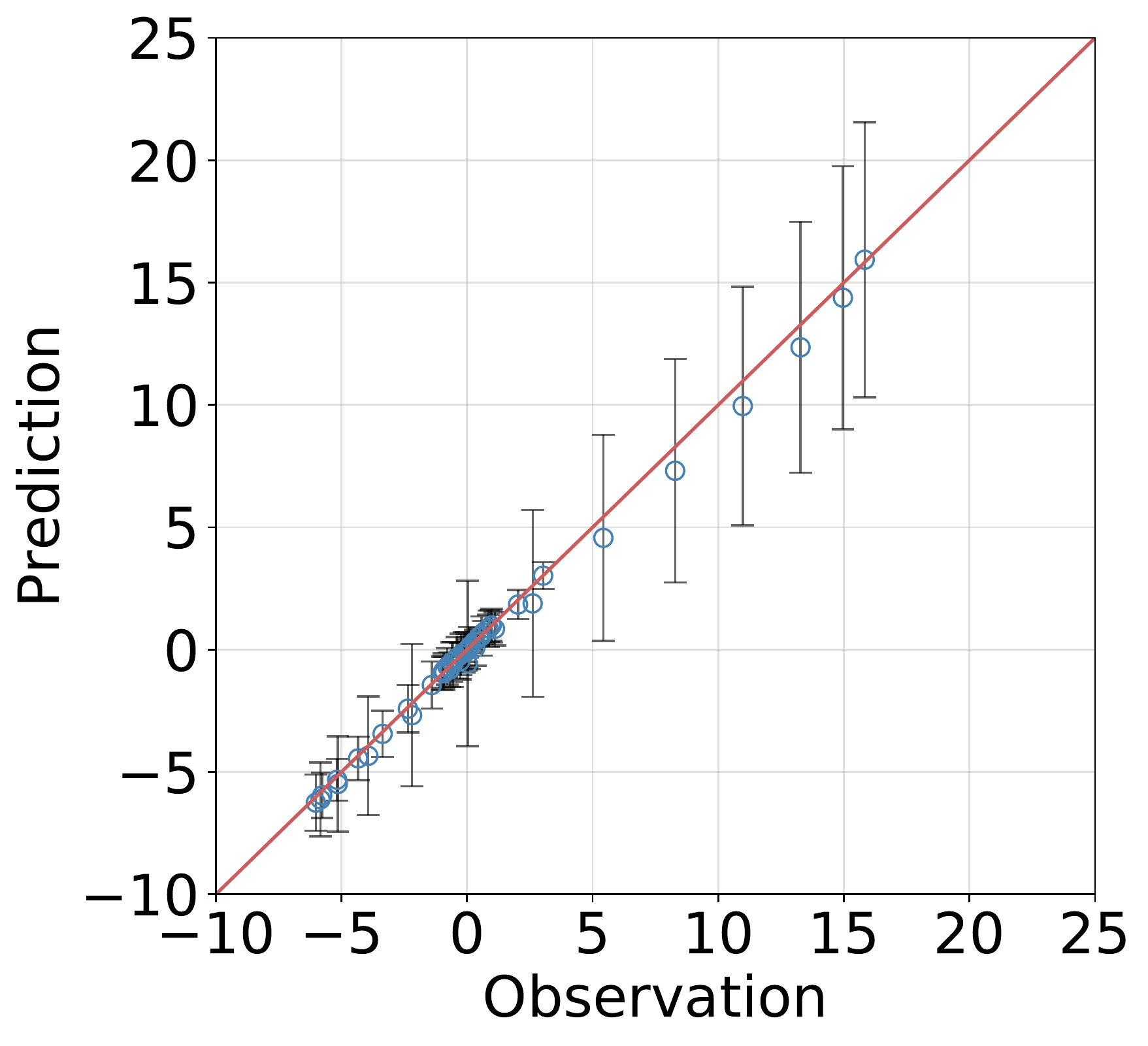} \\
         \includegraphics[scale=0.22]{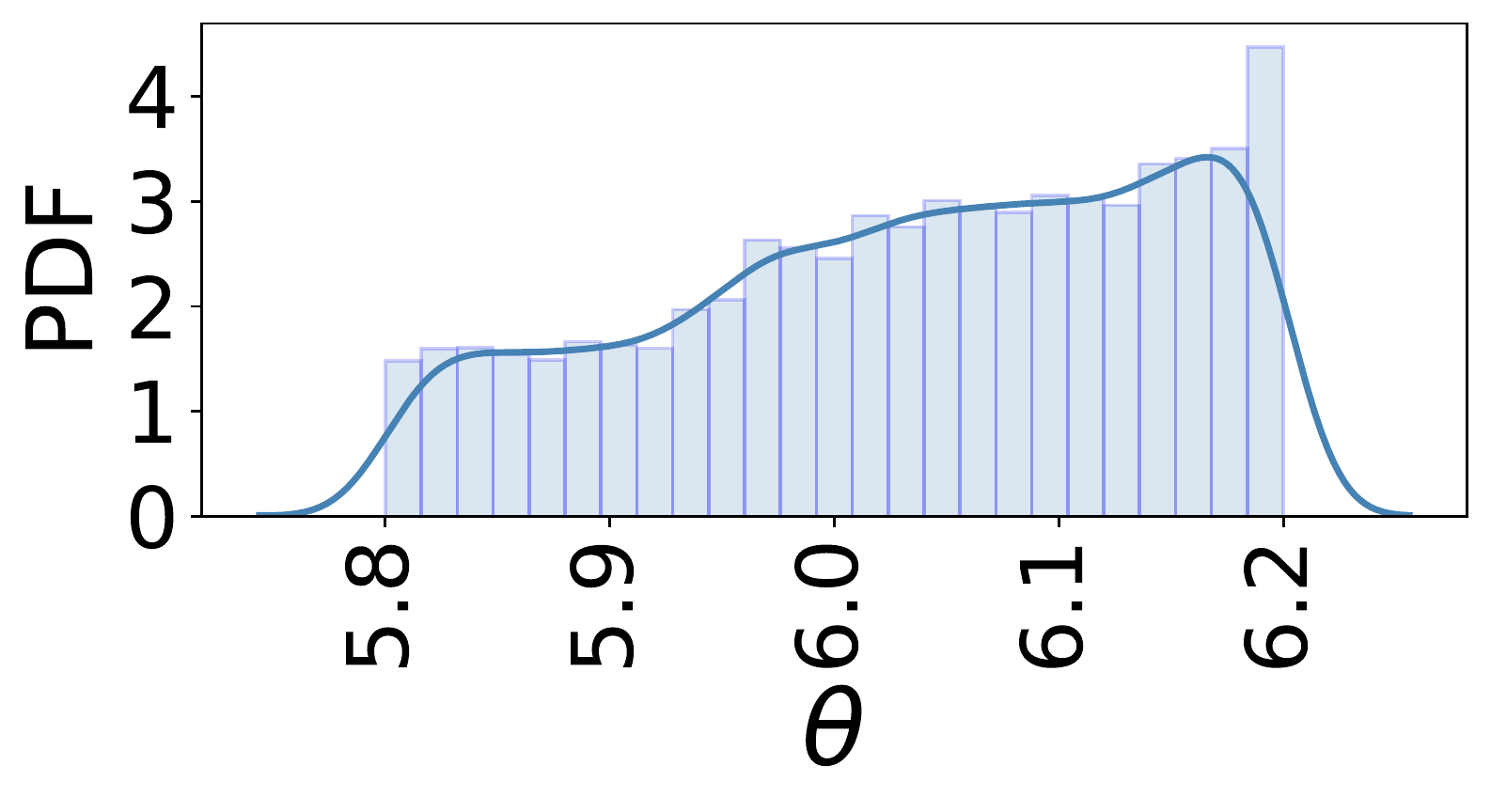} & \hspace{-0.5cm}
         \includegraphics[scale=0.22]{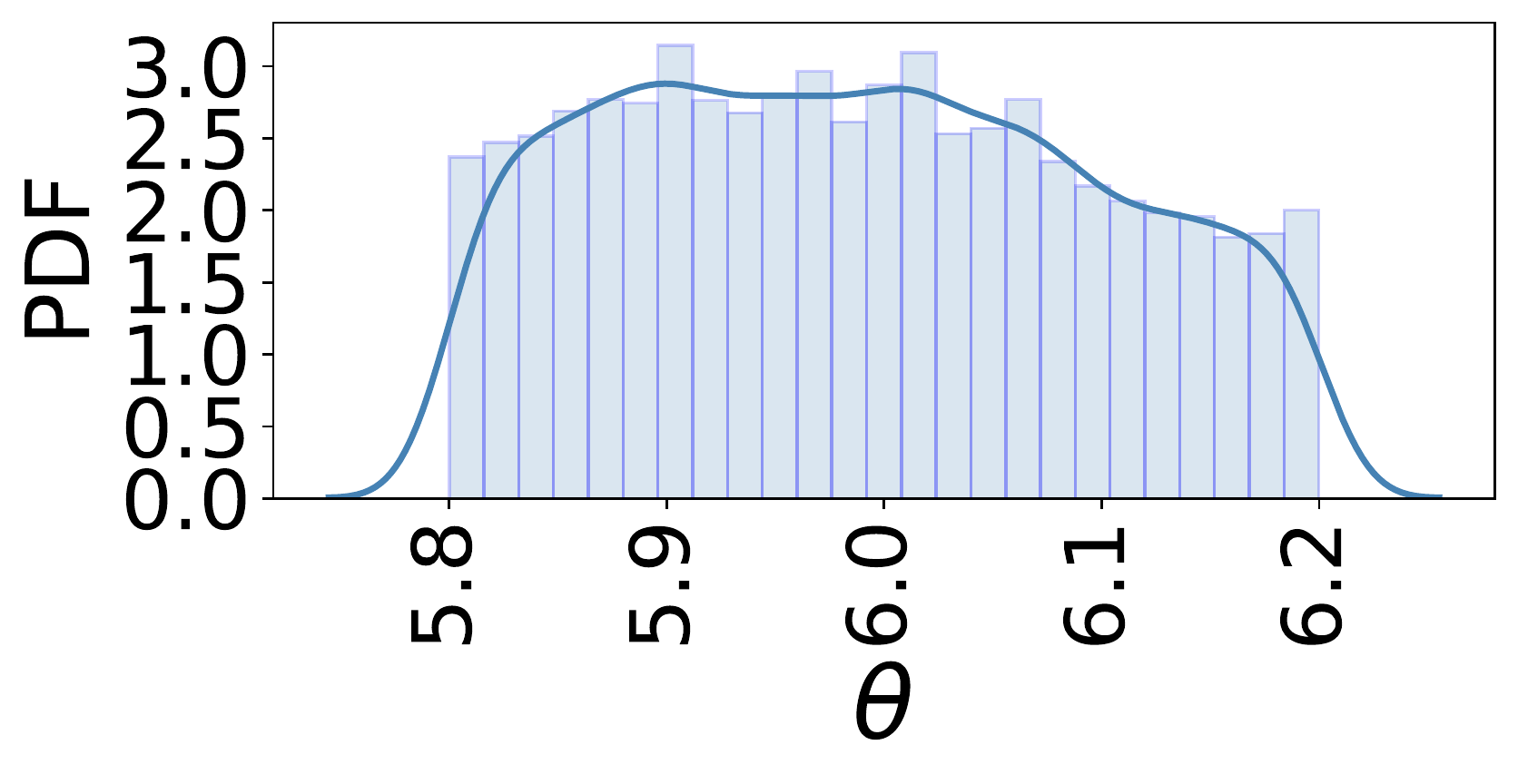} & \hspace{-0.5cm}
         \includegraphics[scale=0.22]{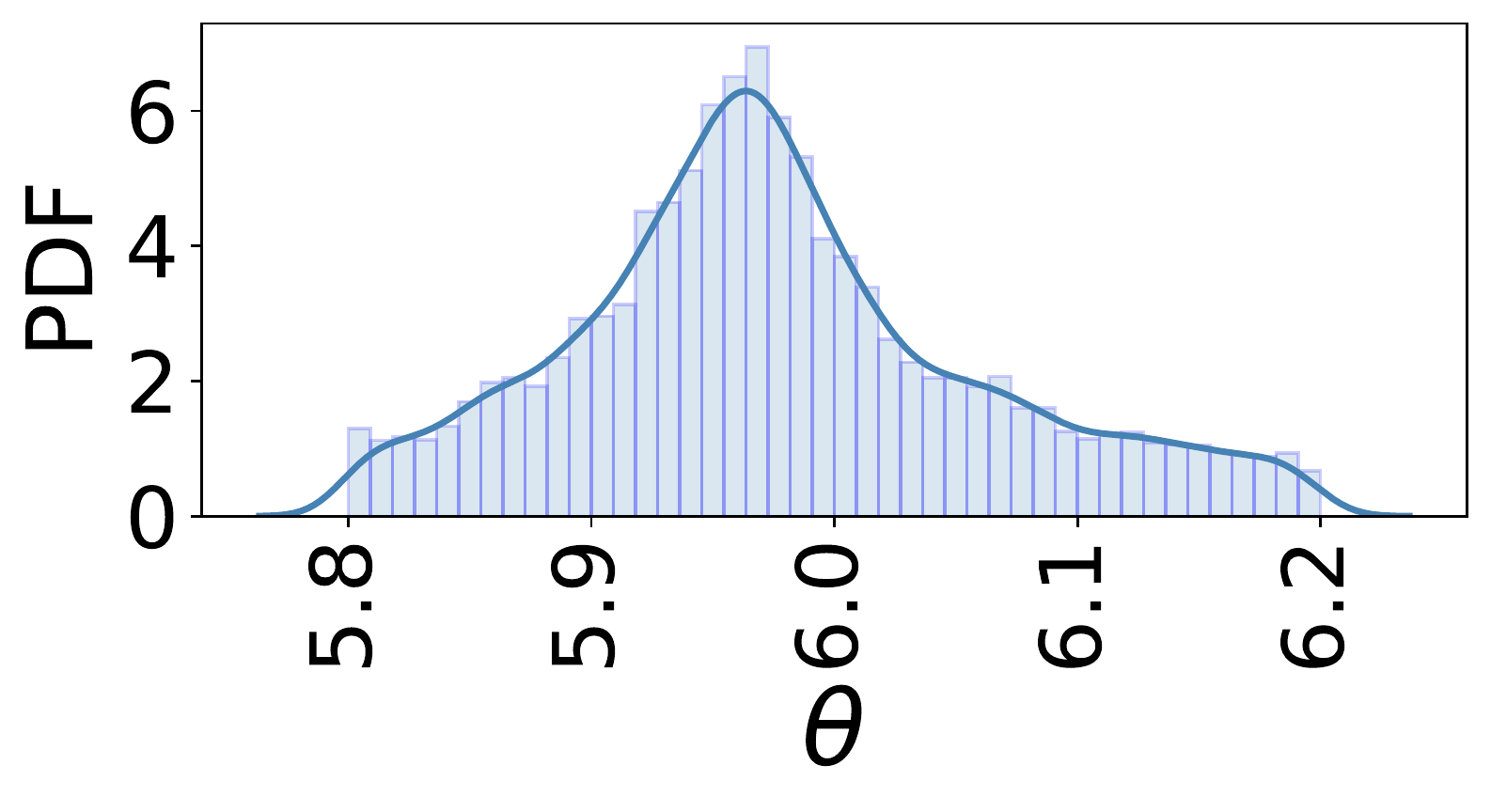} \\
    \end{tabular}
    \caption{(Top) Predicted QoI~$y$ by HC-MFM~(\ref{eq:mfModel}) along with the training and true data, (Middle) predicted~$y$ versus true observations at $50$ test samples of $x\in[0,1]$ with error bars representing $95\%$ CI, (Bottom) posterior probability density function (PDF) of~$\theta$ based on $10^4$ MCMC samples.
    The $y_H$ and $y_L$ training data are generated from \eq~(\ref{eq:ex1}) using $B=C=10$. The training data includes $4$ HF samples combined with (left column) $10$, (middle column) $15$, and (right column) $20$ LF samples. The true data is generated by \eq~(\ref{eq:ex1}) using $\theta=6$. }
    \label{fig:ex1}
\end{figure}

Using the data, the HC-MFM~(\ref{eq:mfModel}) for problem~(\ref{eq:ex1}) is constructed.
The first row in \fig~\ref{fig:ex1} shows the predicted~$y$ with the associated $95\%$ confidence interval (CI) along with the training data and reference true data generated with $\theta=6$.
For all~$n_L$, the predicted~$y$ is closer to HF data than the LF data, however, for $n_L=15$ and $20$, the agreement between the mean of the predicted~$y$ and the true data is significantly improved.
A better validation can be made via the plots in the second row of \fig~\ref{fig:ex1}, where the predicted~$y$ and true values of~$y_H$ at~$50$ uniformly-spaced test samples for $x\in[0,1]$ are plotted.
Clearly, increasing the number of the LF samples while keeping $n_H=4$ fixed, improves the predictions and reduces the uncertainty.
In the third row of \fig~\ref{fig:ex1}, the posterior densities of~$\theta$ are presented.
In all cases, a uniform (non-informative) prior distribution over $[5.8,6.2]$ was considered for $\theta$.
Only for~$n_L=20$, the resulting posterior density of~$\theta$ is high near the true value~$6$.
Therefore, it is confirmed that, as explained by Goh~\et~\cite{goh:13} the main capability of the HC-MFM~(\ref{eq:mfModel}) is in making accurate predictions for~$y$ and only if a sufficient number of training data is available, accurate distributions for the calibration parameters are also obtained.
This is shown here by fixing $n_H$ and increasing $n_L$, which is favorable in practice.
It is also noteworthy that if $\theta$ was known and hence treated as a fixed parameter, then even with $n_L=10$ very accurate predictions for~$y$ could be already achieved (not shown here).

It also should be noted that the mean of the posterior distribution of~$\sigma$, the noise standard deviation, is found to be negligible, as expected. 
This is in fact the case for all other examples in this section.

\subsection{Turbulent Channel Flow}
Turbulent channel flow is one of the most canonical wall-bounded turbulent flows.
The flow develops between two parallel flat walls which are apart by the distance~$2\delta$, and the flow is periodic in the streamwise and spanwise directions.
Channel flow is fully defined by a Reynolds number, for instance the bulk Reynolds number $\rey_b=U_b\delta/\nu$, where~$U_b$ and~$\nu$ specify the streamwise bulk velocity and kinematic viscosity, respectively.
Among different possible QoIs, here we only focus on the friction velocity~$\lut$, as a function of Reynolds number.
This quantity is defined as~$\sqrt{\langle \tau_w \rangle/\rho}$, where~$\tau_w$ and~$\rho$ are the magnitude of the wall-shear stress and fluid density, respectively, and $\langle \cdot \rangle$ represents averaging over time and the periodic directions.
Three fidelity levels are considered: DNS~($M_1$), WRLES~($M_2$), and a reduced-order algebraic model~($M_3$), where the fidelity reduces from the former to the latter.
We use the DNS data of \rfs~\cite{iwamoto:02,lee-moser:15,yamamoto:18}.
The WRLES of channel flow have been performed at different Reynolds numbers without any explicit subgrid-scale model using OpenFOAM~\cite{weller:98} which is an open-source finite-volume flow solver.
For the details of simulations see \rf~\cite{salehPoF:18}, where it was shown that for a fixed resolution in the wall-normal direction, variation of the grid resolutions in the wall-parallel directions could significantly impact the accuracy of the flow QoIs.
Therefore, in the context of the HC-MFM, the calibration parameters for WRLES are taken to be~$\dxp$ and~$\dzp$, which are the cell dimensions in the streamwise and spanwise directions, respectively, expressed in wall-units ($\dxp=\Delta x \,u^\circ_\tau/\nu$ where~$u^\circ_\tau$ is the reference~$u_\tau$ from DNS).
At the lowest fidelity, the following reduced-order algebraic model is considered which is derived by averaging the streamwise momentum equation for the channel flow in the periodic directions and time,
\begin{equation}\label{eq:uTauModel}
    \lut^2/U_b^2=\frac{1}{\rey_b} \frac{\dd}{\dd \eta} \left( (1+\zeta(\eta)) \frac{\dd\lu/U_b}{\dd \eta} \right) \,,
\end{equation}
where~$\eta$ is the distance from the wall normalized by the channel half-height~$\delta$, and~$\zeta(\eta)$ is the normalized eddy viscosity~$\nu_t$.
Reynolds and Tiederman~\cite{rt:67} proposed the following closed form for~$\zeta(\eta)$,
\begin{eqnarray}\label{eq:RT}
   \zeta(\eta)=\frac{\nu_t(\eta)}{\nu}=\frac{1}{2}\left[1+\frac{{\kappa}^2 \rey^2_\tau}{9}\left(1-    (\eta-1)^2\right)^2\left(1+2(\eta-1)^2\right)^2\left(1-\exp(\frac{-\eta \rey_\tau}{A^+})  \right)^2\right]^{1/2}-\frac{1}{2} \,,
\end{eqnarray}
where $\reyt=\lut\delta/\nu$ is the friction-based Reynolds number, and~$\kappa$ and~$A^+$ are two modeling parameters.
At any~$\reyb$ (and given value of~$\kappa$ and~$A^+$), \eq~(\ref{eq:uTauModel}) is integrated over $\eta\in[0,1]$ and is iteratively solved using~\eq~(\ref{eq:RT}) to estimate~$\lut$.
Expressing the channel flow example in the terminology of MFM~(\ref{eq:mfModel}),~$\lut/U_b$ is the QoI~$y$, $x=\rey_b$, $\ft_3=(\kappa,A^+)$, and $\ft_2=(\dxp,\dzp)$.
The training data set consists of the following databases.
For DNS,~$\lut$ is taken from \rfs~\cite{iwamoto:02,lee-moser:15,yamamoto:18} at $\rey_b=5020$, $6962$, $10000$, $20000$, $125000$ and $200400$.
In total,~$16$ WRLES~$\lut$ samples are obtained from a design of experiment based on the prior distributions $\dxp\sim\cU[15,85]$ and $\dzp\sim\cU[9.5,22]$ at $\reyb=5020,6962,10000$, and~$20000$.
Here, we do not consider the observational uncertainty in~$\lut$ which could, for instance, be due to finite time-averaging in DNS and WRLES, but in general the HC-MFM could take such information into account.
The reduced-order model~(\ref{eq:uTauModel}) which is computationally cheap is run for~$10$ values of~$\reyb$ in range~$[2000,200200]$.
For each~$\reyb$,~$9$ joint samples of~$(\kappa,A^+)$ are generated assuming $\kappa\sim\cU[0.36,0.43]$ and $A^+\sim\cU[26.5,29]$ (note that~$\kappa$ is the von K{\'a}rm{\'a}n coefficient).
For all the Gaussian processes in the MFM~(\ref{eq:mfModel}), the exponentiated quadratic covariance function~(\ref{eq:exponKernel}) is used. 
The prior distribution of the hyperparameters are set as the following: $\lambda_f, \lambda_g \sim \mathcal{HC}(\alpha=5)$, $\lambda_{\delta} \sim \mathcal{HC}(\alpha=3)$, $\ell_{f_x}, \ell_{f_{t_3}},\ell_{g_x}, \ell_{g_{t_2}}, \ell_{\delta_x} \sim \Gamma(\alpha=1,\beta=5)$, and the noise standard deviation~$\sigma\sim \mathcal{HC}(\alpha=5)$ (assumed to be the same for all fidelities). 

\begin{figure}[!t]
    \centering
    \begin{tabular}{cccc}
      {\small (a)} &  \hspace{-0.4cm}
       \includegraphics[scale=0.37]{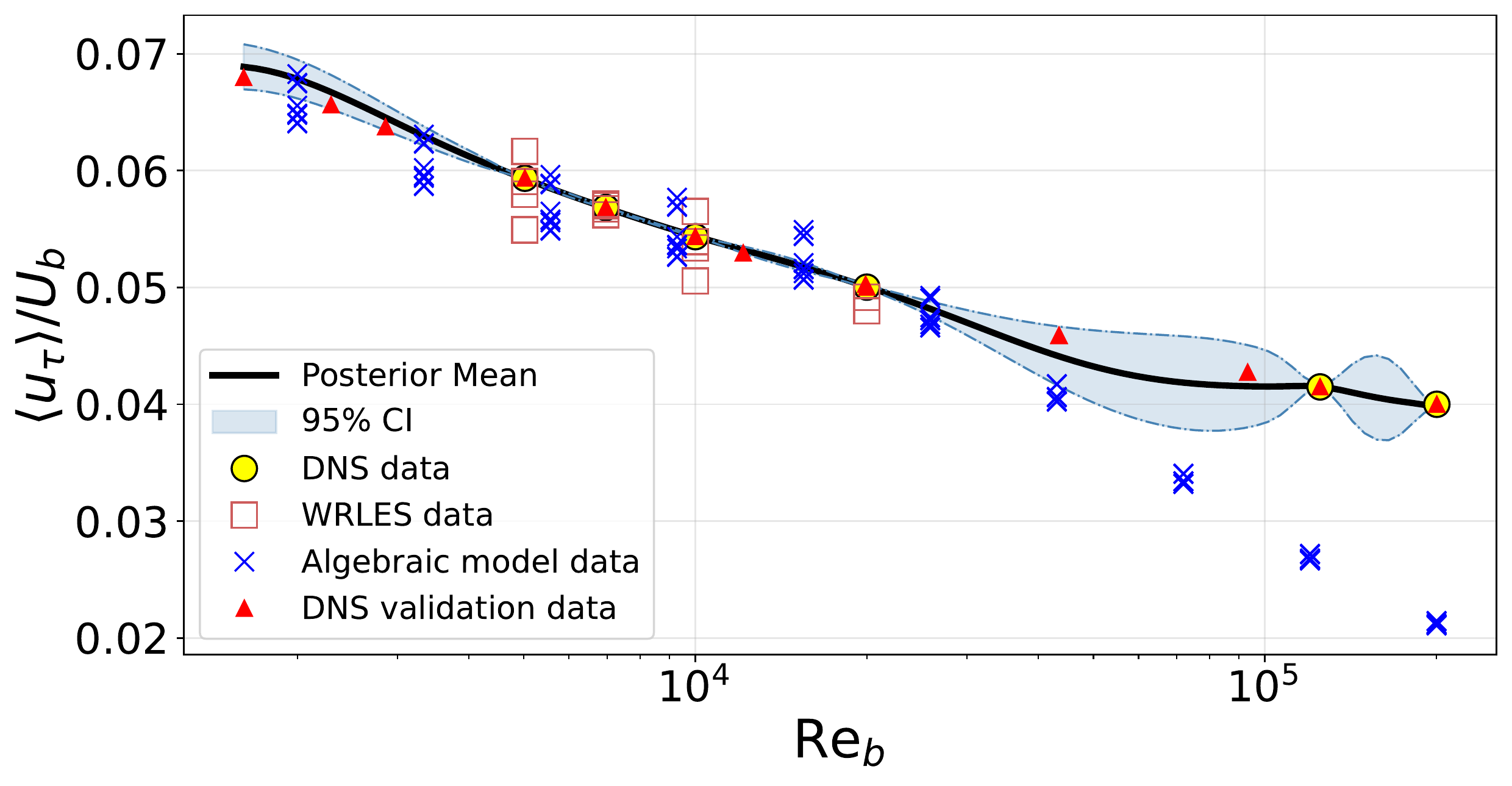} &
      {\small (b)} & \hspace{-0.4cm}
       \includegraphics[scale=0.35]{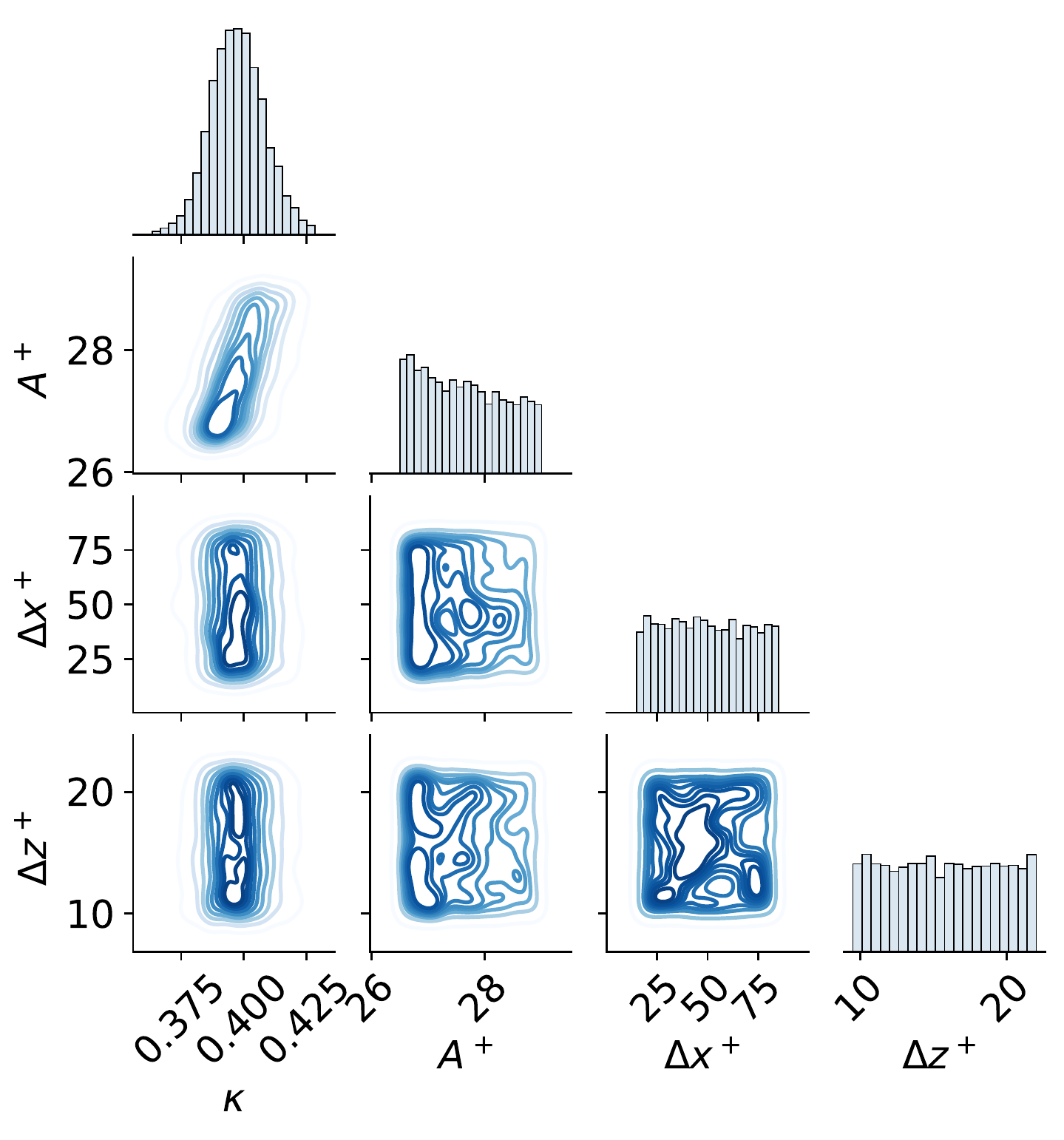} \\
    \end{tabular}
    \caption{(a) Mean prediction of $\lut/U_b$ and associated $95\%$ CI along with the training data and validation data from DNS of \rfs~\cite{iwamoto:02,lee-moser:15,yamamoto:18}, (b) diagonal: posterior density of parameters~$\kappa$,~$A^+$,~$\dxp$, and~$\dzp$; off-diagonal: contour lines of the joint posterior densities of these parameters. The value of the contour lines increases from the lightest to darkest color.}
    \label{fig:chan}
\end{figure}

Using the described training data in the HC-MFM~(\ref{eq:mfModel}) and running the MCMC chain for~$7000$ samples, after an initial~$2000$ samples discarded due to burn-in, the model is constructed.
According to \fig~\ref{fig:chan}(a), the predicted mean of~$\lut/U_b$ follows the trend of the DNS data.
This approximately holds even at high Reynolds numbers, where there is a large systematic error in the algebraic model and no WRLES data is available.
As expected, in this range due to scarcity of the DNS data, the uncertainty in the predictions is high.
The plot in \fig~\ref{fig:chan}(b) shows the joint posterior distributions of the calibration parameters~$\kappa$,~$A^+$,~$\dxp$, and~$\dzp$ along with the histogram of each parameter.
As mentioned above, the prior distributions of all of these parameters were assumed to be uniform and mutually independent.
However, the resulting posterior densities for~$\kappa$ and~$A^+$ are not uniform and the samples of these parameters are correlated.
More interestingly, the peak of the posterior density of~$\kappa$ is close to the value of~$0.4$ that is assumed to be universal across various flows and Reynolds numbers within the turbulence community. 
In contrast, the posteriors of~$\dxp$ and~$\dzp$ are found to be still close to the prior uniform distributions and no correlation between their samples is observed.
This may be at least partially be due to the fact that the number of the WRLES data is limited as they are obtained only at~4 Reynolds numbers.
Nevertheless, over this range of the Reynolds number the posterior prediction of the QoI~$\lut$ is very accurate and has the lowest uncertainty, which seems to indicate that the grid spacing indeed does not lead to a bias for the shear stress computation. This may in fact be an important aspect for building future wall models.

\subsection{Polars for the NACA0015 Airfoil}\label{sec:airfoil}
In this section, the HC-MFM~(\ref{eq:mfModel}) is applied to a set of data for the lift and drag coefficients,~$C_L$ and~$C_D$, respectively, of a wing with a NACA0015 airfoil profile at Reynolds number~$1.6\times 10^6$.
The  angle of attack (AoA) between the wing and the ambient flow is taken to be the design parameter~$x$.
The data comprises of the following sources with respective fidelities in brackets: wind-tunnel experiments by Bertagnolio~\cite{bertagnolio:08}~($M_1$), detached-eddy simulations (DES)~($M2$) and two-dimensional RANS~($M_3$) both by Gilling \et~\cite{gilling:09}.
In their numerical study, Gilling~\et~\cite{gilling:09} investigated the  sensitivity of the DES results to the resolved turbulence intensity (TI) of the fluctuations imposed at the inlet boundary.
The sensitivity was found to be particularly significant near the stall angle.
Therefore, when constructing an MFM, the calibration parameter~$t_2$ in fidelity~$M_2$ is taken to be the~TI.

For the covariance of the Gaussian processes~$\hat{f}(x)$ and~$\hat{g}(x,t_2)$ in HC-MFM~(\ref{eq:mfModel}), the exponentiated quadratic and {Matern-5/2} kernel functions (\ref{eq:exponKernel}) and (\ref{eq:mat52Kernel}) are used, respectively.
The following prior distributions are assumed for the  hyperparameters: 
$\lambda_f , \lambda_g \sim \mathcal{HC}(\alpha=5)$, $\ell_{f_x}\sim\Gamma(\alpha=1,\beta=5)$, 
$\ell_{g_x},\ell_{g_{t_2}}\sim\Gamma(\alpha=1,\beta=3)$, and the noise standard deviation $\sigma\sim \mathcal{HC}(\alpha=5)$ (assumed to be the same for all fidelities). 
To make the model capable of capturing large-scale separation, the stall AoA,~$x_\sta$, is included as a new calibration parameter in the MFM.
Our suggestion is to consider a piecewise kernel function for the covariance of~$\hat{\delta}(x)$ where~$x_\sta$ is the merging point.
If the kernel functions for the AoAs smaller and larger than~$x_\sta$ are denoted by~$k_{\delta_1}(\cdot)$ and~$k_{\delta_2}(\cdot)$, respectively, then the covariance function for~$\hat{\delta}(x)$ may be constructed as, 
\begin{equation}
    \Sigma_{\delta_{ij}} = 
    \lambda_{\delta_1}^2 \varphi(x_i) k_{\delta_1}(\bar{d}_{\delta_{ij}}) \varphi(x_j)
    +
    \lambda_{\delta_2}^2 \varphi(x_i) k_{\delta_2}(\bar{d}_{\delta_{ij}}) \varphi(x_j)    \,,
\end{equation}
where,~$\bar{d}_{\delta_{ij}}$ is defined similar to those in \eqs~(\ref{eq:euclid1}) and~(\ref{eq:euclid2}) but only in~$x$, and~$\varphi(x)$ is a function to smoothly merge the two covariance functions. 
In particular, we use the logistic function, 
\begin{equation}
    \varphi(x)=\left[1+\exp(-\alpha_\sta(x-x_\sta))\right]^{-1} \,,
\end{equation}
where $\alpha_\sta$ is a new hyperparameter.
The kernel functions~$k_{\delta_1}(\cdot)$ and~$k_{\delta_2}(\cdot)$ are both modeled by the {Matern-5/2} function~(\ref{eq:mat52Kernel}). 
As the prior distributions, we assume $\lambda_{\delta_1}\sim\mathcal{HC}(\alpha=3)$, $\lambda_{\delta_2}\sim\mathcal{HC}(\alpha=5)$, $\ell_{\delta_1}\sim \Gamma(\alpha=1,\beta=5)$, $\ell_{\delta_2}\sim \Gamma(\alpha=1,\beta=0.5)$, 
$\alpha_{\sta}\sim\mathcal{HC}(\alpha=2)$, and $x_\sta\sim\cN(14,0.2)$.

\begin{figure}[!t]
    \centering
    \begin{tabular}{cccc}
    {\small (a)} &
    \includegraphics[scale=0.32]{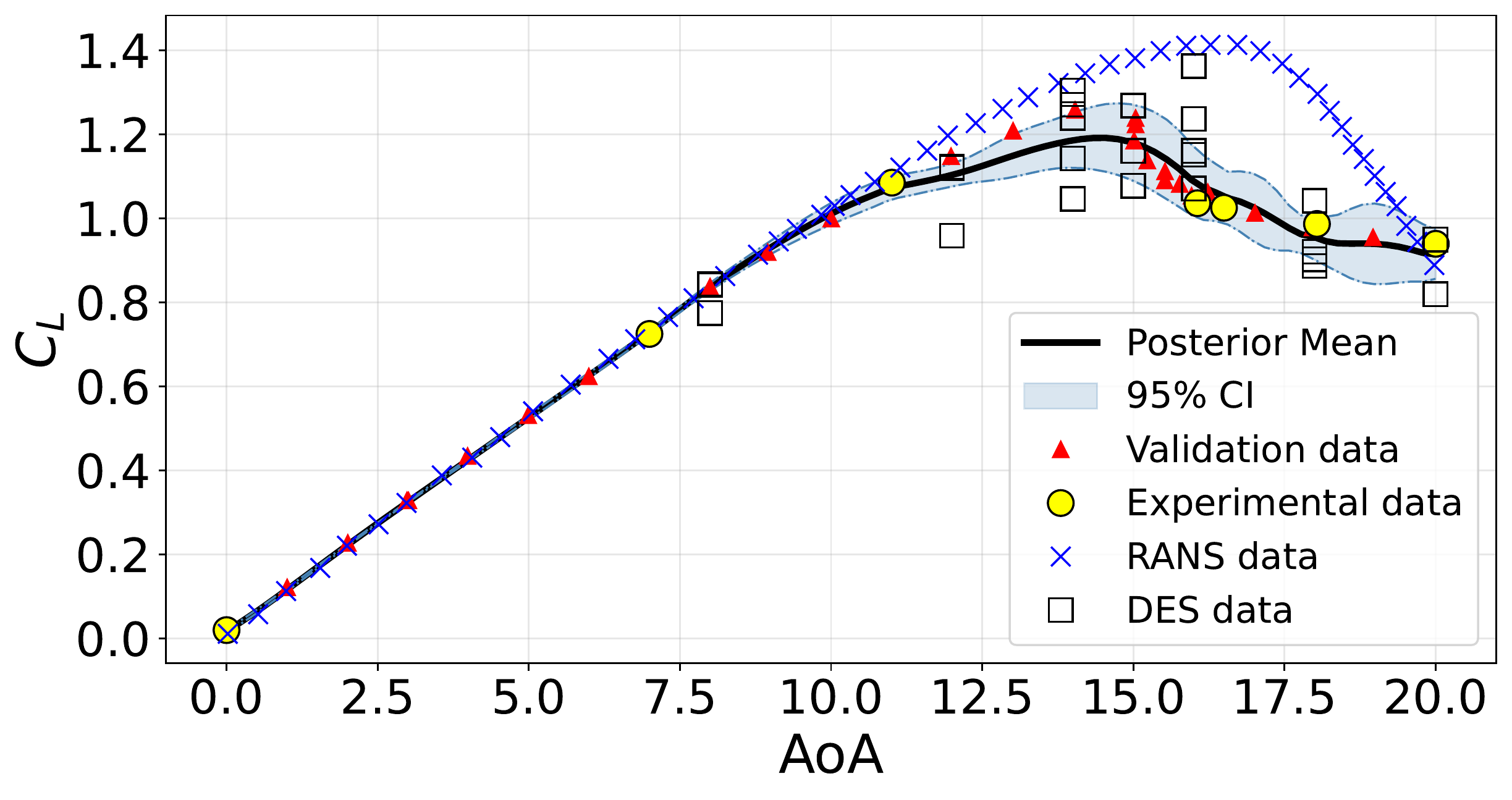} &
    {\small (b)} &
    \includegraphics[scale=0.3]{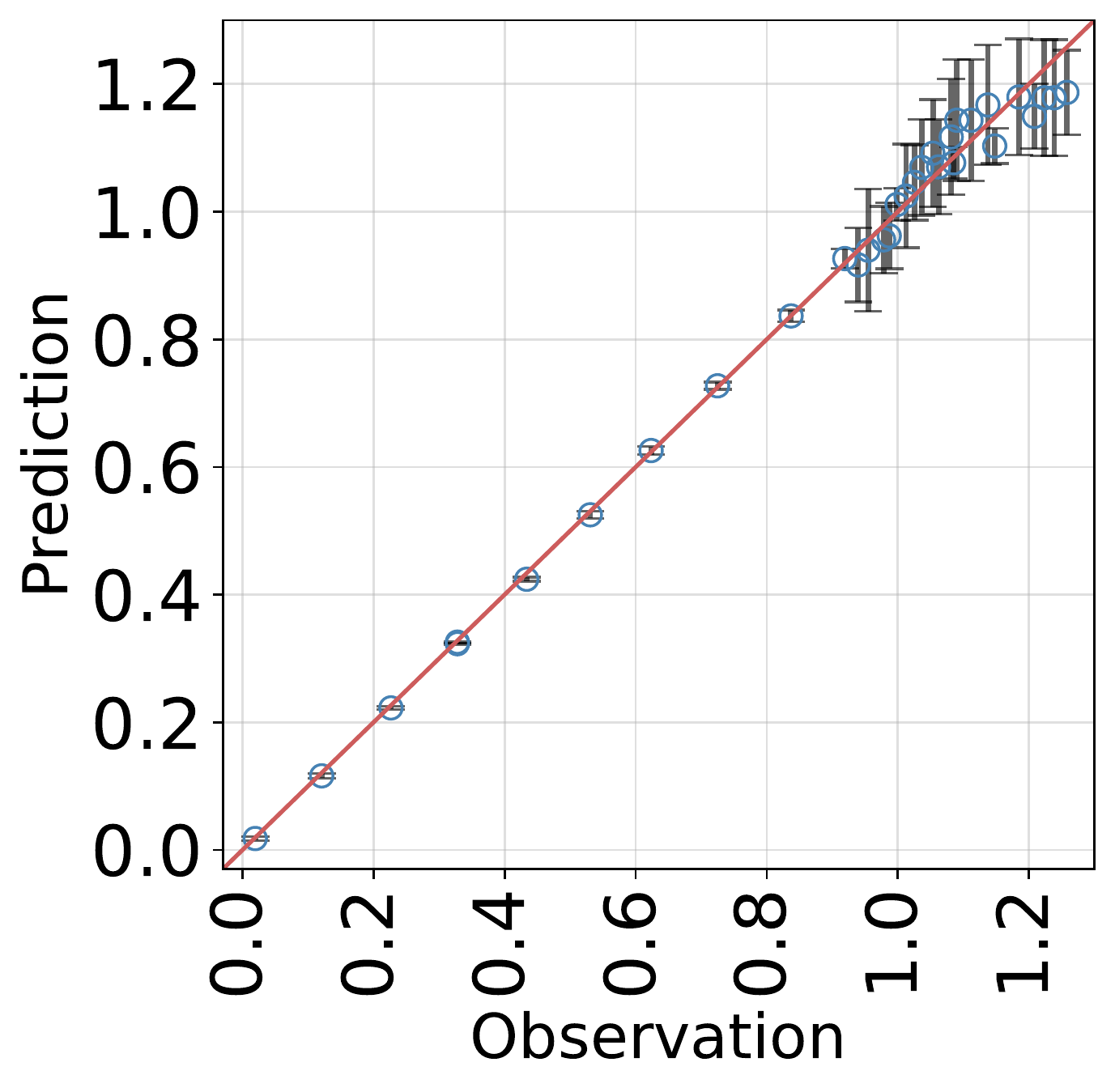} \\
    {\small (c)} &
    \includegraphics[scale=0.32]{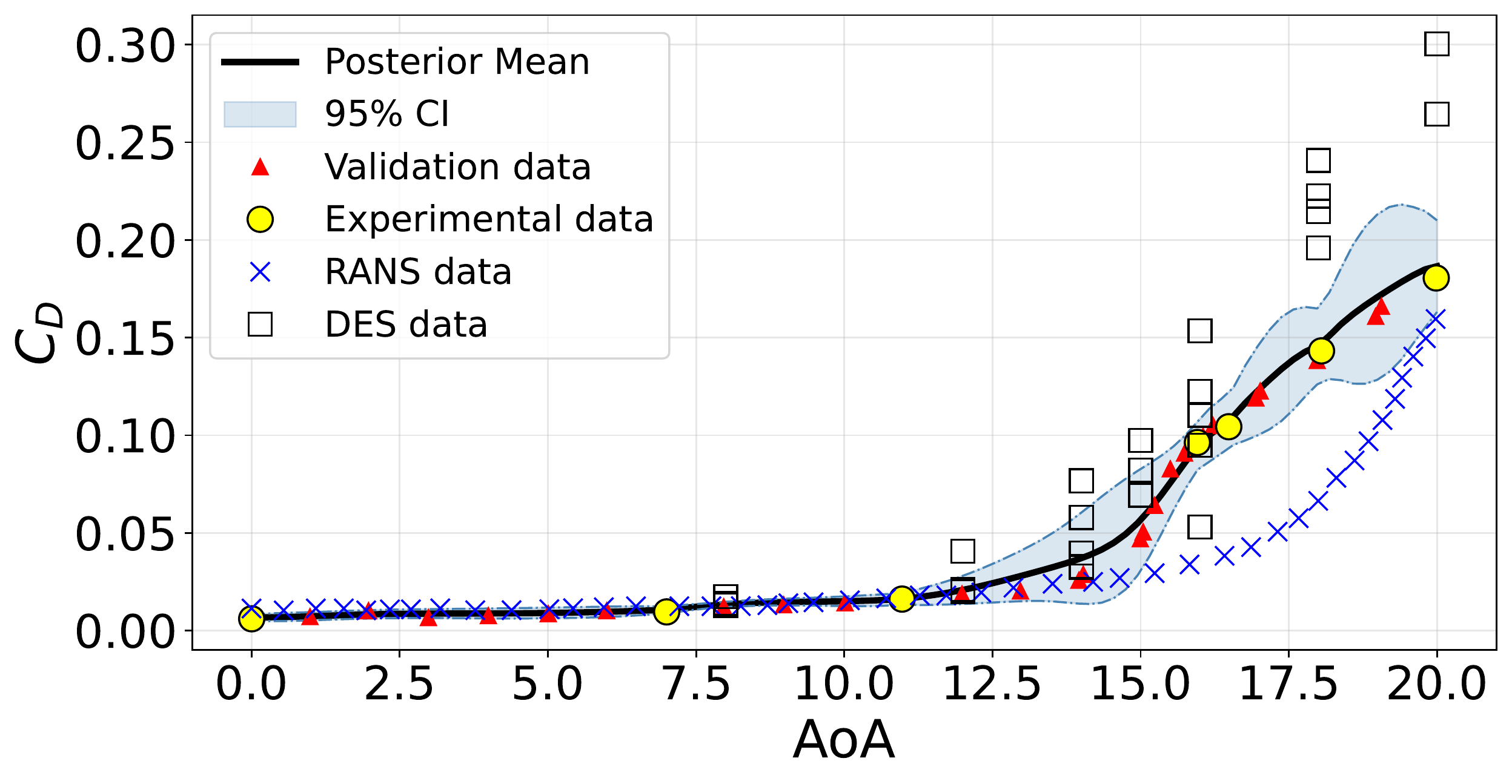} &
    {\small (d)} &
    \includegraphics[scale=0.3]{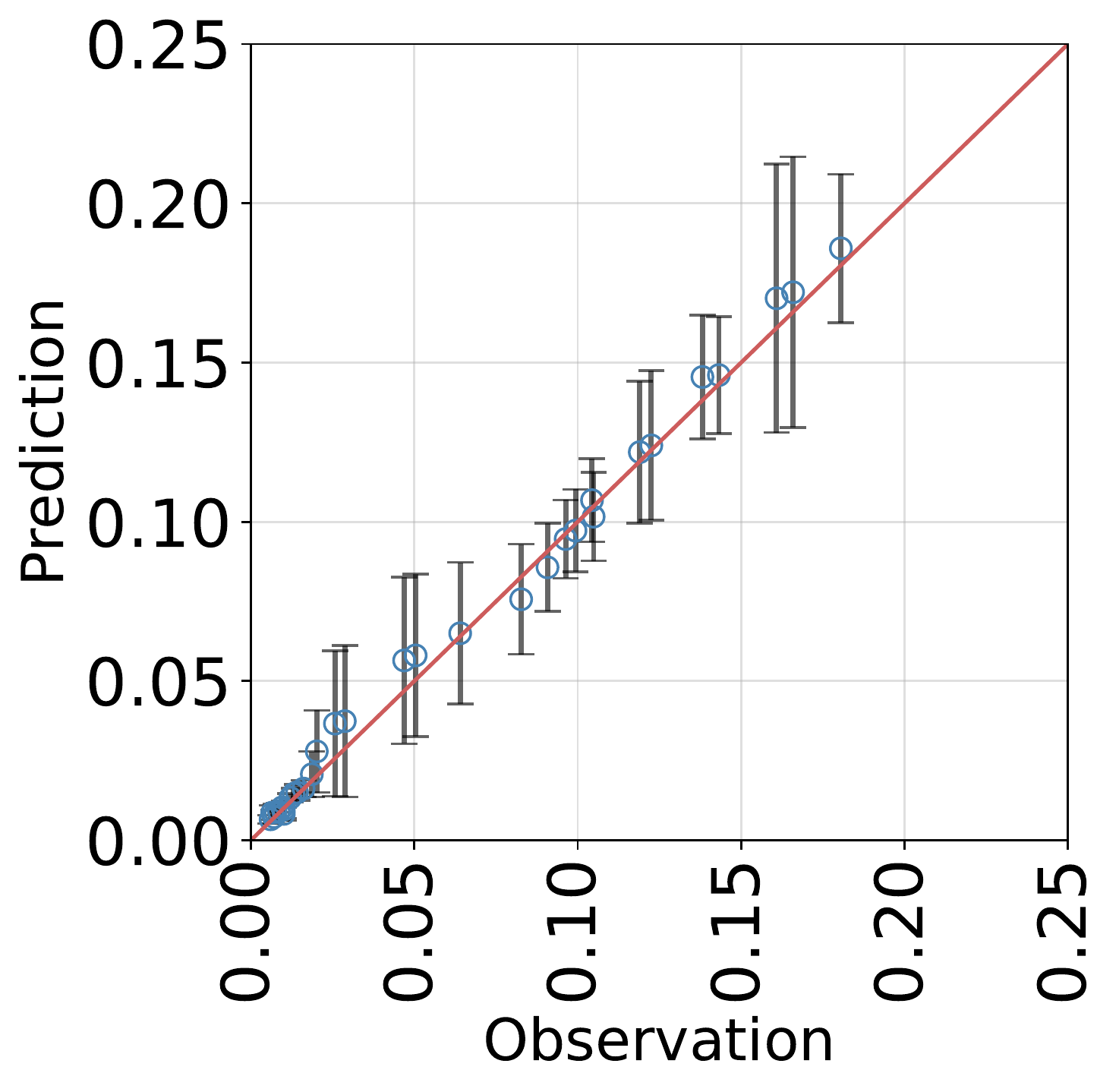} \\
    \end{tabular}
    \caption{(a) Lift coefficient $C_L$ and (c) drag coefficient $C_D$ plotted against the angle of attack (AoA):
    the HC-MFM~(\ref{eq:mfModel}) is trained by the experimental data of \rf~\cite{bertagnolio:08} (yellow circles), as well as the DES (squares) and RANS (crosses) data by Gilling \et~\cite{gilling:09}.
    The DES are performed in \rf~\cite{gilling:09} with the resolved turbulence intensities $\text{TI}=0\%,0.1\%,0.5\%,1\%$, and $2\%$ at the inlet.
    The validation data (red triangles) are also taken from the experiments of \rf~\cite{bertagnolio:08}.
    The mean prediction by the HC-MFM~(\ref{eq:mfModel}) is represented by the solid line along with associated $95\%$ confidence interval.
    (b) $C_L$, (d) $C_D$ predictions by HC-MFM plotted against the validation data. The error bars represent the $95\%$ CI.}
    \label{fig:cl_cd}
\end{figure}

The admissible range of~$x=\text{AoA}$ is assumed to be~$[0^\circ,20^\circ]$ over which the experimental and RANS data are available, see~\rfs~\cite{bertagnolio:08,gilling:09}.
The training HF data are taken to be a subset of size~$7$ of the experimental data of \rf~\cite{bertagnolio:08}.
The rest of the experimental data are used to validate the predictions of the MFM.
For the purpose of examining the capability of the MFM in a more challenging situation, the training HF samples are explicitly selected to exclude the range of AoAs where the stall happens.
The DES data of Gilling~\et~\cite{gilling:09} are available at~$7$  $\text{AoA}\in[8^\circ,19^\circ]$ and~$5$ different values of~TI.
Employing these training data in the HC-MFM~(\ref{eq:mfModel}) and drawing~$10^4$ MCMC samples after excluding an extra~$5000$ initial samples for burn-in, the predictions for~$C_L$ and~$C_D$ shown in \fig~\ref{fig:cl_cd}(a,c) are obtained.
The expected value of the predictions has a trend similar to that of the experimental validation data of \rf~\cite{bertagnolio:08} and is not diverging towards either the physically-invalid RANS data or scattered DES data at $\text{AoA}\gtrsim 10^\circ$.
A more elaborate comparison is made through scatter plot of the MFM predictions against the validation data in \fig~\ref{fig:cl_cd}(b,d).
For both~$C_L$ and~$C_D$, the agreement between the predicted mean values with the validation data at lower AoAs (before stall) is excellent and for most of the higher AoAs, even near and after the stall, is very good.
Due to the scarcity of the HF training data and also systematic error in the RANS and DES data, the error bars at the predicted values can be relatively large, as more evident in the~case~of~$C_D$~in~\fig~\ref{fig:cl_cd}(d).

\begin{figure}[!h]
    \centering
    \begin{tabular}{ccccc}
    \vspace{-0.2cm}
       {} &
       \includegraphics[scale=0.22]{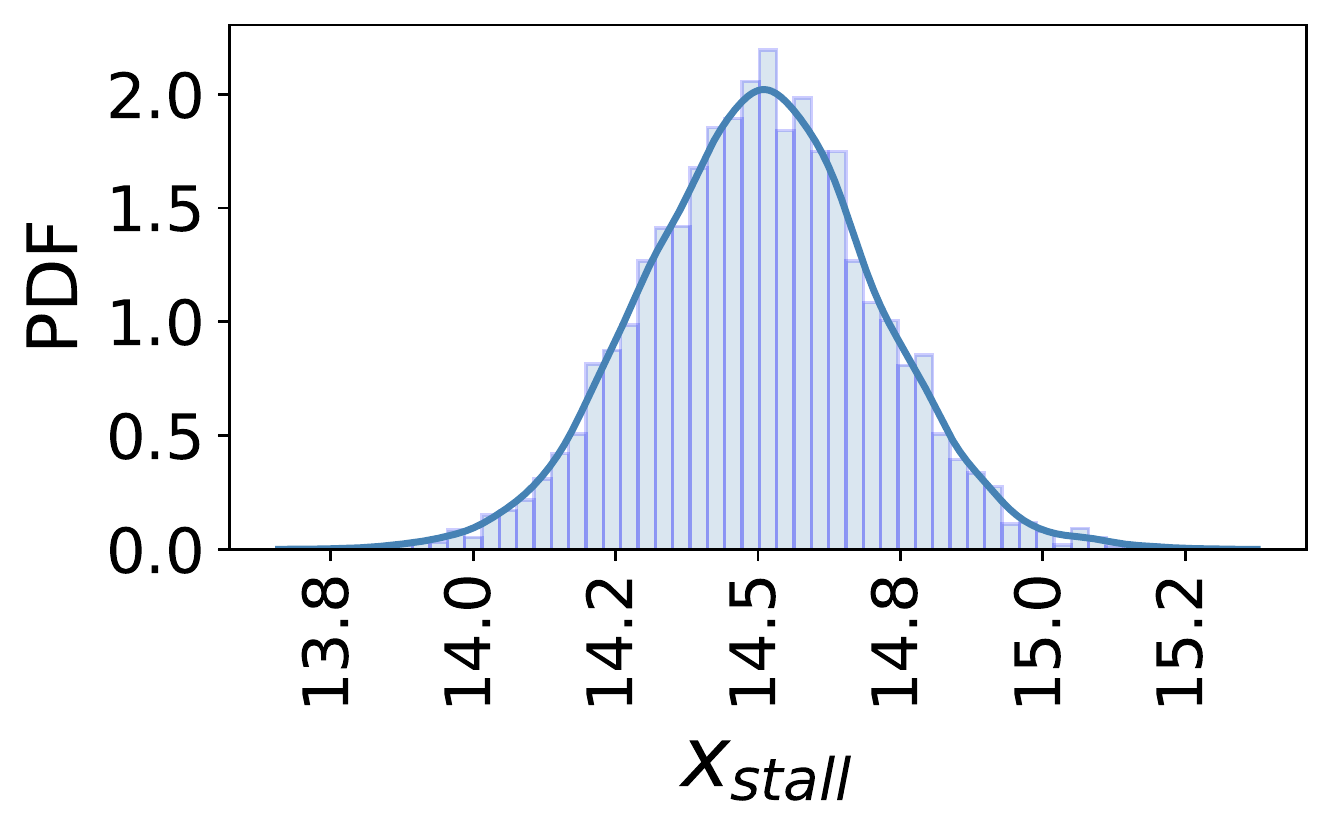}    & \hspace{-0.3cm}
       \includegraphics[scale=0.22]{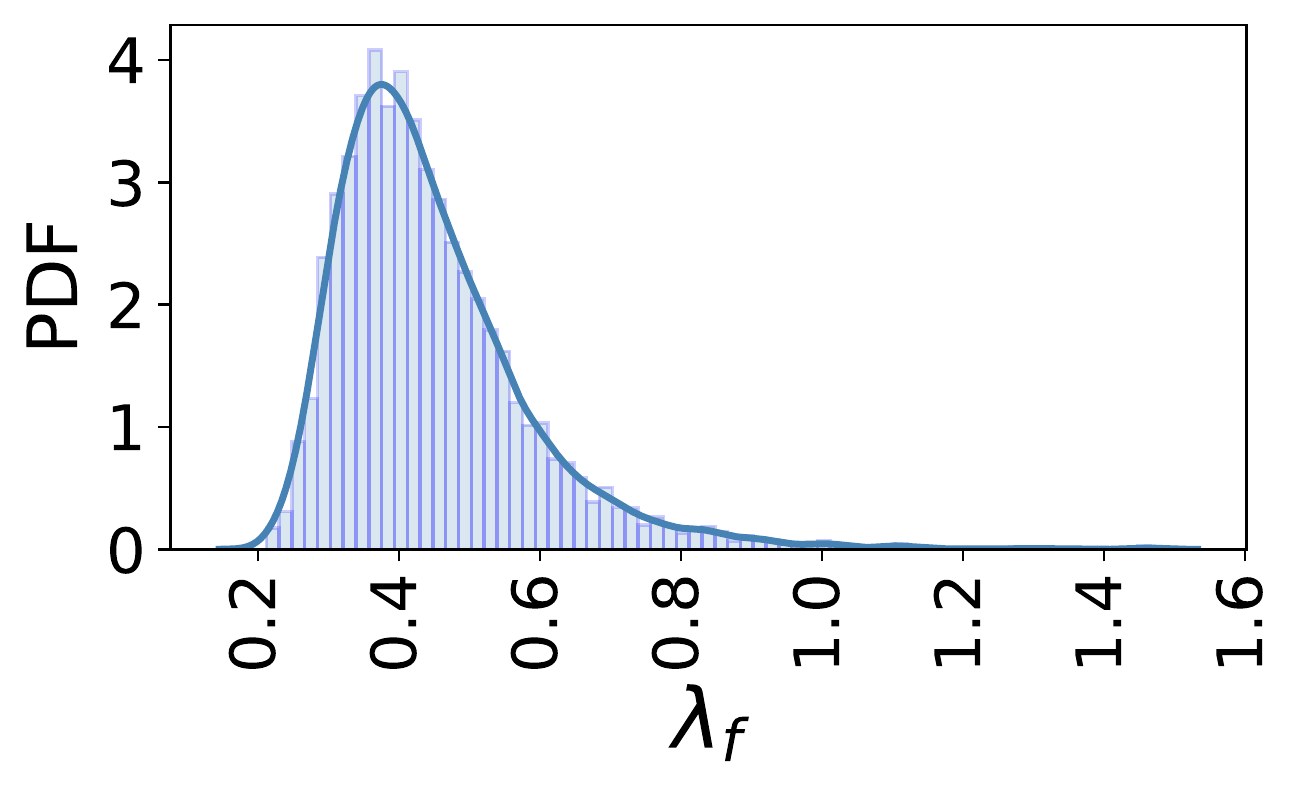}     & \hspace{-0.3cm}
       \includegraphics[scale=0.22]{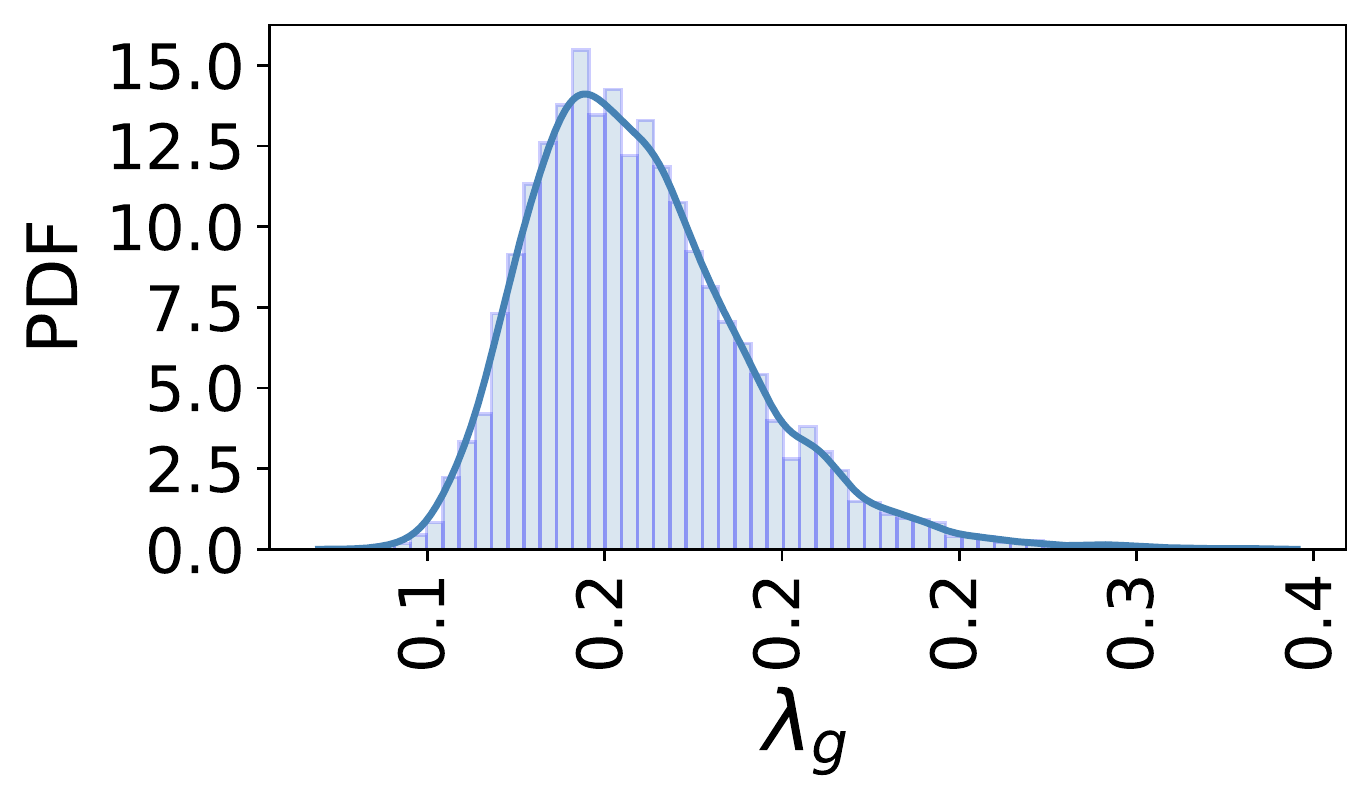}& \hspace{-0.3cm}
       \includegraphics[scale=0.22]{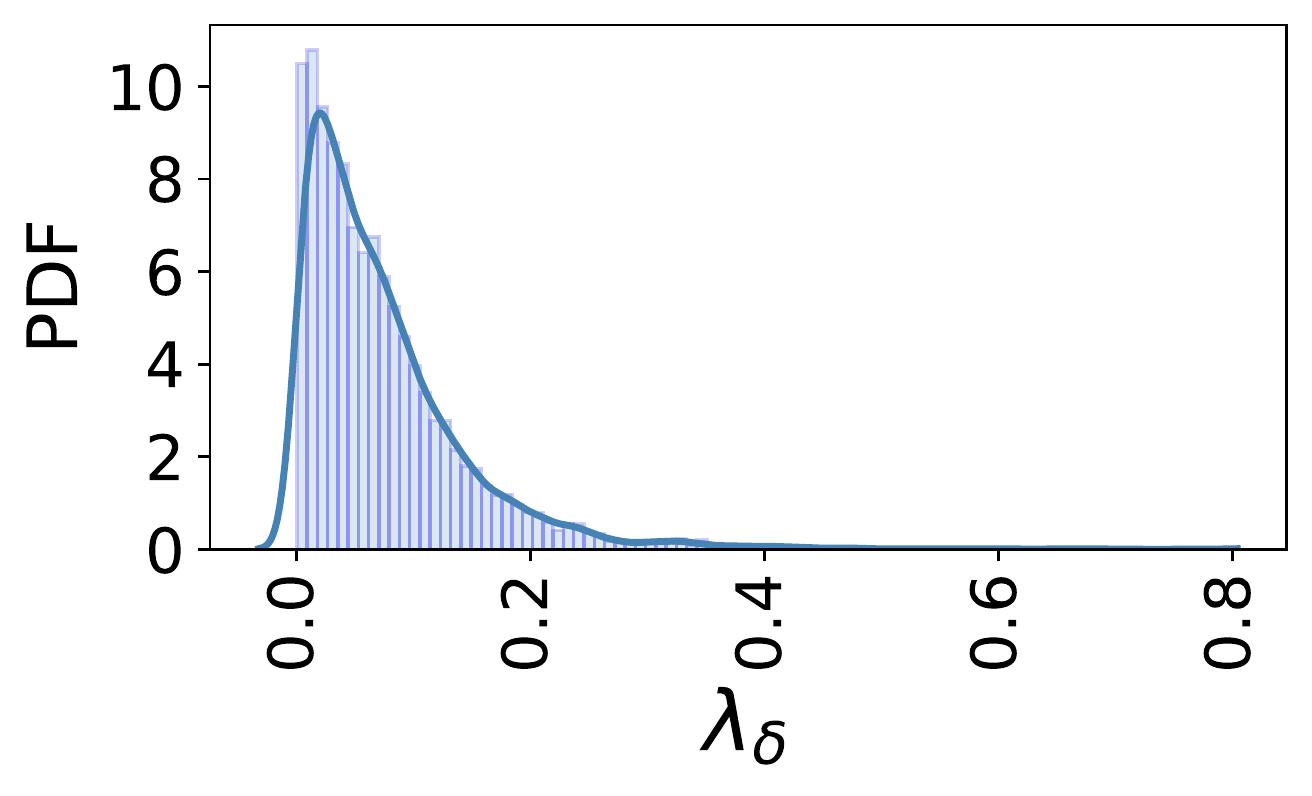}      \\
       {\small (a)}&
       \includegraphics[scale=0.22]{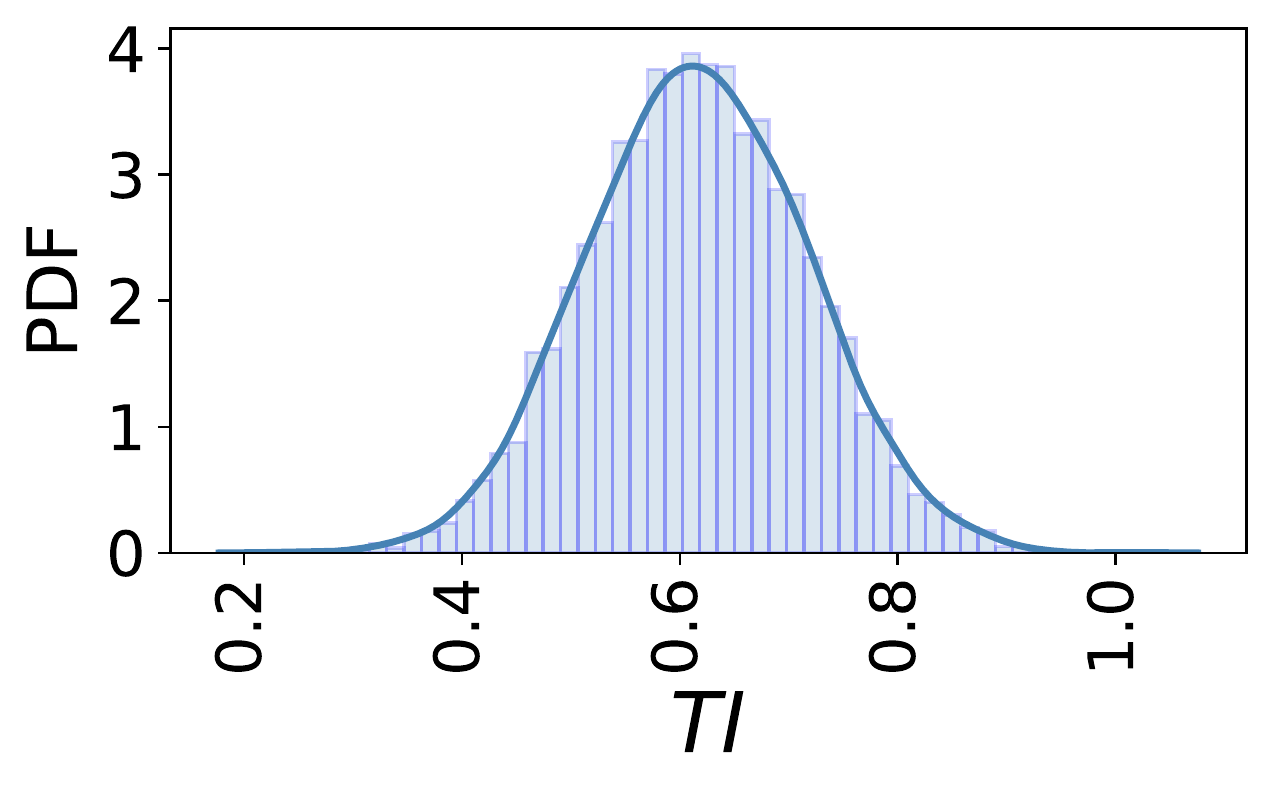}    & \hspace{-0.3cm}
       \includegraphics[scale=0.22]{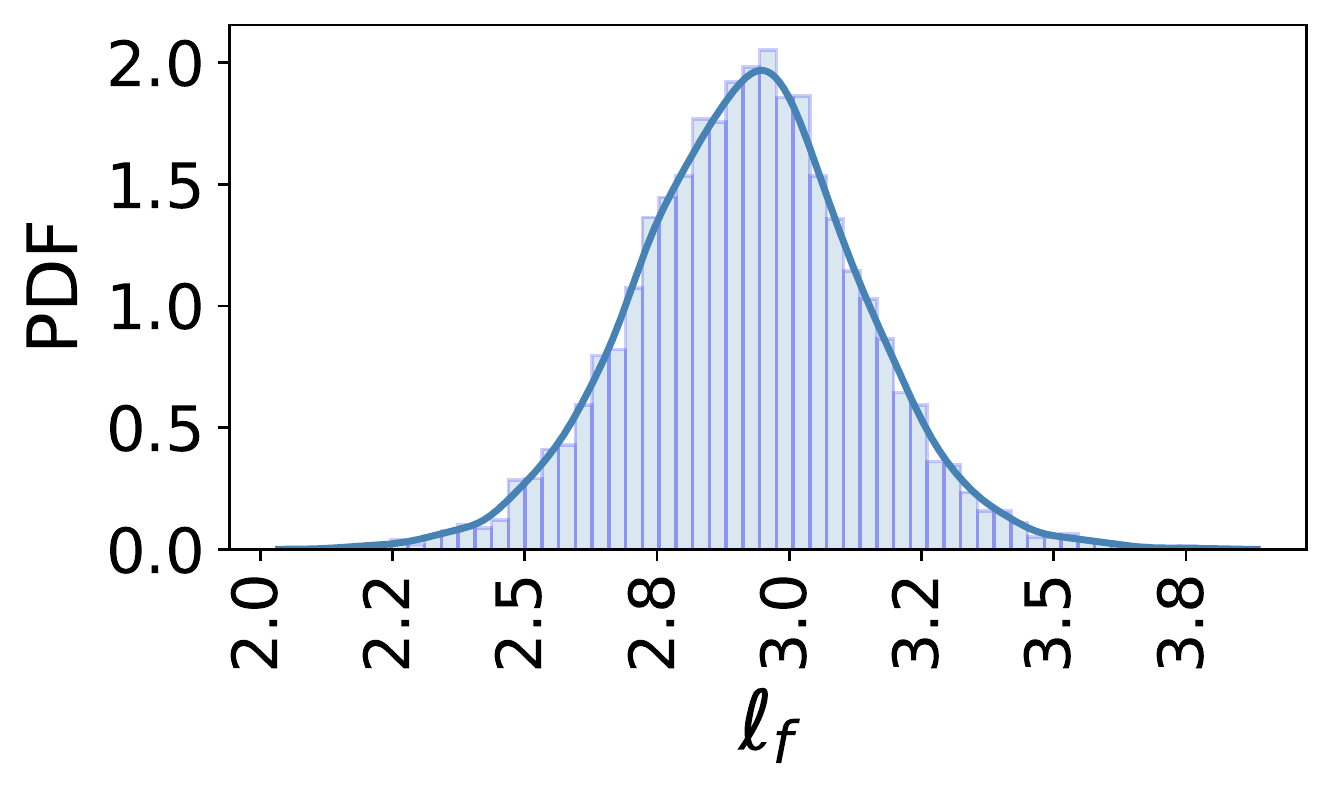}     & \hspace{-0.3cm}
       \includegraphics[scale=0.22]{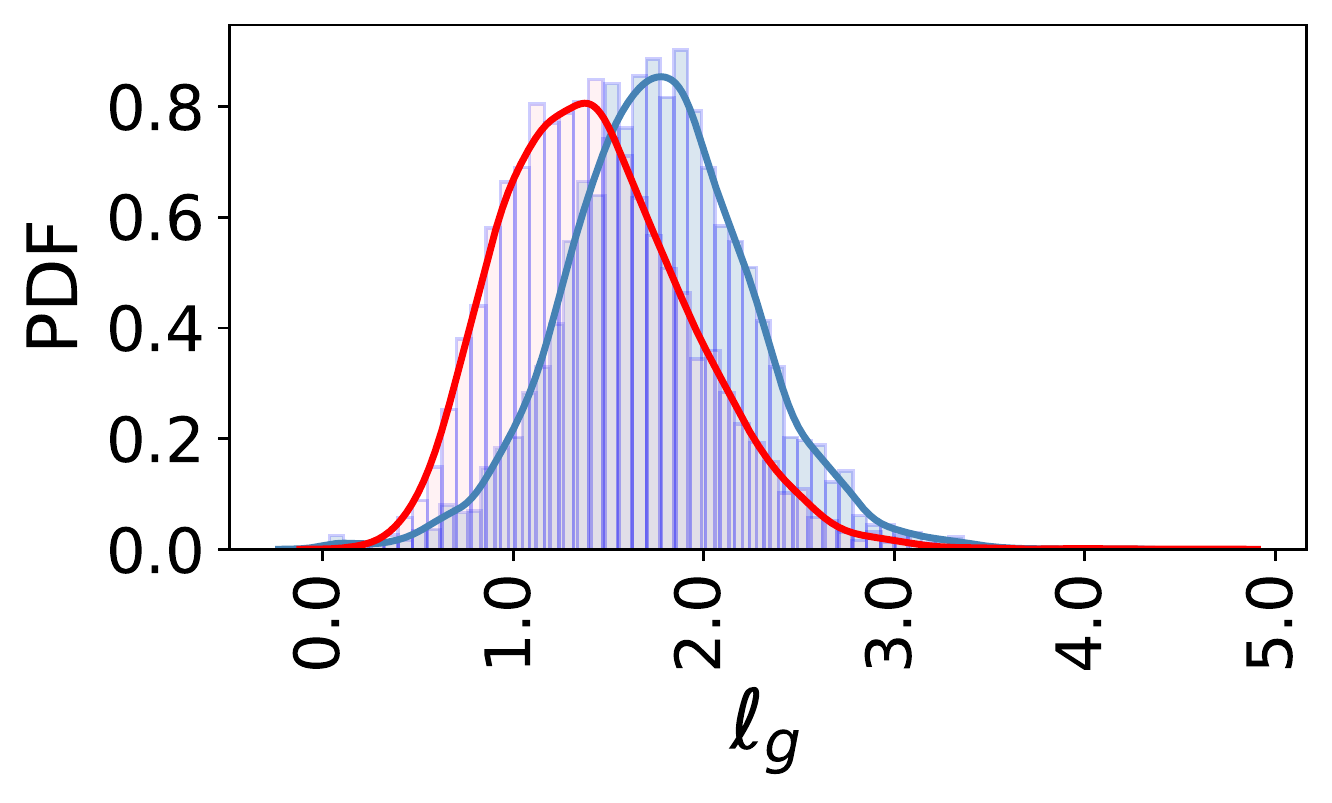}& \hspace{-0.3cm}
       \includegraphics[scale=0.22]{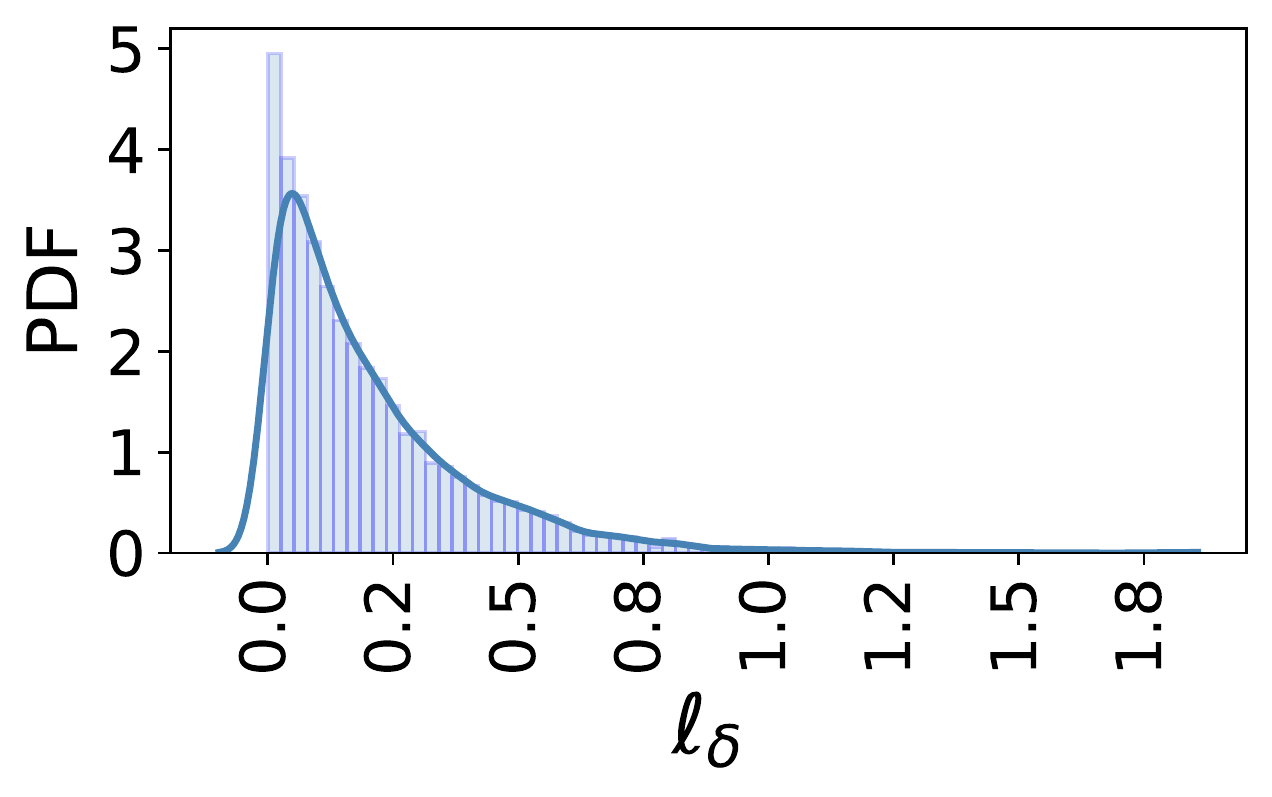}      \\  \vspace*{-0.2cm}
       {} &
       \includegraphics[scale=0.22]{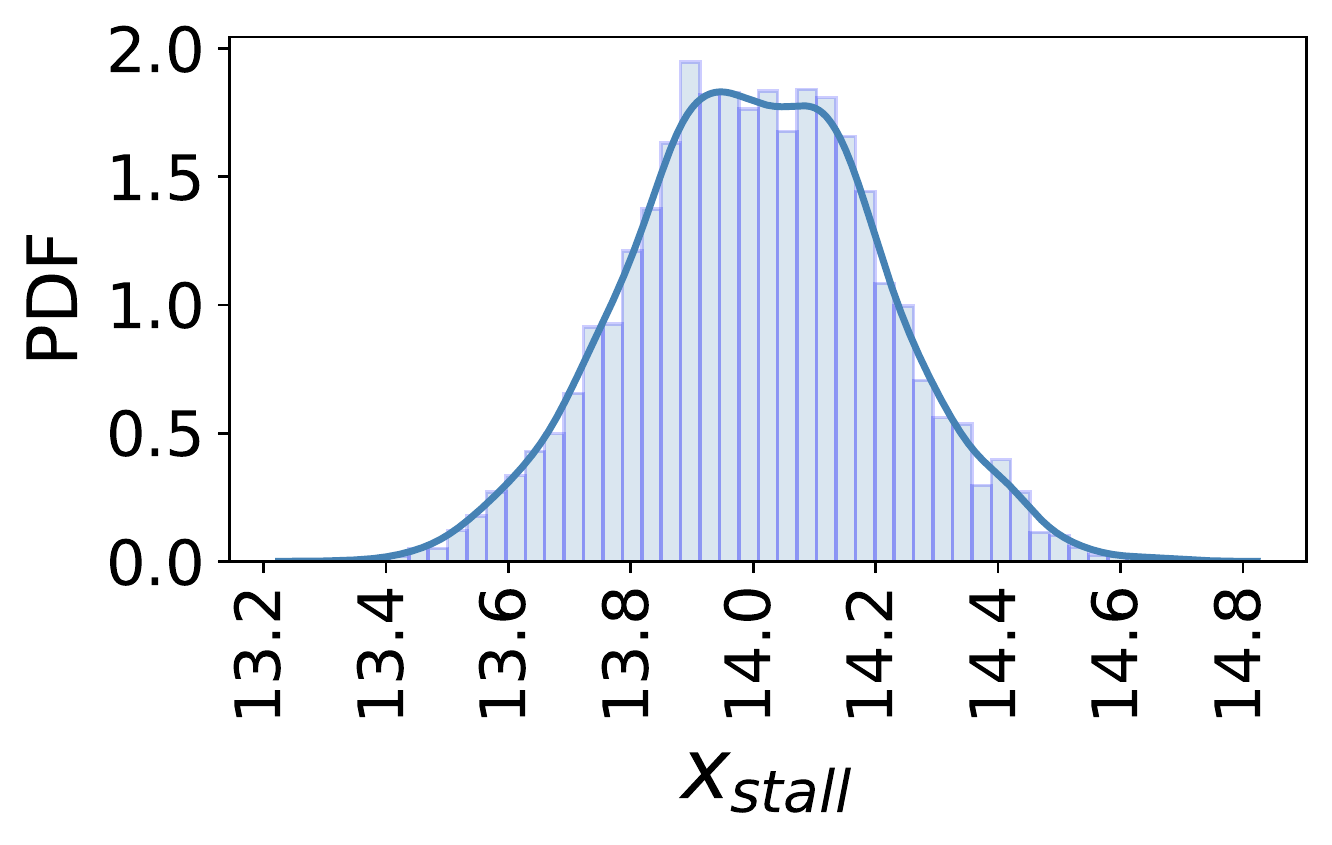}    & \hspace{-0.3cm}
       \includegraphics[scale=0.22]{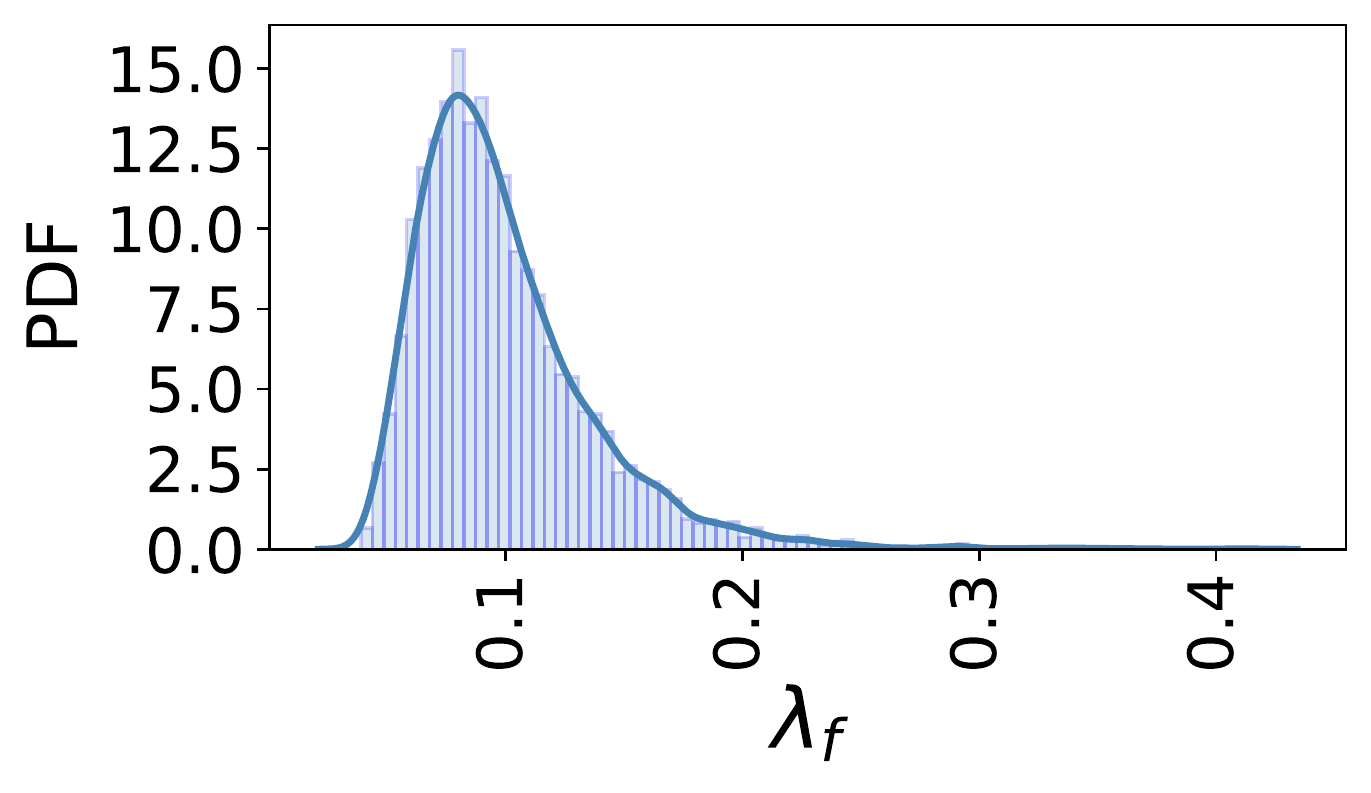}     & \hspace{-0.3cm}
       \includegraphics[scale=0.22]{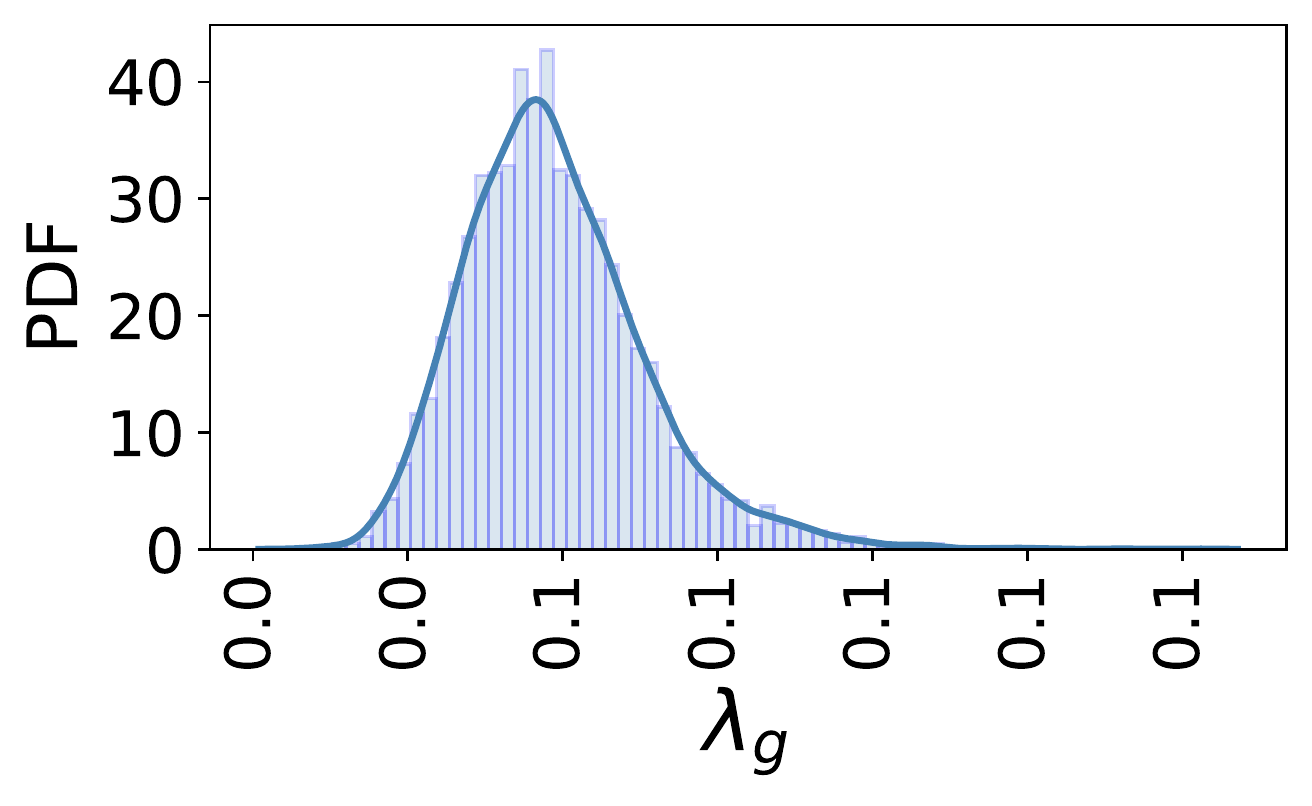}& \hspace{-0.3cm}
       \includegraphics[scale=0.22]{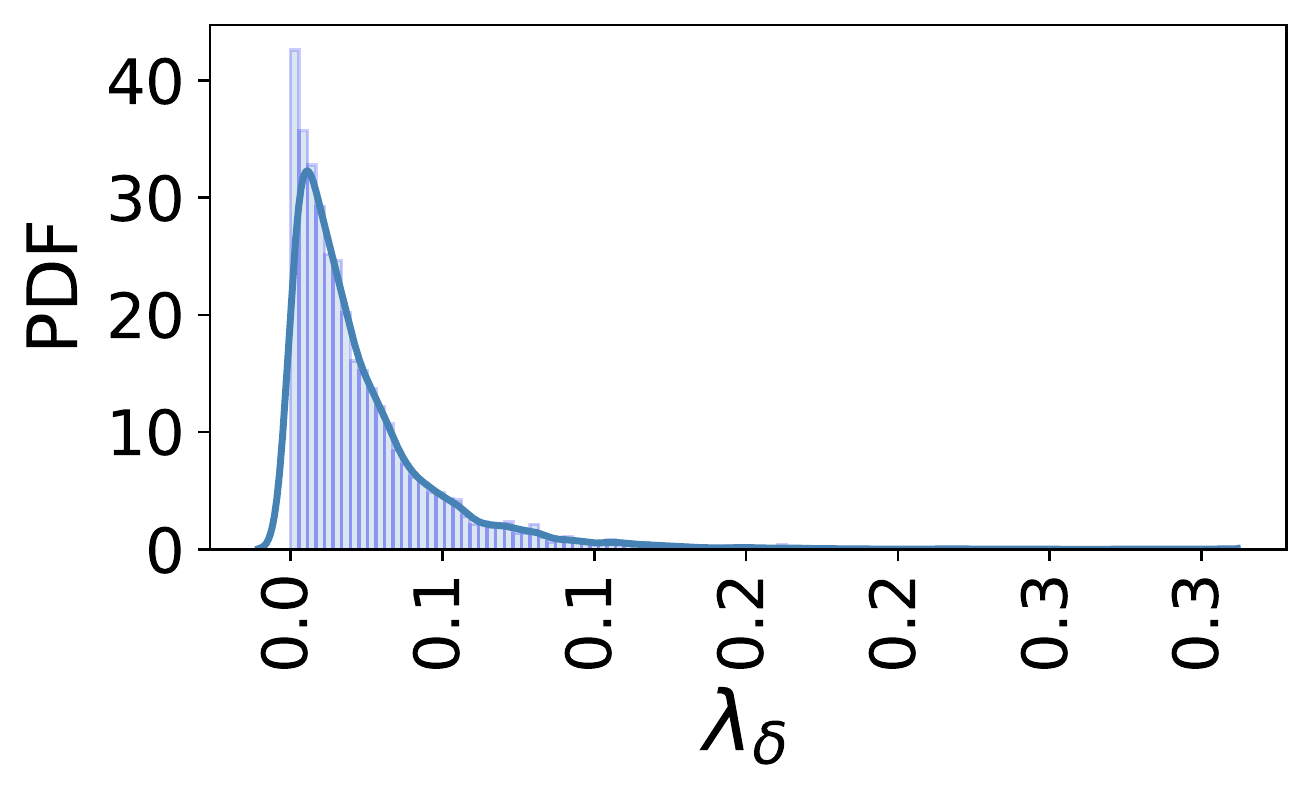}      \\
       {\small (b)}&
       \includegraphics[scale=0.22]{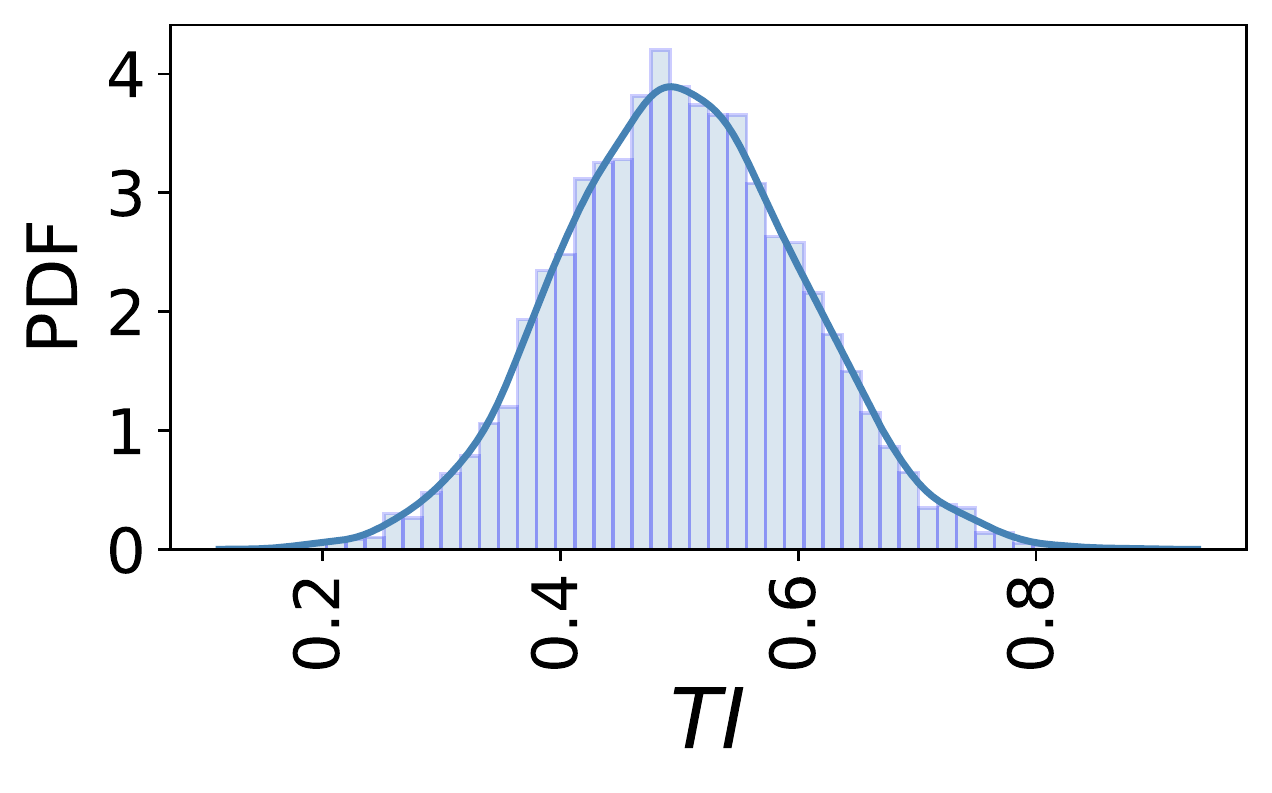}    & \hspace{-0.3cm}
       \includegraphics[scale=0.22]{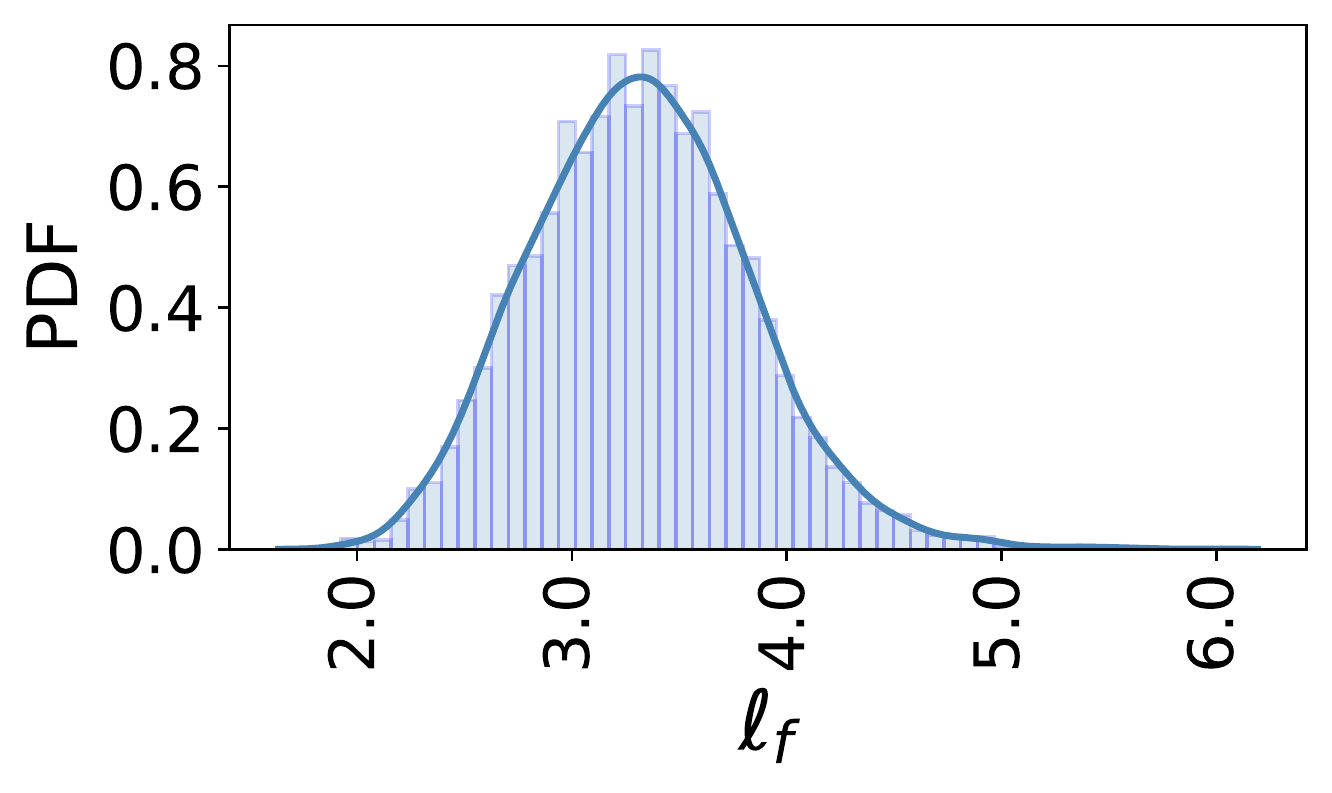}     & \hspace{-0.3cm}
       \includegraphics[scale=0.22]{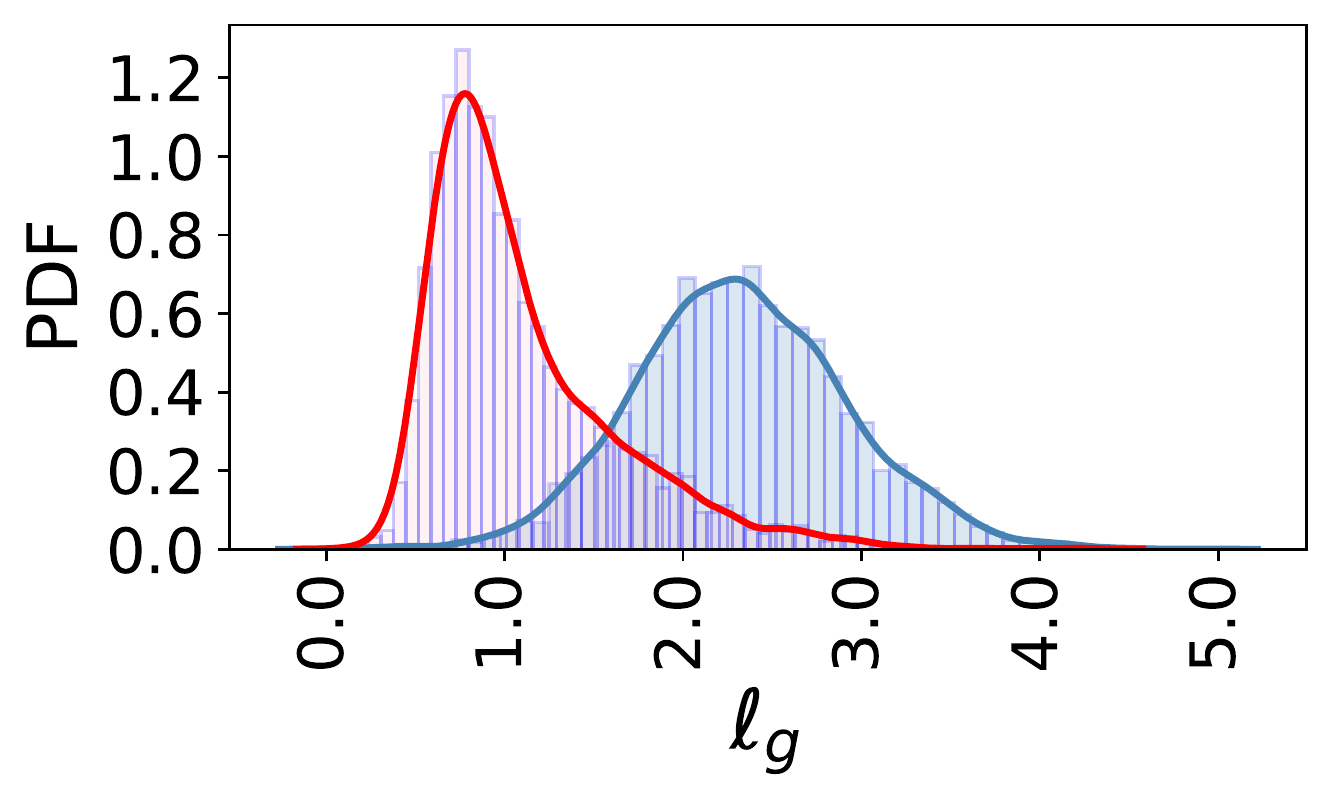}& \hspace{-0.3cm}
       \includegraphics[scale=0.22]{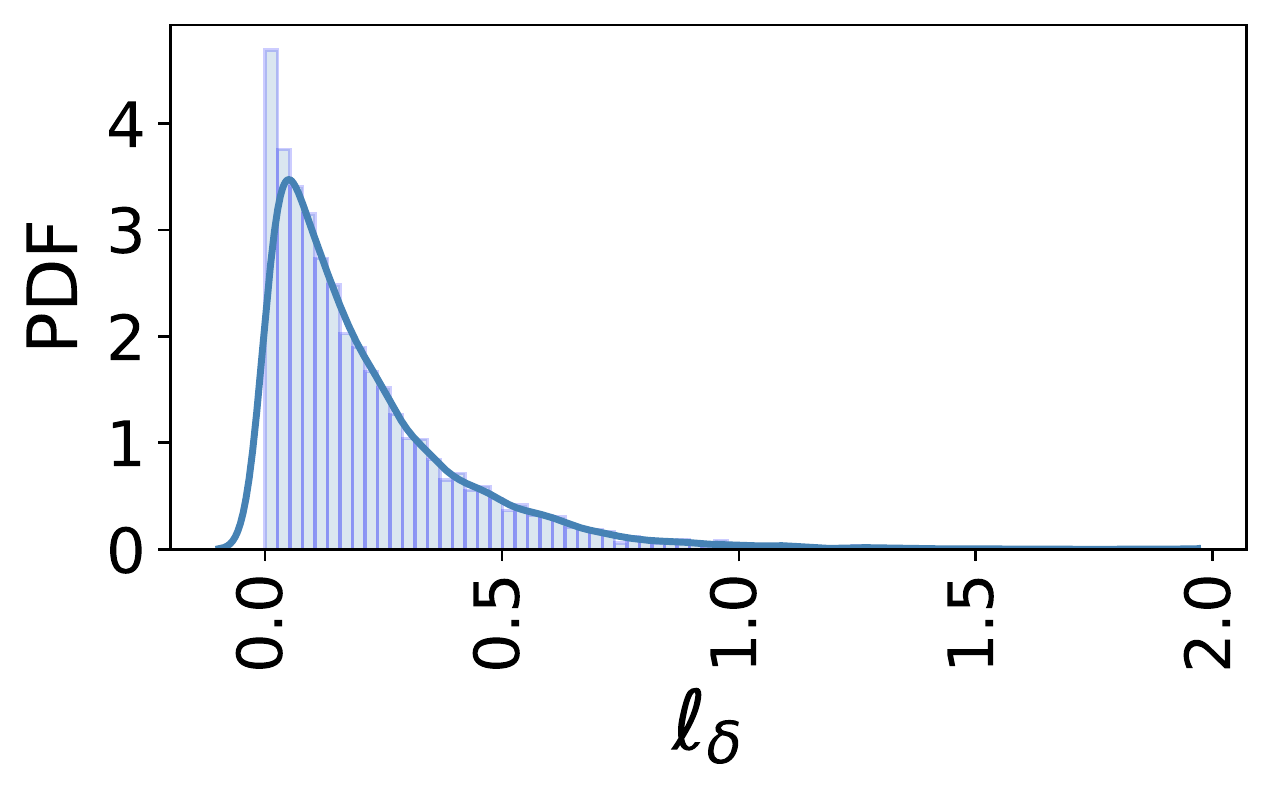}      \\
    \end{tabular}
\caption{Posterior PDFs of the calibration parameters and hyperparameters of the GPs appearing in the HC-MFM~(\ref{eq:mfModel}) for (a)~$C_L$ and (b)~$C_D$. Associated training data and predictions are shown in \fig~\ref{fig:cl_cd}. In the plots of~$\ell_g$, the blue and red histograms are corresponding to the AoA and TI, respectively.}
\label{fig:cl_cd_pars}

\end{figure}

\fig~\ref{fig:cl_cd_pars} shows the posterior densities of different parameters appearing in the MFM constructed for~$C_L$ and~$C_D$.
As expected, the distribution of the kernels' hyperparameters varies between the two QoIs.
But more importantly, the posterior distributions of~$x_\sta$ and calibrated TI are also dependent on the QoI.
This clearly shows the suitability of the present class of MFMs in which calibration of the parameters of different fidelities is performed as a part of constructing the MFM for a given QoI. 
The alternative strategy, which is common in practice (e.g.\ for co-Kriging models without calibration), is to calibrate the LF models against the HF data of one of the QoIs and then run the calibrated LF model to make realizations of all~QoIs. However, given the present results, this strategy seems clearly  less efficient and leads to inferior quality of prediction.

As a general goal, an MFM constructs a surrogate for the QoIs in the space of the design/controlled parameters aiming for the surrogate outputs to be as close as possible to the HF data.
In this regard, the MFMs facilitate applying different types of sample-based UQ techniques and optimization, see~\eg~\cite{smith}.
In connection with the present example, consider a UQ forward problem to estimate the stochastic moments of~$C_L$ and~$C_D$ due to the variation of the AoA.
For instance, assume~$\text{AoA}\sim\mathcal{N}(15,0.1)$ degrees.
This results in the following estimations for the expectation and variance of~$C_L$ and~$C_D$ with the associated~$95\%$ confidence intervals: $\BE_x[C_L]=1.1775 \pm 0.1037$, $\BV_x[C_L]=2.6675 \times 10^{-5} \pm 4.5096\times 10^{-5} $, $\BE_x[C_D]=5.6891\times 10^{-2} \pm 2.5722\times 10^{-2}$, and $\BV_x[C_D]=1.0618 \times 10^{-5} \pm 9.5994\times 10^{-6}$.
Note that without the HC-MFM, and only based on the data of RANS or/and DES, such estimations would be at best inaccurate, but in general impossible to make.

\subsection{Effect of geometrical uncertainties in the periodic hill flow}
\begin{figure}[!h]
	\centering
	\includegraphics{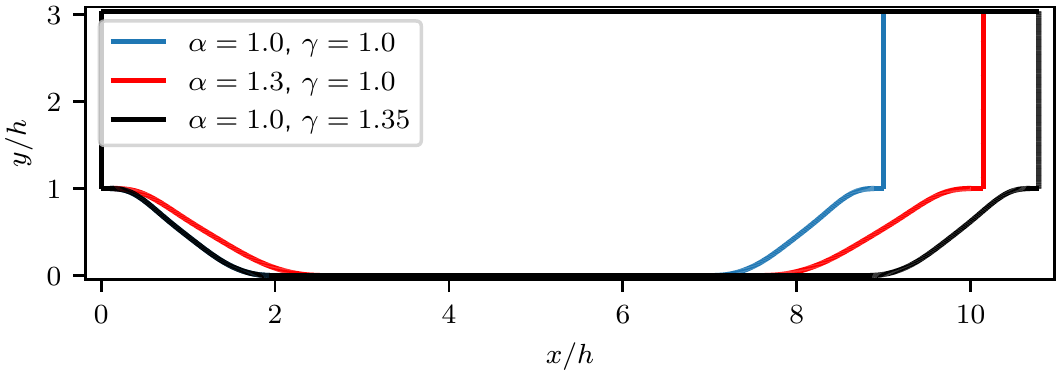}
	\caption{The geometry of the periodic hill simulations, illustrating the effect of parameters~$\alpha$ and~$\gamma$ using three sets of values for them.}
	\label{fig:phill_geom}
\end{figure}

In this last flow case, we consider the turbulent flow over periodic hills with geometrical uncertainties, see the sketch in \fig~\ref{fig:phill_geom}.
The outline in blue corresponds to the configuration studied in several prior works (hereafter, baseline geometry), for example by Fröhlich et al.~\cite{frohlich:05} and more recently by Gao et al.~\cite{gao:20}.
The latter reference can be consulted for a good overview of other previous efforts.
The shape of the hill is defined by six segments of third-order polynomials, see \eg~\cite{xiao:20} for the exact definition.
The flow Reynolds number is~$5600$.

In \rf~\cite{xiao:20}, a parameterization of the geometry was introduced by scaling the length of the hill.
Particularly, the authors performed a series of DNS for $L_{x}/h = 3.858 \alpha + 5.142$, where~$L_x$ is the length of the geometry, $h$ is the height of the hill, and~$\alpha$ is a parameter.
The value $\alpha=1$ corresponds to the baseline geometry.
The corresponding DNS data set for several values of~$\alpha$ has been made publicly on Github, which was extended in 2021 with additional data introducing a new parameter~$\gamma$:
\begin{equation}
	L_{x}/h = 3.858 \alpha + 5.142 \gamma\ .
\end{equation}
The effect of~$\alpha$ and~$\gamma$ on the geometry is illustrated in \fig~\ref{fig:phill_geom} with red and black curves, respectively.
The purpose of the present example is to demonstrate how the HC-MFM can be used to economically assess the effect of uncertain parameters~$\alpha$ and~$\gamma$ on the flow QoIs.
To that end we combine the DNS data discussed above from \rf~\cite{xiao:20} with data from RANS simulations performed by us using ANSYS Fluent v19.5~\cite{ansys}.
Particularly, for the DNS we use the data for $\alpha \in \{0.5, 1.0, 1.5\}$ and $\gamma \in \{0.4166, 1.0, 1.5834\}$.
The same values are also used for the RANS, complemented by two additional samples for both~$\alpha$ and~$\gamma$ for which DNS results are not available.
These are selected based on the Gauss-Legendre quadrature rule and are equal to $\alpha = \{0.702, 1.297\}$ and $\gamma = \{0.653,  1.347\}$.
Therefore, there is a total of $5\times 5$ samples over the space of~$\alpha-\gamma$ corresponding to which RANS simulations are performed. 
For the uncertainty propagation and sensitivity analysis, see below, we assume~$\alpha$ and~$\gamma$ to be independent, and $\alpha\sim \cU[0.448,1.552]$ and $\gamma\sim\cU[0.356,1.644]$.

The standard~$k$~-~$\omega$ turbulence model is used in the RANS simulations, as defined in~\cite{ansys} based on the work of Wilcox~\cite{wilcox:06}.
The available low-Reynolds-number correction to the model was not used.
The model depends on a number of parameters which are listed in~\cite[p.~61]{ansys}.
It can be shown~\cite[p.~134]{wilcox:06} that the parameters~$\alpha_\infty$,~$\beta^*_\infty$, $\sigma$ and~$\beta_i$ are coupled to the von Karm{\'a}n coefficient~$\kappa$ through the following equation:
\begin{equation} \label{eq:kappa_beta}
	\alpha_\infty = \beta_i/\beta_\infty^* - \sigma \kappa^2/\sqrt{\beta_\infty^*}.
\end{equation}
To illustrate the automatic calibration capability of the HC-MFM, we assume~$\kappa$ to be uncertain and perform simulations for~5 sample values of $\kappa \in \{ 0.348, 0.367, 0.4   , 0.433, 0.452\}$, which follow the Gauss-Legendre quadrature rule. 
To prescribe the desired value of~$\kappa$, we set~$\alpha_\infty = 0.52$,~$\sigma = 0.5$, $\beta_i = 0.0708$ (as suggested by Wilcox~\cite[p.~135]{wilcox:06}), and manipulate~$\beta_\infty^*$ according to~\eqref{eq:kappa_beta}.
Note that the considered training samples include the standard choice $\beta_\infty^* = 0.09$, corresponding to~$\kappa = 0.4$.
Using~$5$ samples for each of~$\alpha$,~$\gamma$ and~$\kappa$ and using a tensor-product rule, a total of~125 RANS simulations were performed for this study.

For RANS simulations, quadrilateral cells were used to discretize the computational domains, with the total grid size ranging from $\approx 150 \cdot 10^3$ to $\approx 500 \cdot 10^3$ cells, depending on the domain size as defined by~$\alpha$ and~$\gamma$.
Since the selected turbulence model requires accurate resolution of the boundary layer, the selection of the mesh size was guided by the  discretization in the corresponding DNS case~\cite{xiao:20}. Specifically, we adopted the same number of cells in the streamwise and wall-normal directions as in the DNS, and applied size grading in the wall-normal direction to ensure low values of~$y^+$.
This means that in the streamwise direction the mesh may be unnecessarily fine for RANS, but since these simulations are two-dimensional and steady-state, they are still negligibly cheap compared to the DNS.

Second-order numerical schemes were used to discretize the RANS equations in ANSYS Fluent~\cite{ansys}.
Specifically, for the convective fluxes a second-order upwind scheme was used.
The coupled solver was used for pressure-velocity coupling, which did an excellent job at converging the simulations.

\subsubsection{Creating ground truth and verifying the model}
\begin{figure}[t]
    \centering
    \begin{tabular}{cc}
         \includegraphics[scale=0.4]{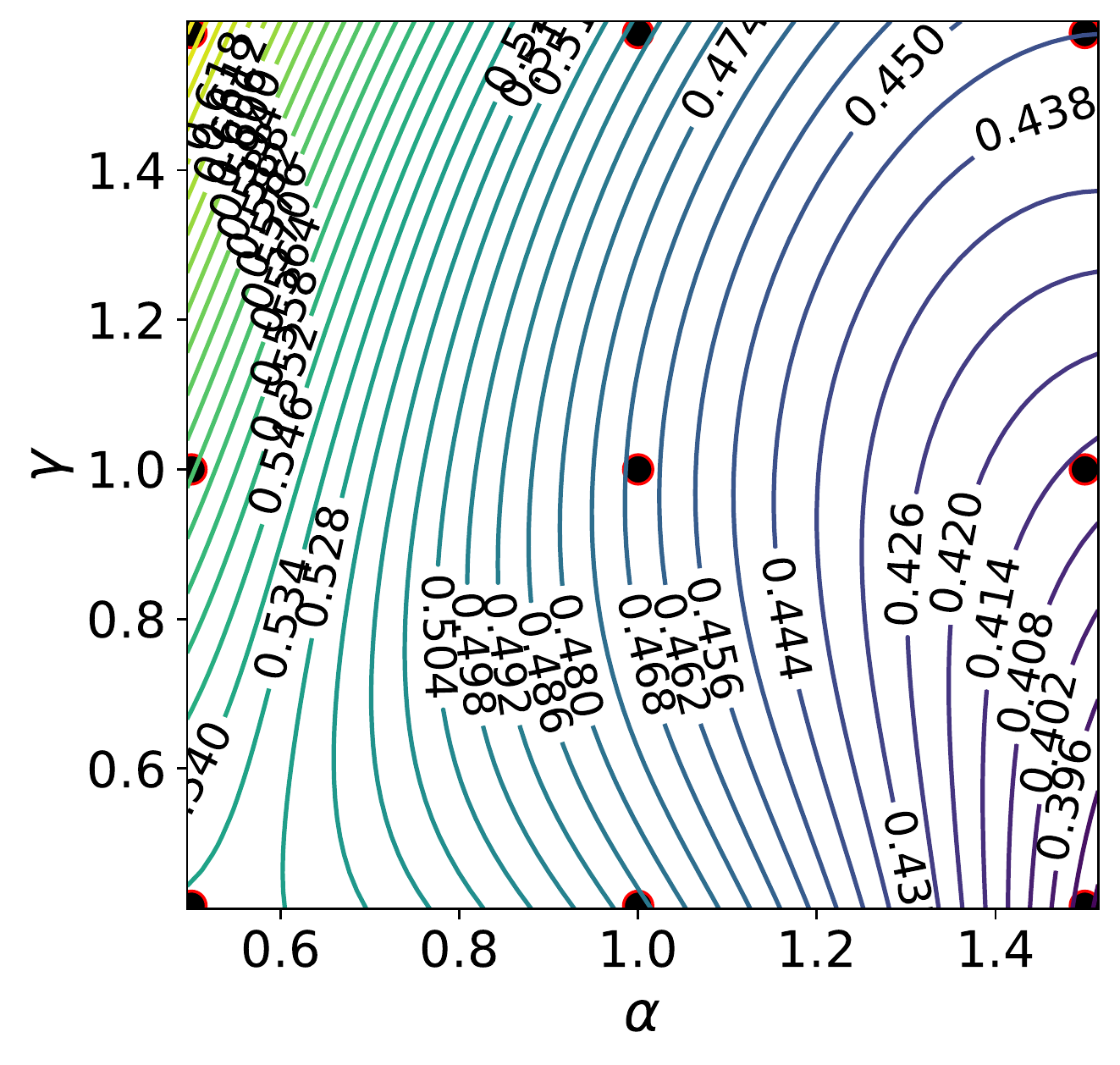}& 
         \includegraphics[scale=0.5]{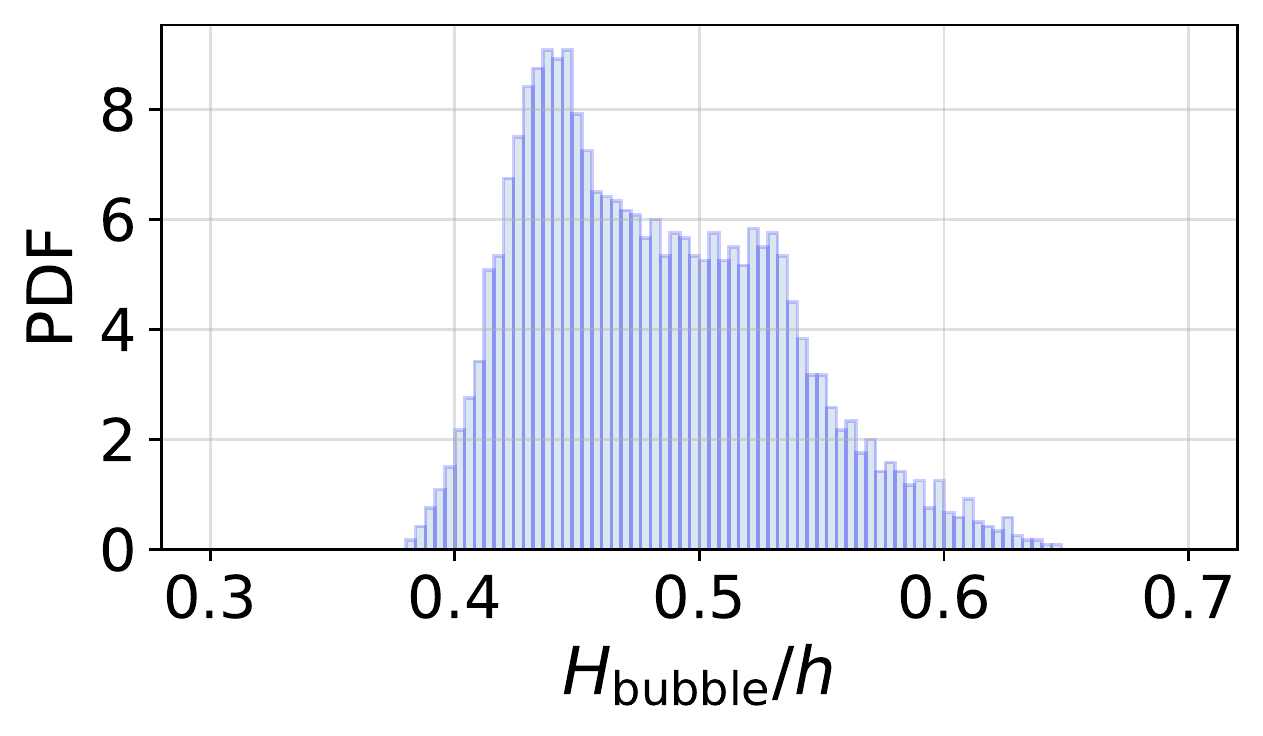}\\
    \end{tabular}
    \caption{(Left) Isolines of the response surface and (Right) PDF of $H_{\rm bubble}/h$ at $x/h=2.5$ due to the variation of~$\alpha$ and~$\gamma$ using all the~$9$ DNS data of \rf~\cite{xiao:20} (represented by the symbols in the left plot). These plots are considered as ground truth or reference for evaluate the performance of the multifidelity model.}
    \label{fig:phill_ground2.5}
\end{figure}

Using all the~$9$ DNS data sets available from \rf~\cite{xiao:20}, the response surface of the QoIs at the space of $\alpha-\gamma$, and associated PDF of the QoIs can be estimated.
These will be used as the ground truth or reference to evaluate the performance of the HC-MFM in the following analyses.
The interpolation from the DNS samples to an arbitrary mesh covering the whole admissible range of~$\alpha$ and~$\gamma$ can be done using polynomial-based methods such as PCE (used here) or Lagrange interpolation, as well as GP regression.
Based on the data available from \rf~\cite{xiao:20} and the performed RANS simulations, different QoIs can be considered. 
Hereafter, to demonstrate the power of the method, we take the normalized height of the separation bubble, $H_{\rm bubble}/h$ at the streamwise location~$x/h=2.5$ as the QoI. 
Alternatively other locations~$x/h$ as well as different flow quantities could be considered. 
The response surface of the QoI and associated PDF are illustrated in \fig~\ref{fig:phill_ground2.5}. 
Based on the pattern of the isolines, we can observe that the parameter~$\alpha$ exhibits a stronger influence on the QoI than~$\gamma$. 
This can be quantitatively confirmed via the values of the total Sobol indices~\cite{sobol:01} as reported in \tab~\ref{tab:phill_UQresults}.
The resulting PDF is bimodal with two peaks around the most probable observed values of~$H_{\rm bubble}/h$ at $x/h=2.5$.

\begin{figure}
    \centering
    \begin{tabular}{cc}
         \includegraphics[scale=0.4]{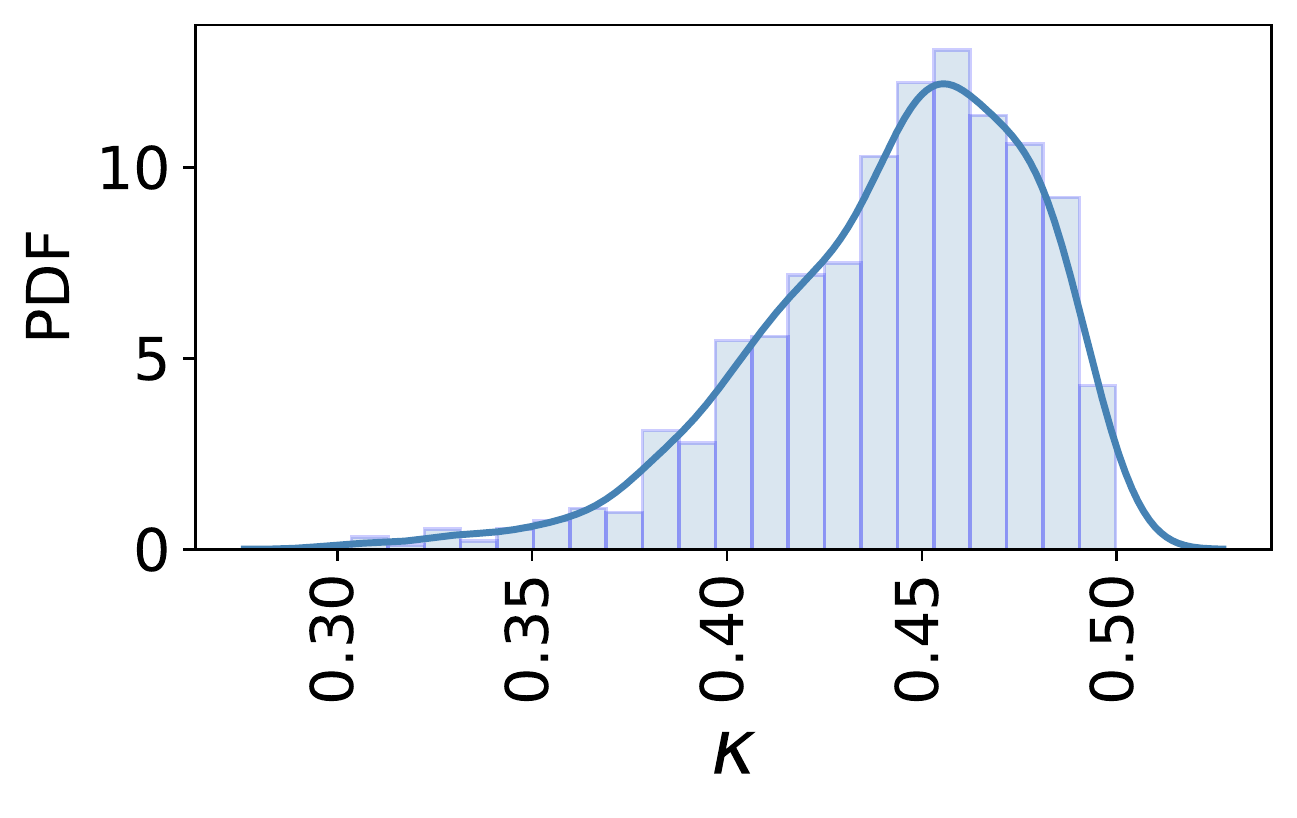} &
         \includegraphics[scale=0.4]{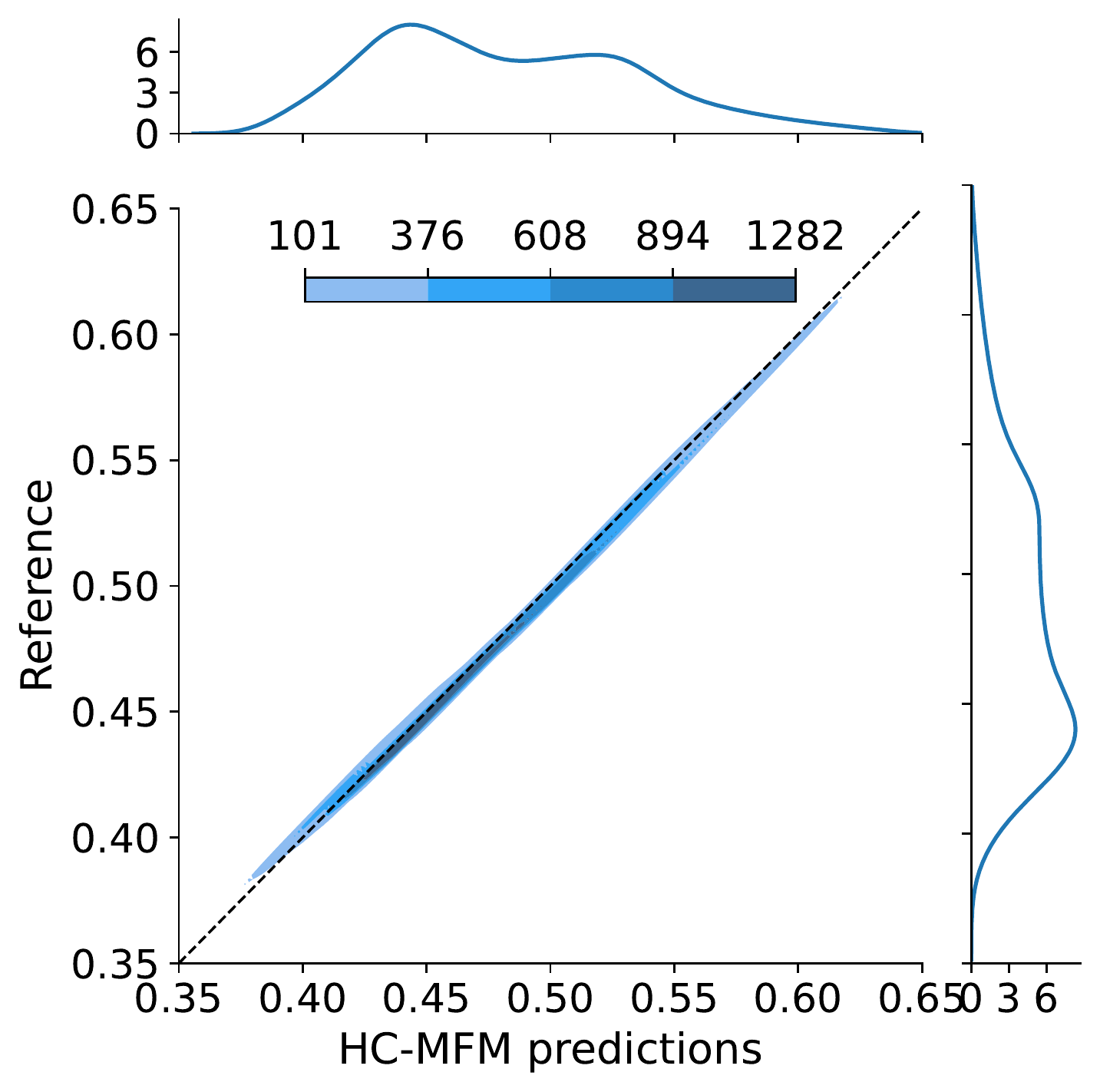}\\
    \end{tabular}
    \caption{(Left) The sample posterior PDF of~$\betas$ and (right) sample joint PDF of $H_{\rm bubble}/h$ at $x/h=2.5$ obtained from the HC-MFM using all~$9$ DNS data sets of \rf~\cite{xiao:20} along with the~$125$ RANS simulations performed in the present study. The RANS simulations are performed using~$5$ samples of $\kappa$ equal to $0.348$, $0.367$, $0.4$, $0.433$, and $0.452$. The marginal PDFs on the top and right axes are found using the kde (kernel density estimation) method.}
    \label{fig:phill_refPostBeta}
\end{figure}

The reference posterior distribution of~$\betas$ as the RANS calibration parameter can now be inferred. 
To this end, the HC-MFM described in \sect~\ref{sec:phill_MFM} is constructed using~$9$ DNS data sets of \rf~\cite{xiao:20} and~$125$ RANS simulations. 
The prior distribution of~$\betas$ is taken to be uniform over the range of~$[0.3,0.5]$. 
This non-informative prior distribution removes any bias towards any particular value in the distribution of~$\betas$. 
Through a Bayesian inference via an MCMC method, the sample posterior distribution of~$\betas$ shown in \fig~\ref{fig:phill_refPostBeta}~(left) is obtained. 
Note that this calibration is in fact a pure UQ inverse problem, see \eg~\cite{smith}, where all the RANS data are utilized to construct a surrogate for~$\betas$, and the DNS data are used as training data to infer the distribution of~$\betas$.
The estimated mean and standard deviation of the posterior distribution of~$\betas$ are~$0.47087$ and~$0.05387$, respectively. 
As compared to the standard value~$0.4$ being used in the literature, the estimated mean is somewhat larger. However, from a physical point of view, one would not expect an accurate value of~$\kappa$ for this type of flow due to the separated nature and the relatively low Reynolds number.

Another advantage of using all the available RANS and DNS data in the HC-MFM is that the implementation (algorithm and coding) of the HC-MFM can be verified.
As shown in \fig~\ref{fig:phill_refPostBeta}~(Right), the prediction of the HC-MFM constructed by combining all DNS and RANS data sets completely agree with the predictions of the single-fidelity model based on only the DNS data. 
Note that the predicted marginal PDF of the QoI in the HC-MFM is the same as the reference PDF in \fig~\ref{fig:phill_ground2.5}.

\subsubsection{Application of the HC-MFM}\label{sec:phill_MFM}
Adopting the general notation of \sect~\ref{sec:method}, the HC-MFM for the present example can be written as, 
\begin{eqnarray}\label{eq:mfModel_phill}
  \begin{cases}
   y_{M_1}(\fx_i) \quad\,\, = {\hat{f}(\fx_i,\theta_{2_i})} +
   {\hat{\delta}(\fx_i)} + \varepsilon_{1_i} \,&,\quad i=1,2,\cdots,n_1 \\
   y_{M_2}(\fx_{i}) = {\hat{f}(\fx_i,t_{2_i})}+ \varepsilon_{2_i}  \,&,\quad i=1+n_1,2+n_1,  \cdots , n_2+n_1
  \end{cases} \,,
\end{eqnarray}
where~$M_1$ and~$M_2$ denote DNS and RANS, respectively, the design parameters are~$\fx_i=(\alpha_i,\gamma_i)$, and,~$t_{2}$ and~$\theta_2$ refer to the simulated and calibrated instances of~$\betas$, respectively.
The kernel of $\hat{f}(\fx,t_2)$ is taken to be the exponentiated quadratic function~(\ref{eq:exponKernel}), while for~$\hat{\delta}(\fx)$, the {Matern-5/2} function~(\ref{eq:mat52Kernel}) is employed. 
For the hyperparameters, the following prior distributions are considered: $\kappa\sim\cU[0.3,0.5]$, $\lambda_f \sim\mathcal{HC}(\alpha=5)$, $\ell_{f_\fx},\ell_{f_{t_2}}\sim \Gamma(\alpha=1,\beta=5)$, $\lambda_\delta \sim\mathcal{HC}(\alpha=1)$, and
$\ell_{\delta_\fx} \sim \Gamma(\alpha=1,\beta=1)$.
In this example, the uncertainty in the DNS and RANS data is neglected. 
Despite this, when implementing the model in \texttt{PyMC3}~\cite{pymc3}, the prior of the Gaussian noise standard deviation is set to be $\sigma\sim \mathcal{HC}(\alpha=5)$ (same for both fidelities). 
But as expected, the mean and standard deviation of the posterior distribution of~$\sigma$ are obtained to be approximately zero.

\begin{figure}
    \centering
    \begin{tabular}{cc}
         \includegraphics[scale=0.35]{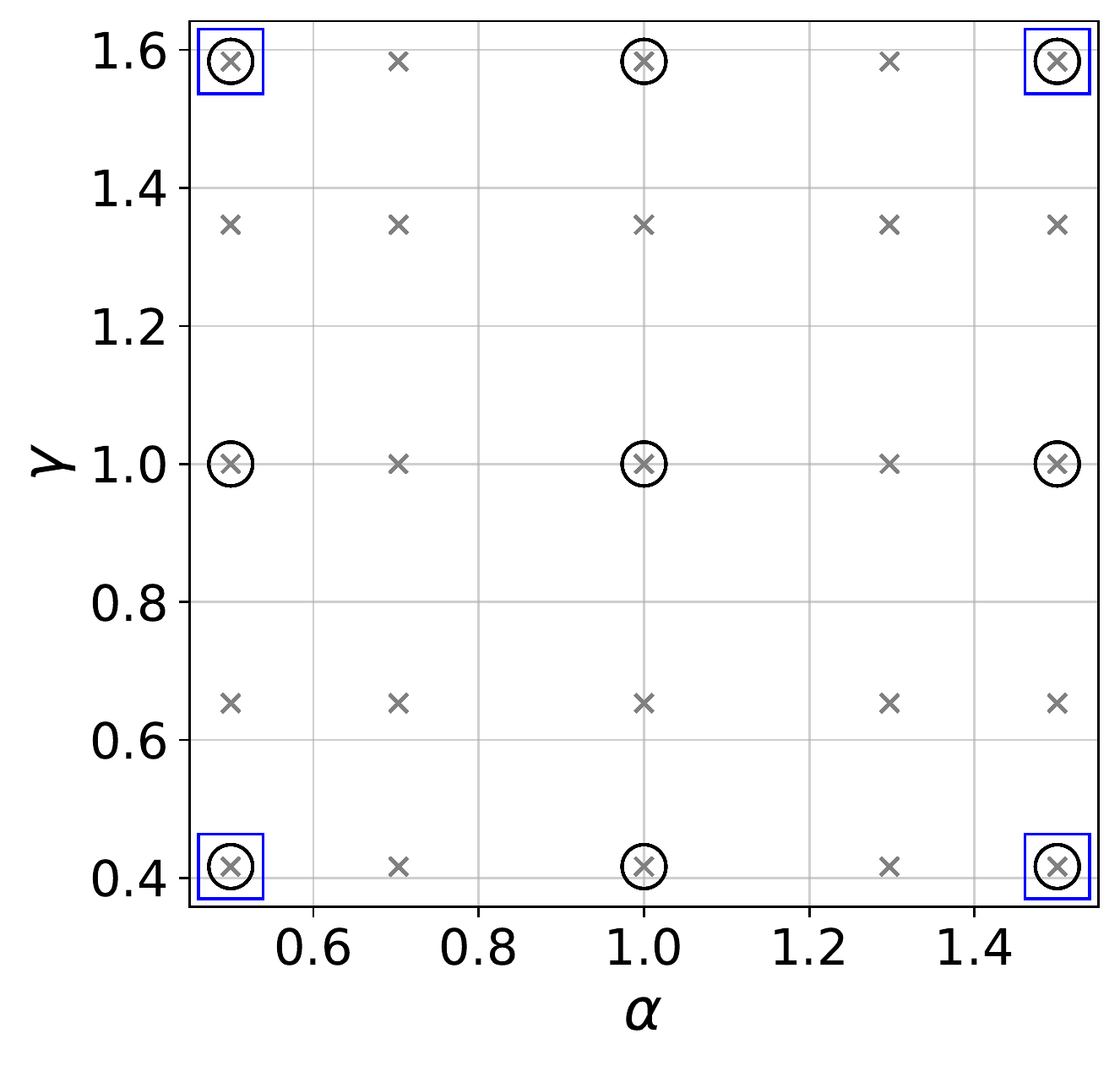} &
         \includegraphics[scale=0.35]{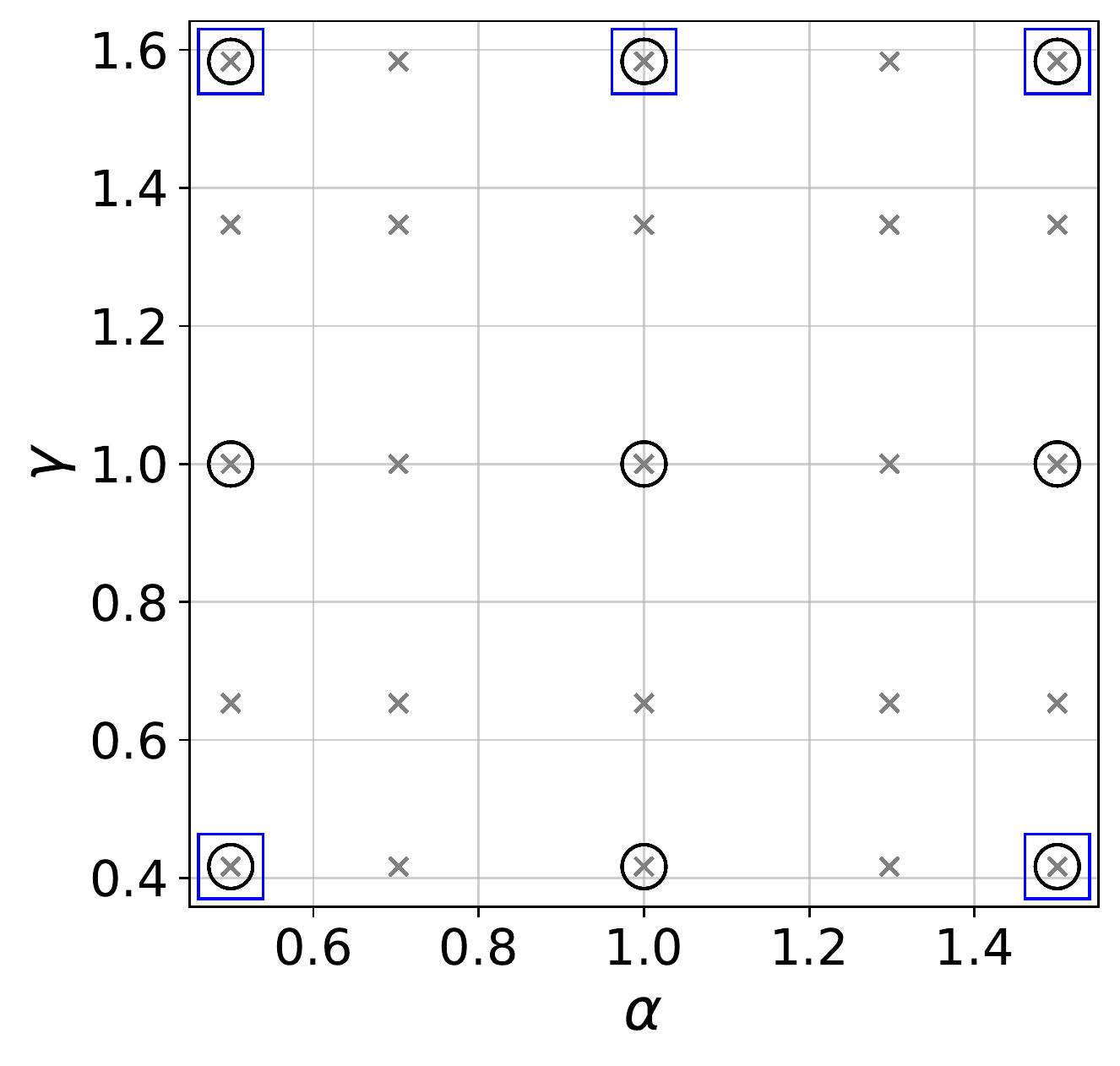} \\
    \end{tabular}
    \caption{Schematic representation of the samples from $\alpha$ and $\gamma$ corresponding to the low-fidelity (LF),~$\times$,  high-fidelity (HF),~$\Box$, and all available DNS data from \rf~\cite{xiao:20},~$\circ$. In the text, the left and right plots are referred to as Case-A and Case-B, respectively. Note that for both cases, there are~5 samples for~$\betas$ associated to each of the LF samples represented here.}
    \label{fig:phill_caseABSamples}
\end{figure}

The hyperparameters will be inferred from the combined set of~$n_1$ DNS and~$n_2$ RANS data.
In the analyses to follow, we use all the RANS simulations as low-fidelity (LF) data, therefore~$n_2=125$.
Two subsets of the DNS data of \rf~\cite{xiao:20} with size~$n_1=4$ and~$5$ are taken to be the high-fidelity (HF) data. 
Combining these two HF data sets with the LF data, Case-A and Case-B data sets are obtained for multifidelity modeling.
\fig~\ref{fig:phill_caseABSamples} represents the samples of these two cases in the space of~$\alpha-\gamma$ parameters.

\begin{figure}
    \centering
    \begin{tabular}{ccc}
         \includegraphics[scale=0.35]{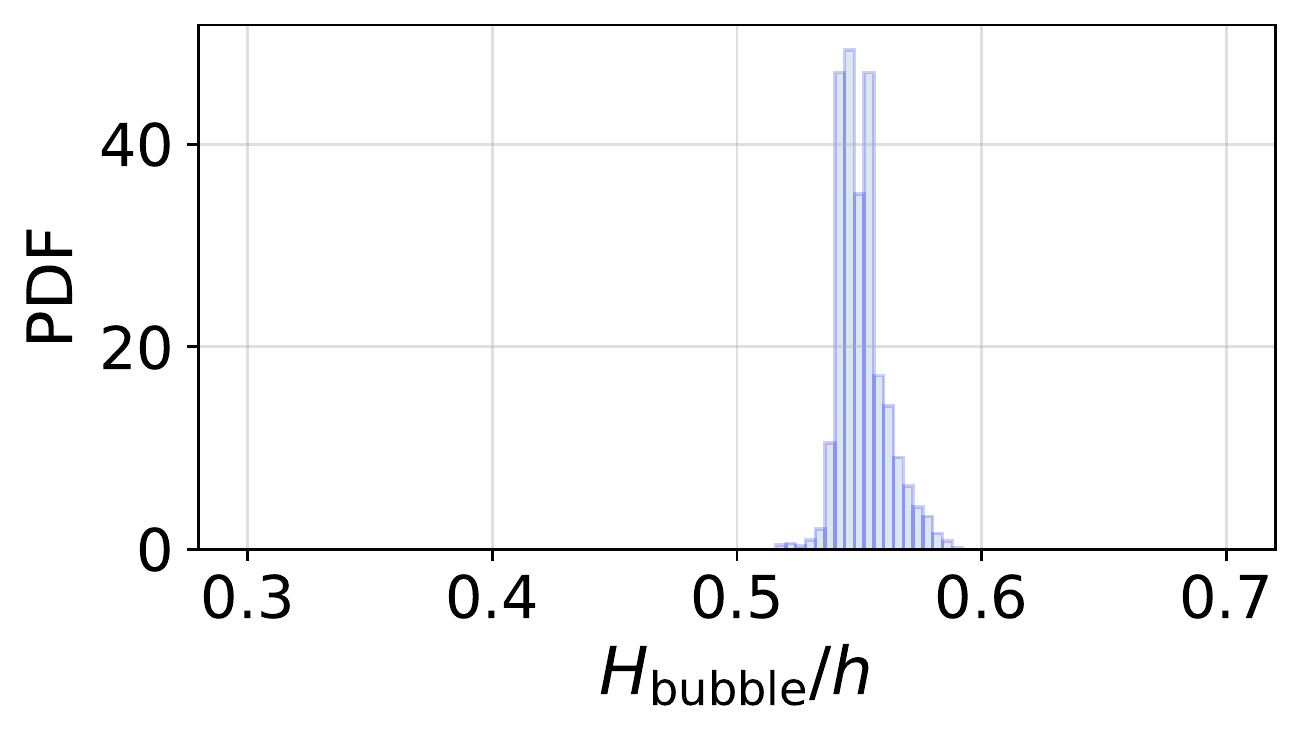}  &
         \includegraphics[scale=0.35]{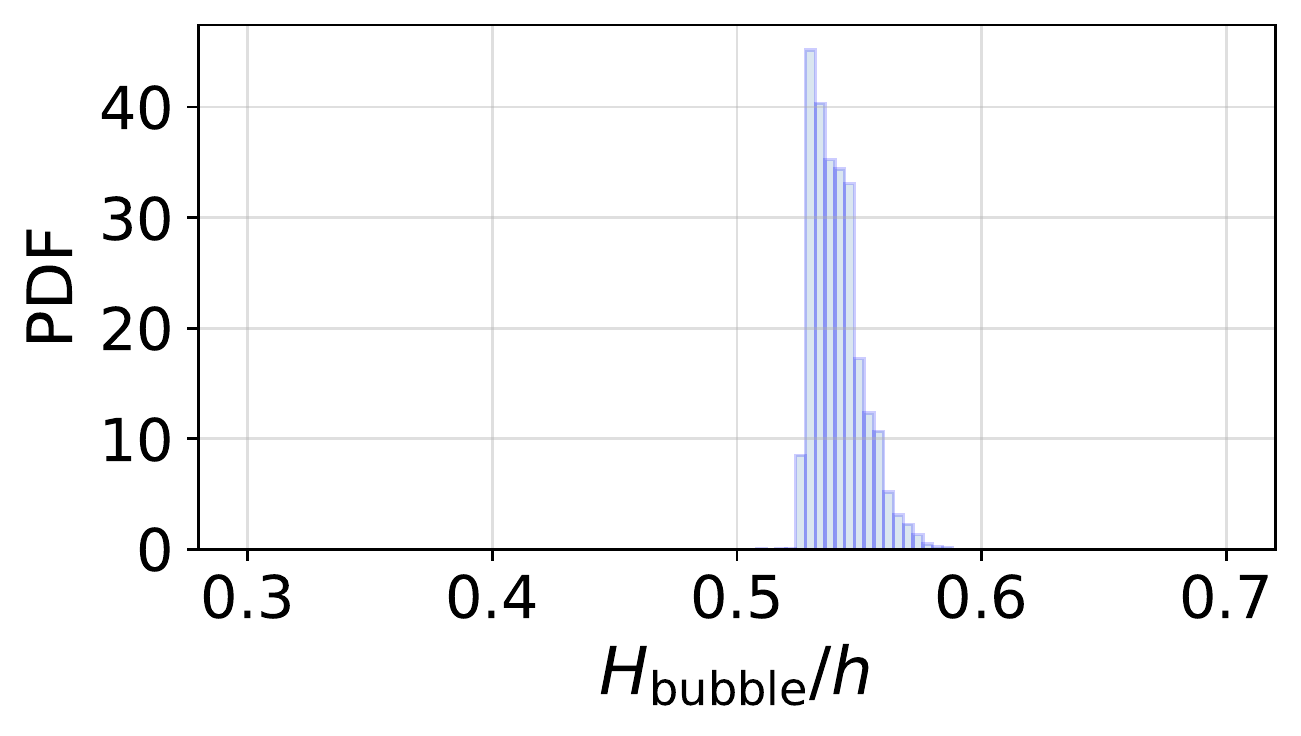}  &
         \includegraphics[scale=0.35]{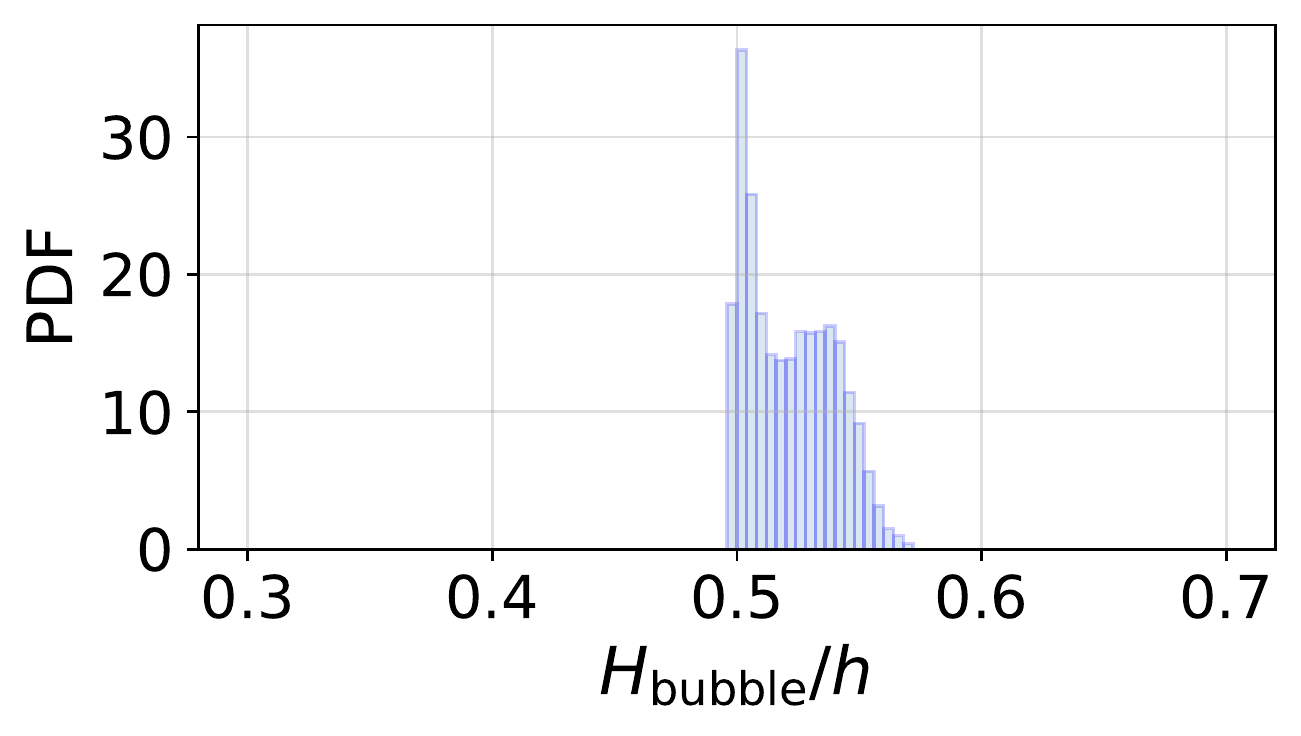} \\
         {\small (a)} & {\small (b)} & {\small (c)} \\
         \includegraphics[scale=0.35]{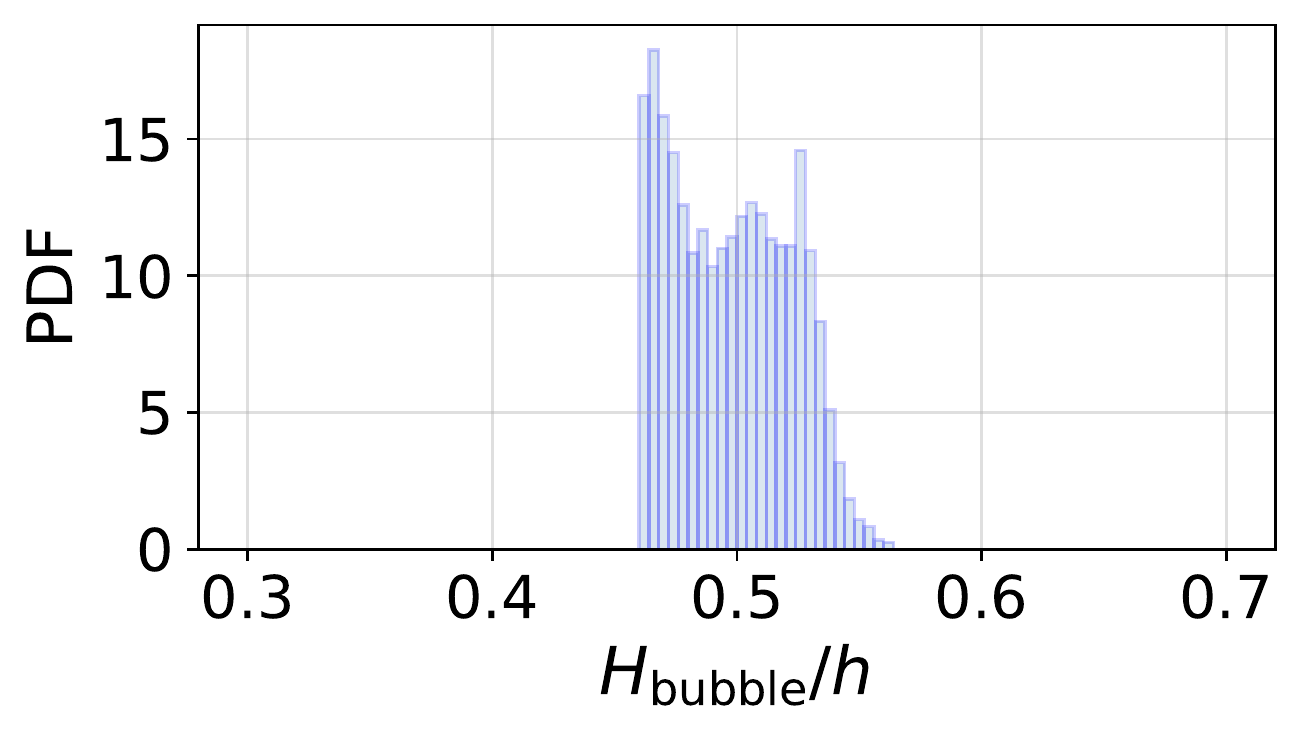}  &
         \includegraphics[scale=0.35]{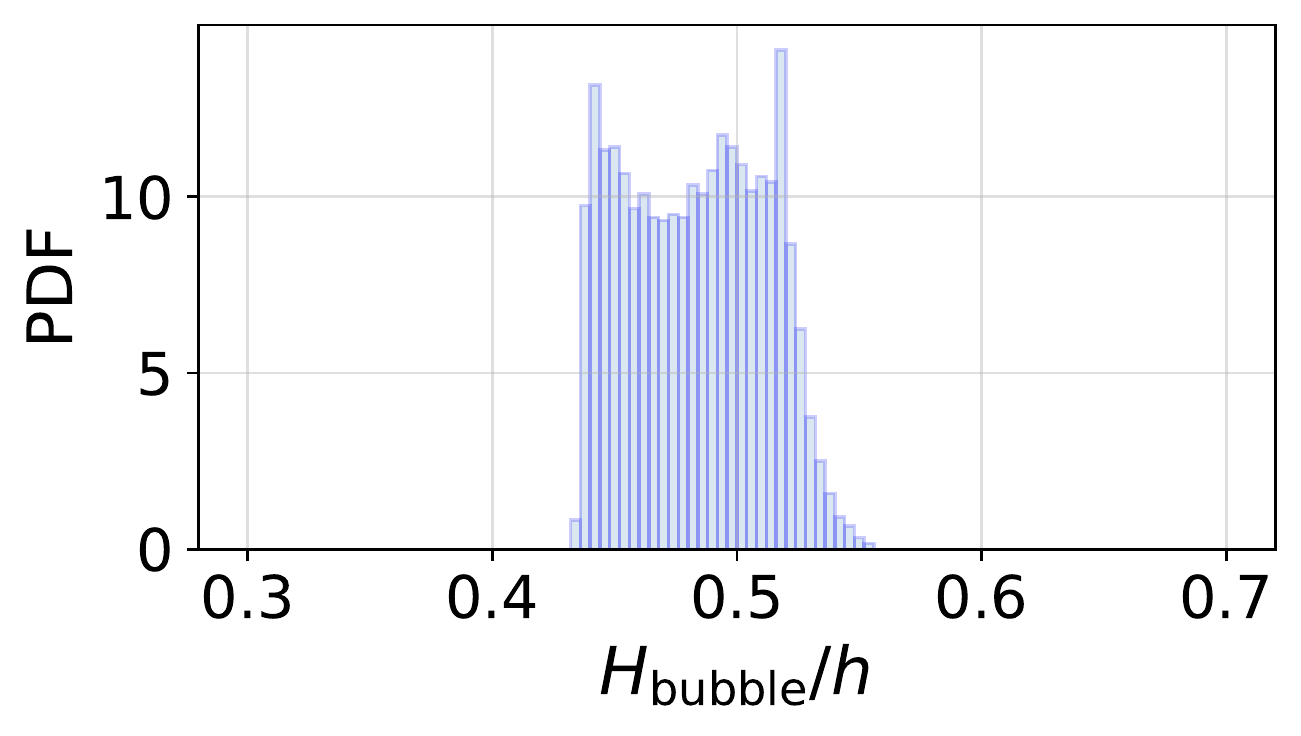} &
         \includegraphics[scale=0.35]{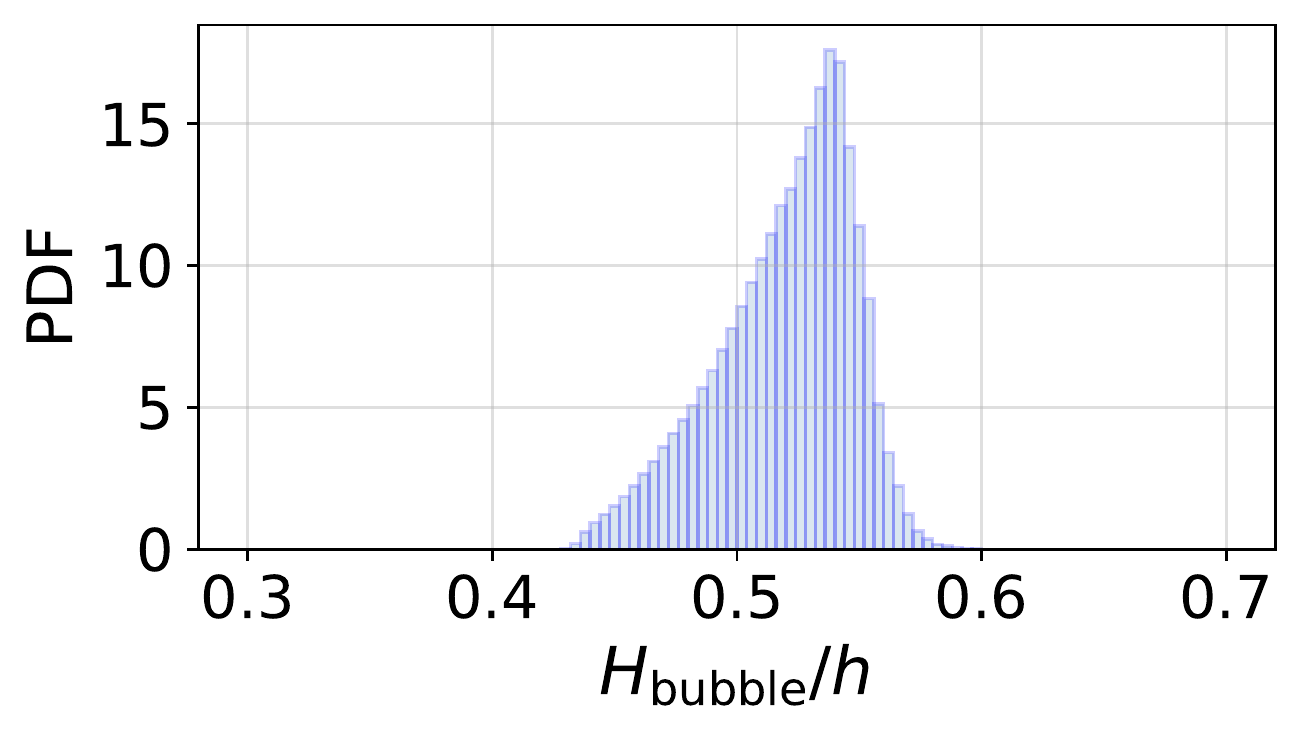} \\
         {\small (d)} & {\small (e)} & {\small (f)} \\
    \end{tabular}
    \caption{(a-e) PDF of $H_{\rm bubble}/h$ at $x/h=2.5$ due to the variation of~$\alpha$ and~$\gamma$ using the RANS data simulated with~$\betas$ equal to $0.348$, $0.367$, $0.400$, $0.433$, $0.452$, respectively. Note that $5\times 5$ samples are taken from the $\alpha-\gamma$ space at each of these constant-$\betas$ simulations. The PDF in (f) is obtained using all the $5 \times 5 \times 5$ samples from~$\alpha$,~$\gamma$ and~$\betas$.}
    \label{fig:phill_pdf_RANS}
\end{figure}

Before constructing the HC-MFM, it is important to look at the LF data. 
In \fig~\ref{fig:phill_pdf_RANS}, the PDF of the QoI due to the variation of~$\alpha$ and~$\gamma$ for different training samples of~$\betas$ is represented. 
The two expected yet important observations are that the PDF of the QoI is significantly influenced by the value of~$\betas$ used in the RANS simulations, and the fact that the PDFs are much different from the ground truth PDF shown in \fig~\ref{fig:phill_ground2.5}. 
The larger influence of~$\betas$ compared to~$\alpha$ and~$\gamma$ is also reflected in the associated Sobol indices~\cite{sobol:01}, as reported in \tab~\ref{tab:phill_UQresults}.
Note that the PDF of the QoI considering the simultaneous variation of~$\alpha$,~$\gamma$, and~$\betas$ is shown in \fig~\ref{fig:phill_pdf_RANS}(f).

\begin{figure}[!h]
    \centering
    \begin{tabular}{cc}
         \includegraphics[scale=0.35]{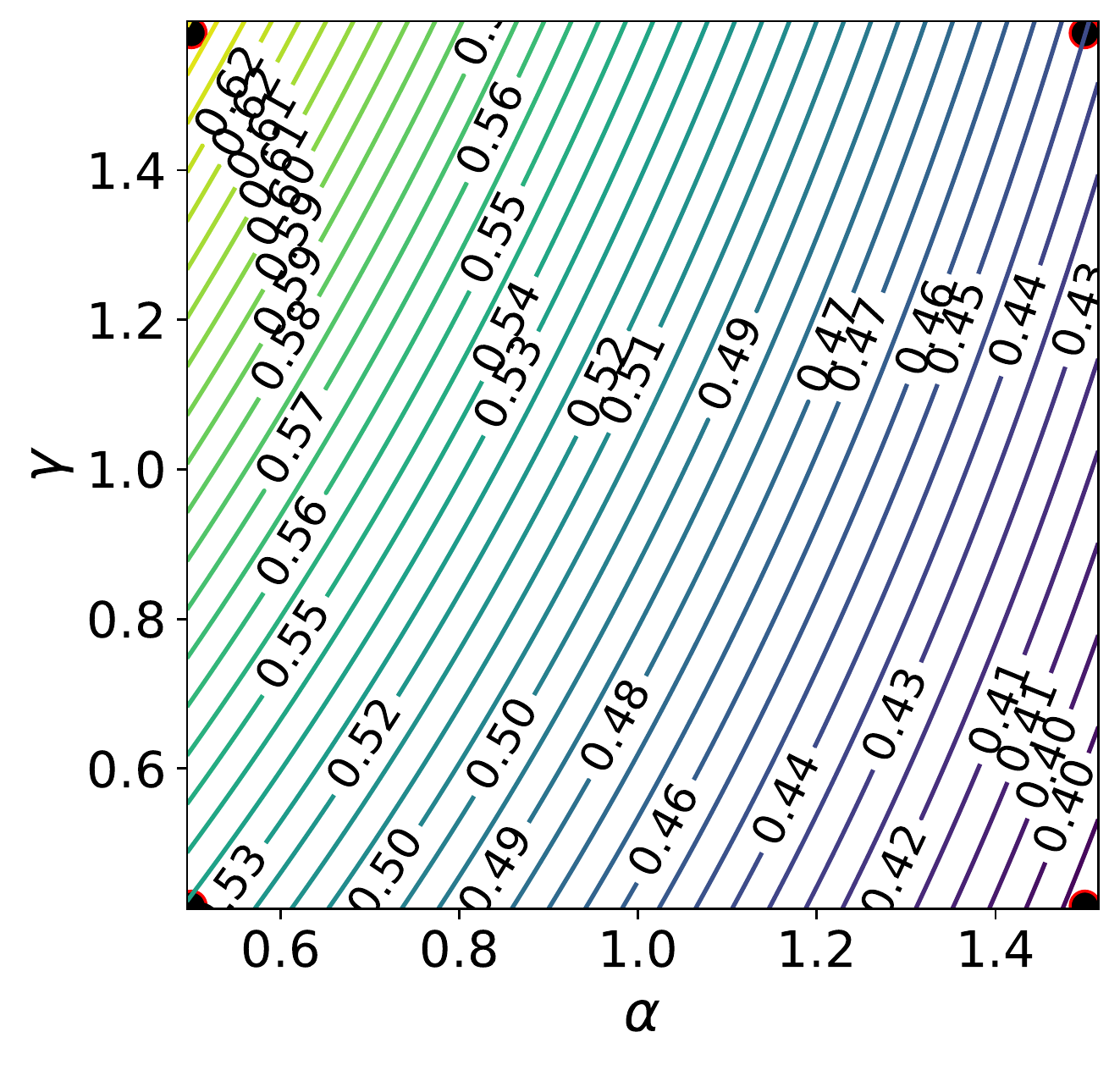} &
         \includegraphics[scale=0.35]{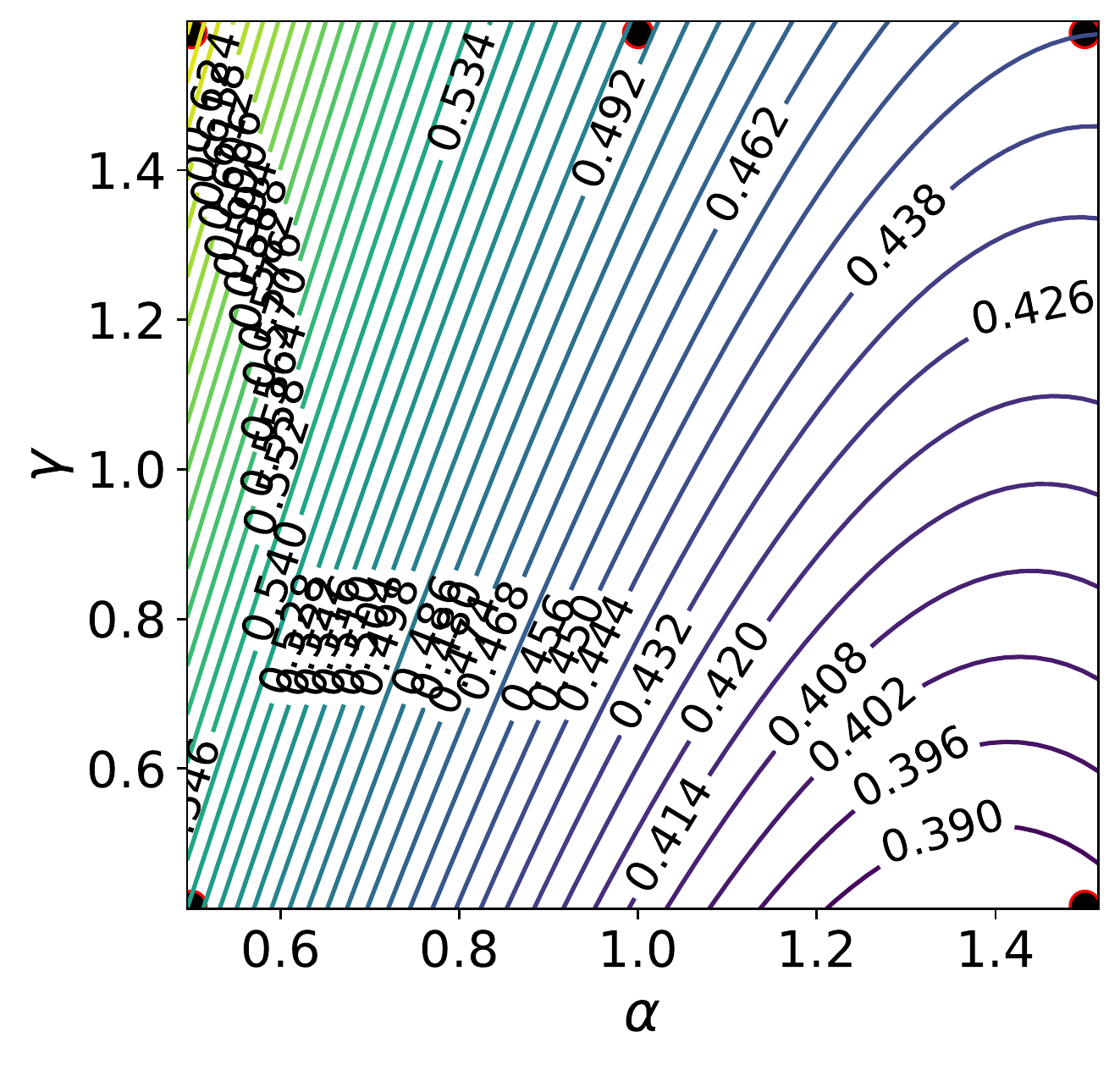} \\
         \includegraphics[scale=0.35]{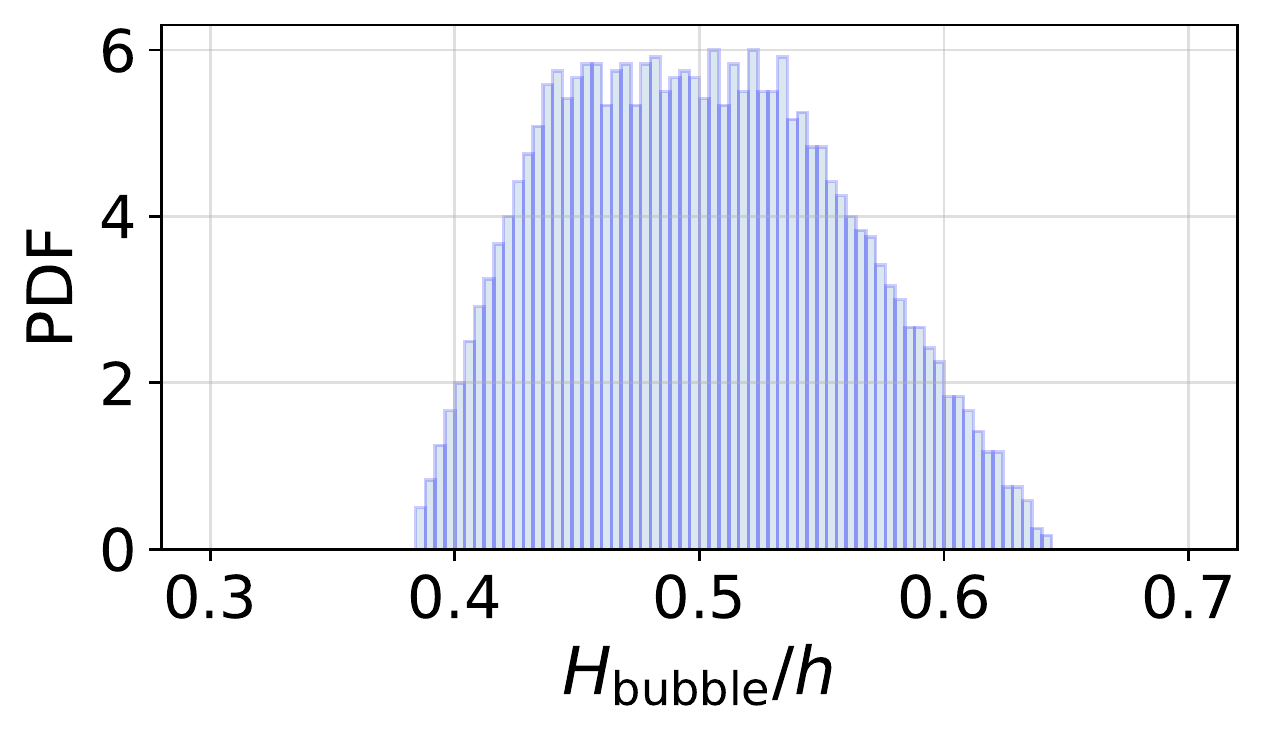} &
         \includegraphics[scale=0.35]{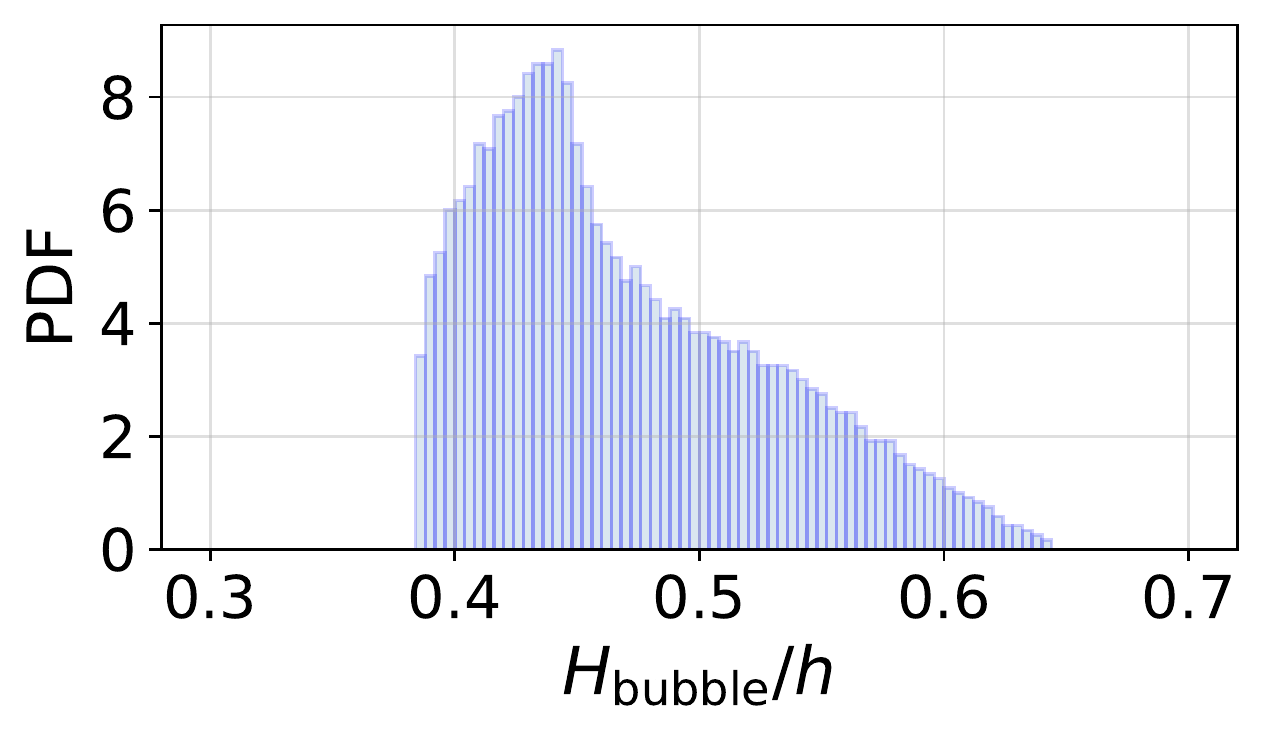} \\
    \end{tabular}
        \caption{(Top) Isolines of the response surface and (bottom) PDF of~$H_{\rm bubble}/h$ at~$x/h=2.5$ due to the variation of~$\alpha$ and~$\gamma$ using the HF data of (left) Case-A and (right) Case-B. The data are taken from the DNS of \rf~\cite{xiao:20} and are specified by dots in the top plots. }
    \label{fig:phill_pdfiso_HF}
\end{figure}

\begin{figure}[!h]
    \centering
    \begin{tabular}{cc}
         \includegraphics[scale=0.35]{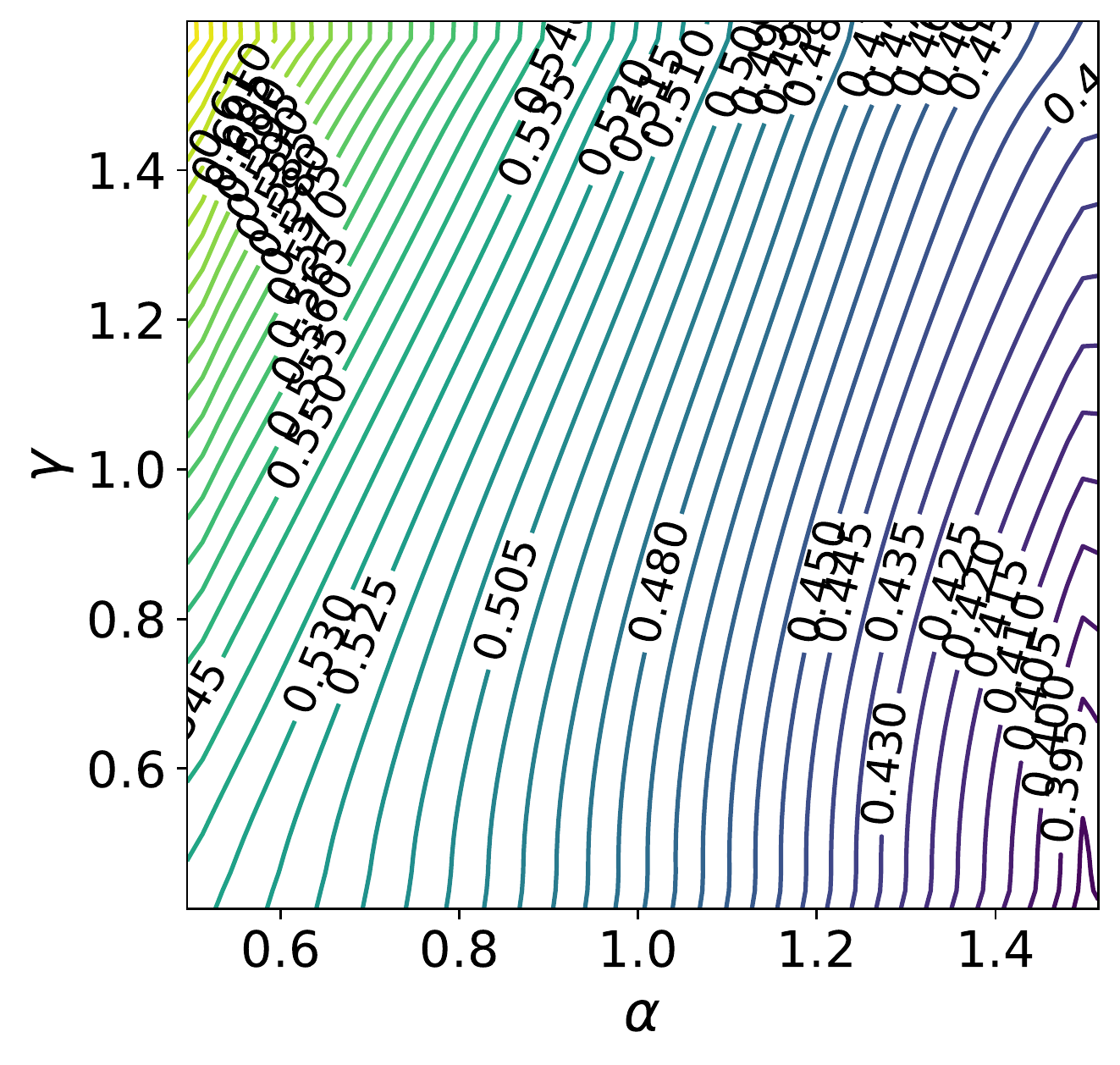} &
         \includegraphics[scale=0.35]{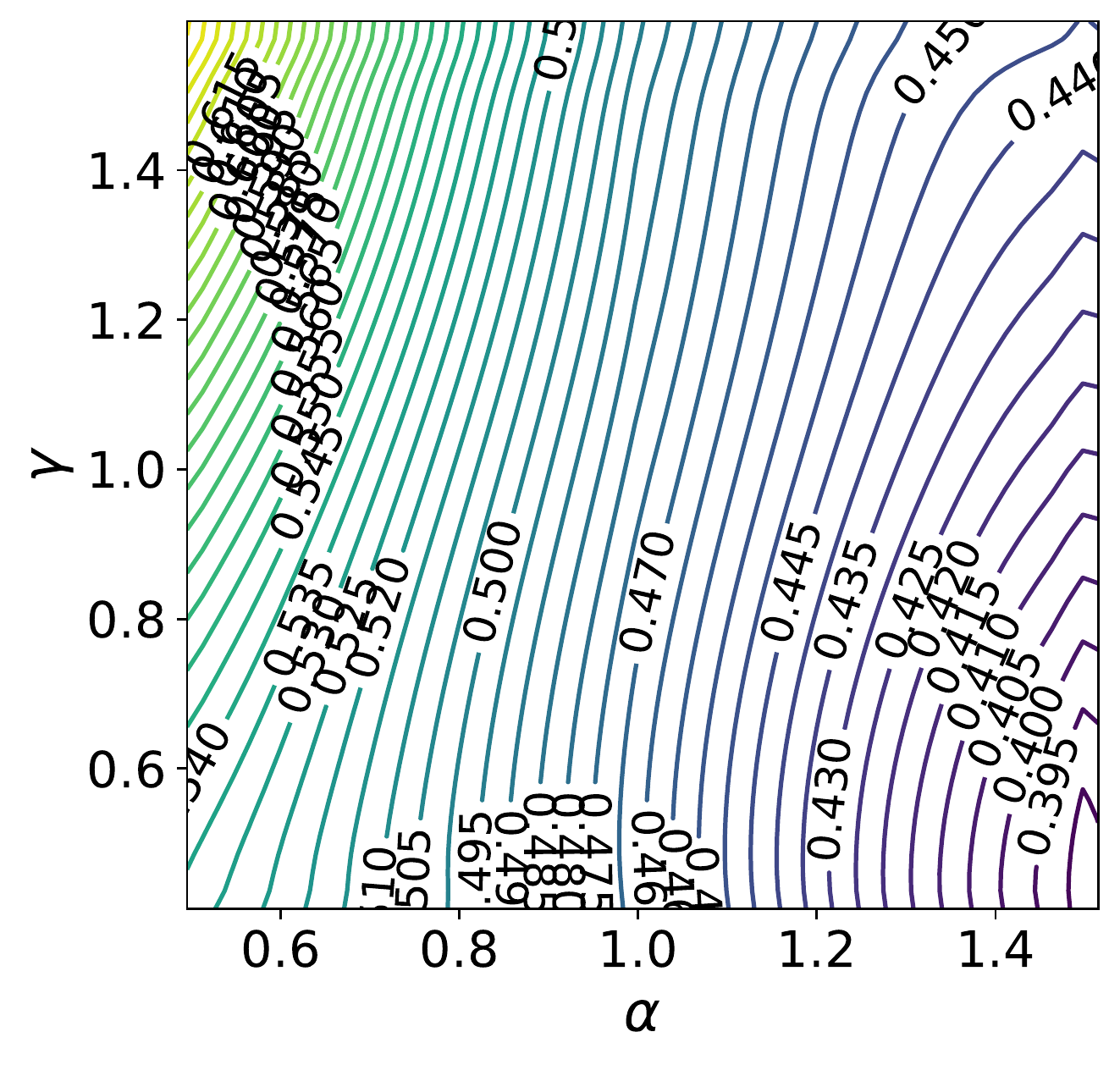} \\
         \includegraphics[scale=0.35]{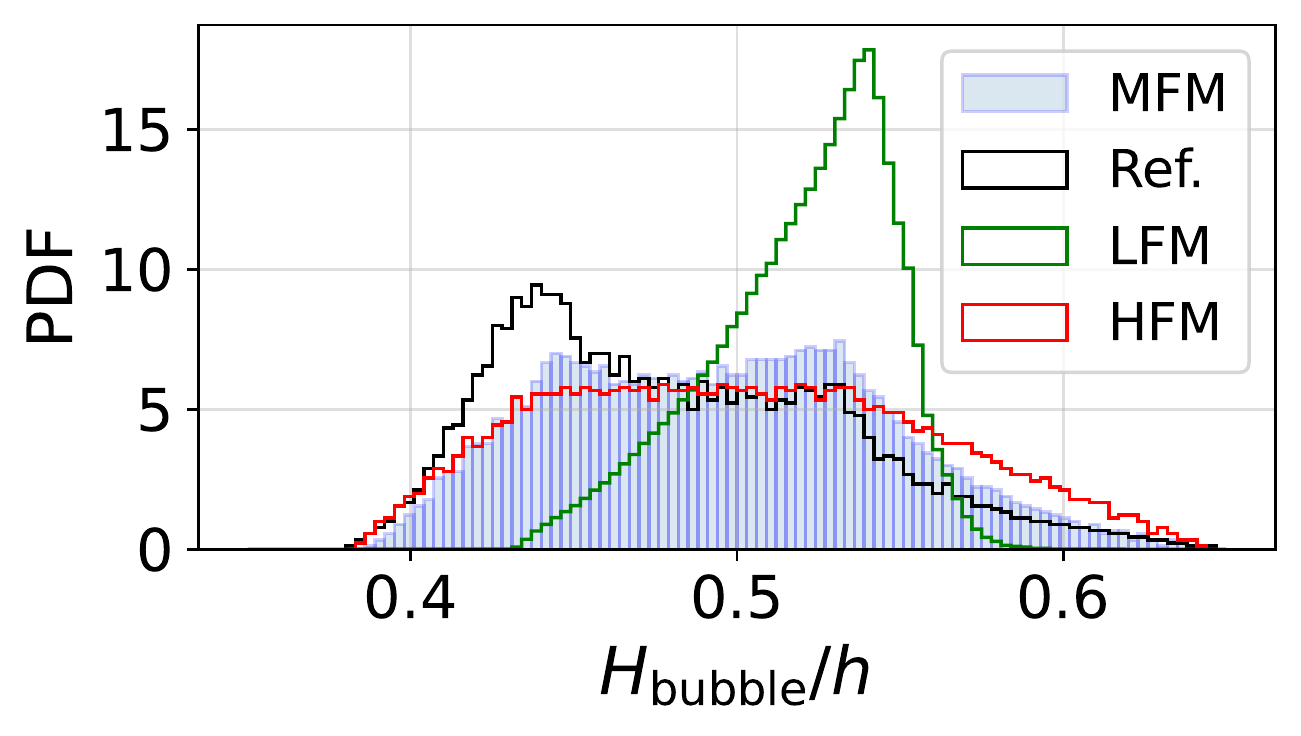} &
         \includegraphics[scale=0.35]{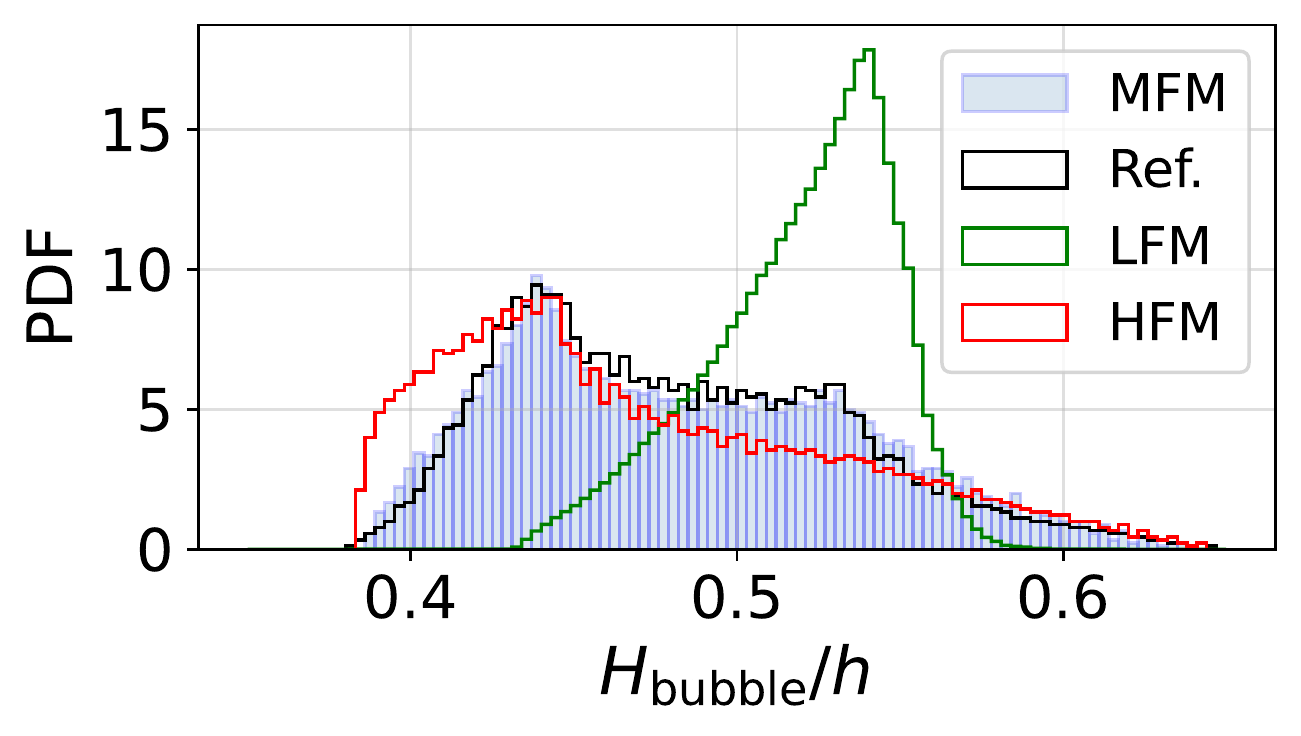} \\
    \end{tabular}
    \caption{(Top) Isolines of the response surface and (bottom) PDF of~$H_{\rm bubble}/h$ at~$x/h=2.5$ due to the variation of~$\alpha$ and~$\gamma$ obtained from the HC-MFM with the data of (left) Case-A and (right) Case-B. In each of the plots in the bottom row, the PDF resulting from the HC-MFM is compared to the PDFs of the ground truth (Ref., \fig~\ref{fig:phill_ground2.5}), low-fidelity data (LFM, \fig~\ref{fig:phill_pdf_RANS}), and high-fidelity data (HFM, \fig~\ref{fig:phill_pdfiso_HF}).}
    \label{fig:phill_pdfiso_MF}
\end{figure}

\begin{figure}[!h]
    \centering
    \begin{tabular}{cc}
    \includegraphics[scale=0.4]{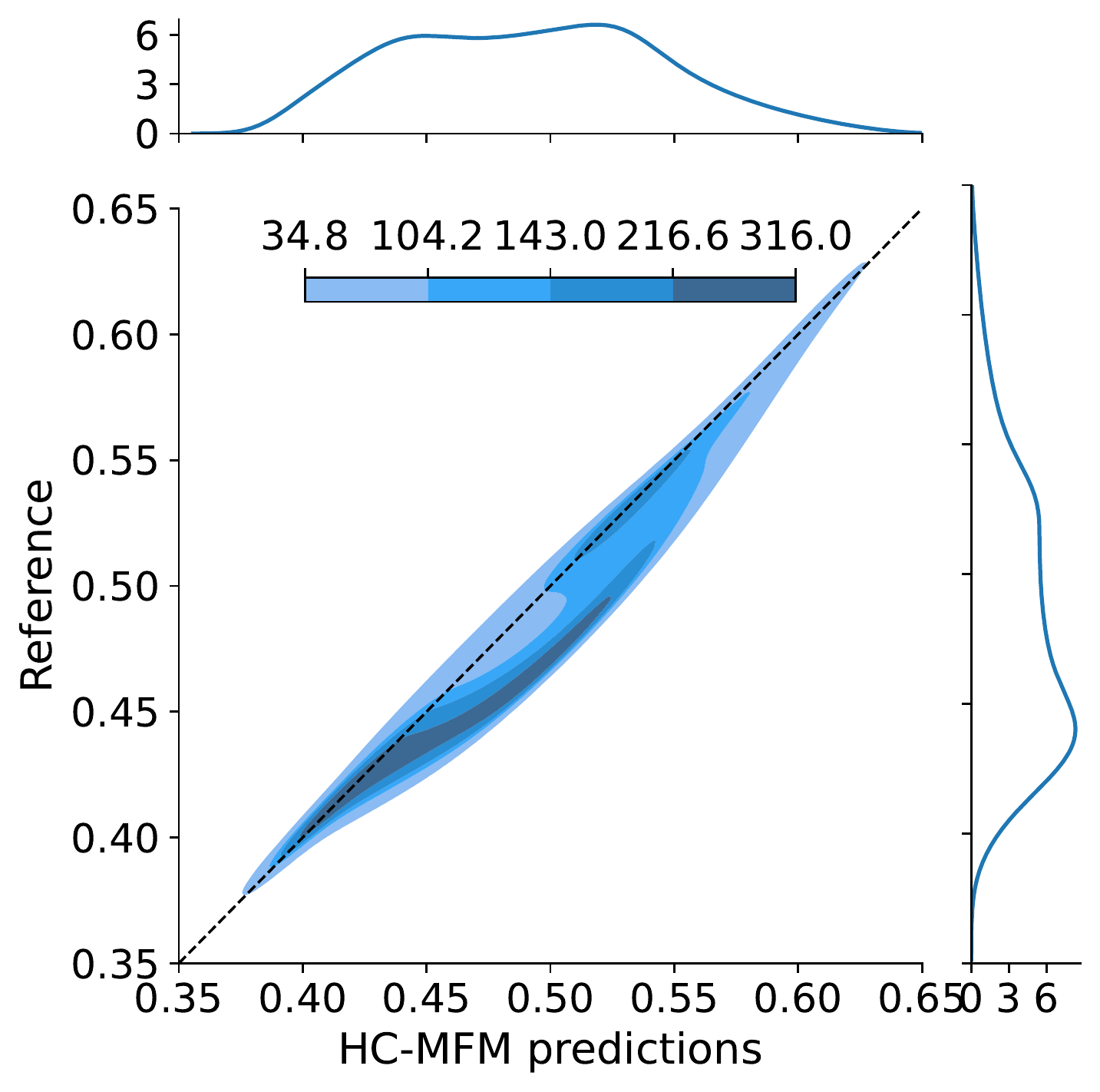} &     
    \includegraphics[scale=0.4]{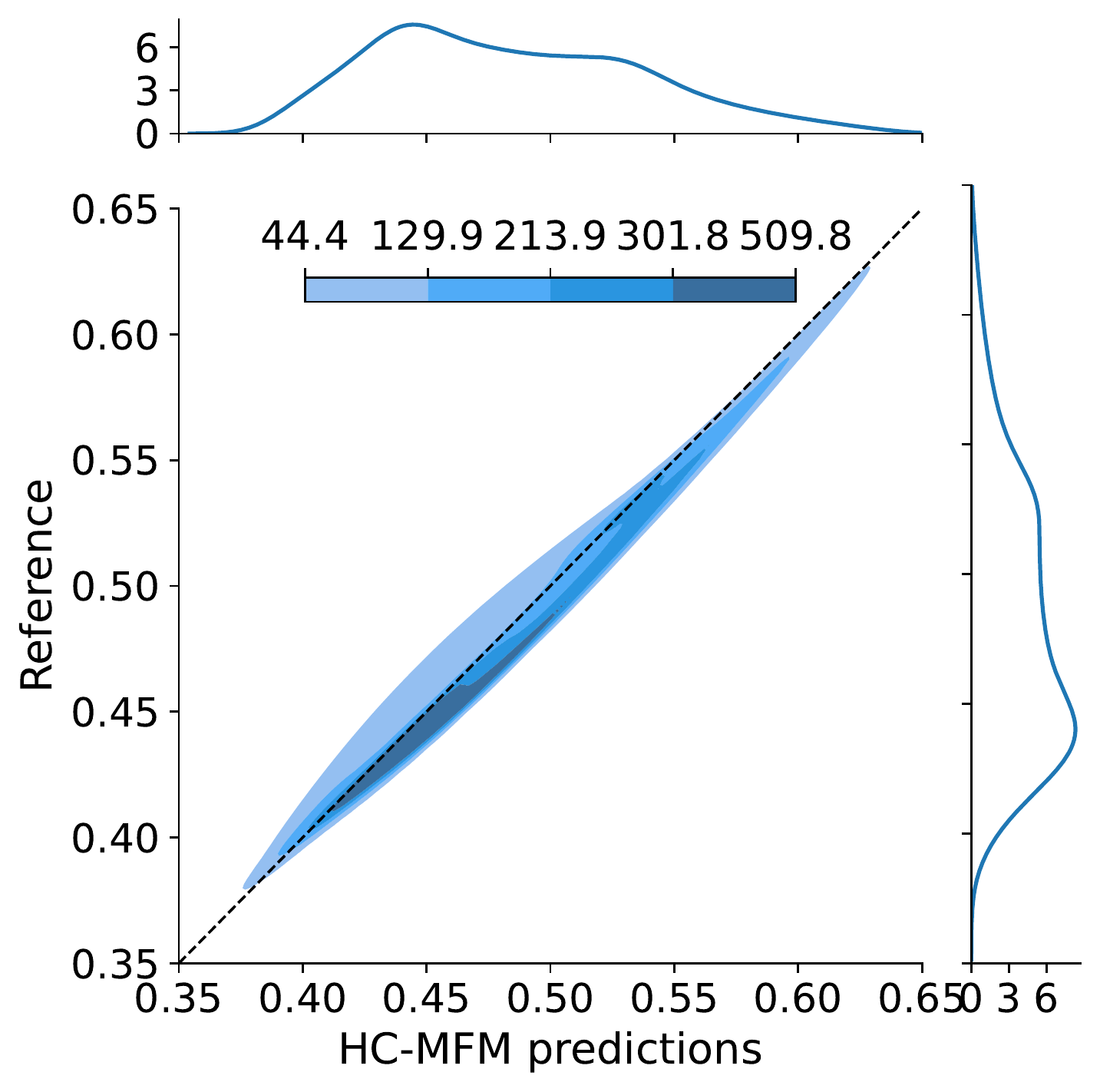} \\
    \includegraphics[scale=0.4]{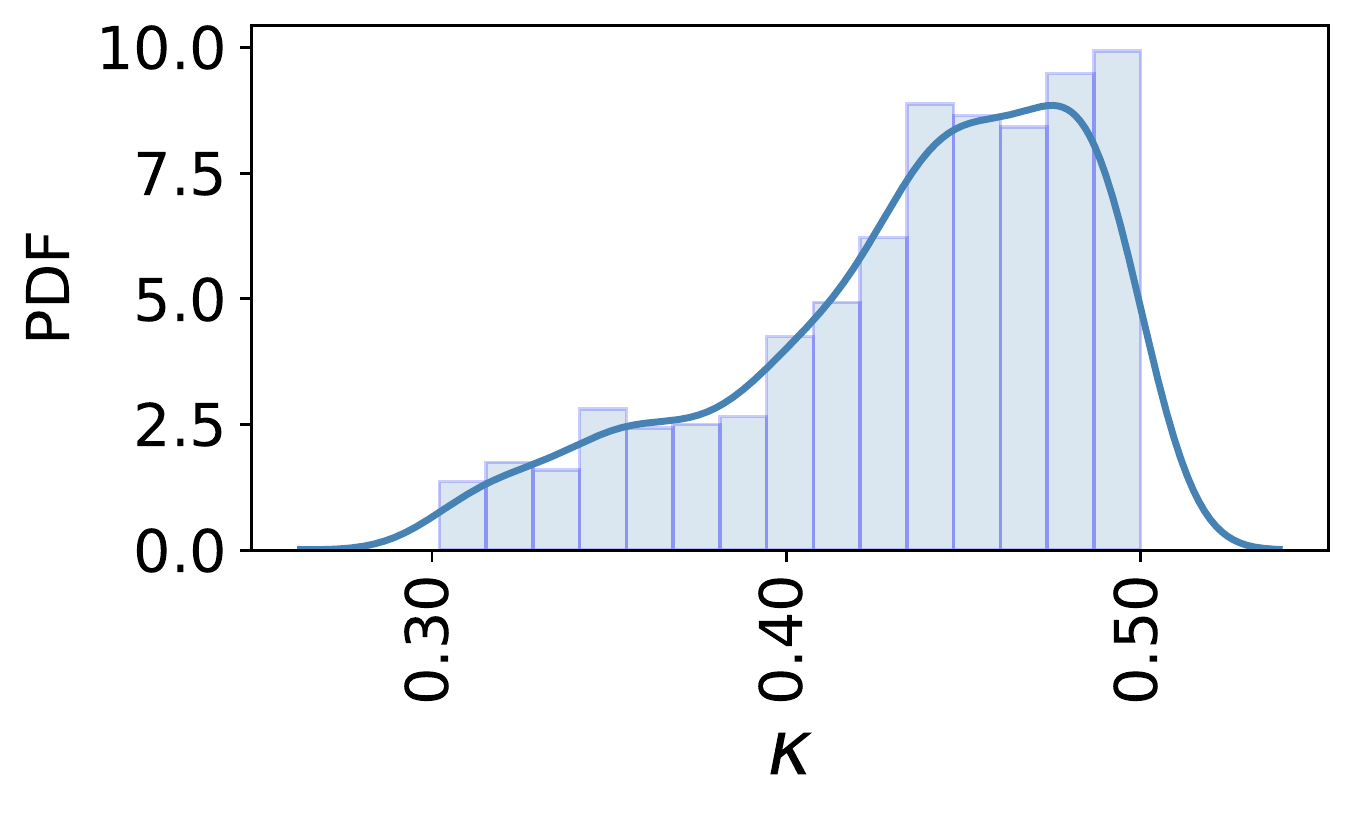} &  
   \includegraphics[scale=0.4]{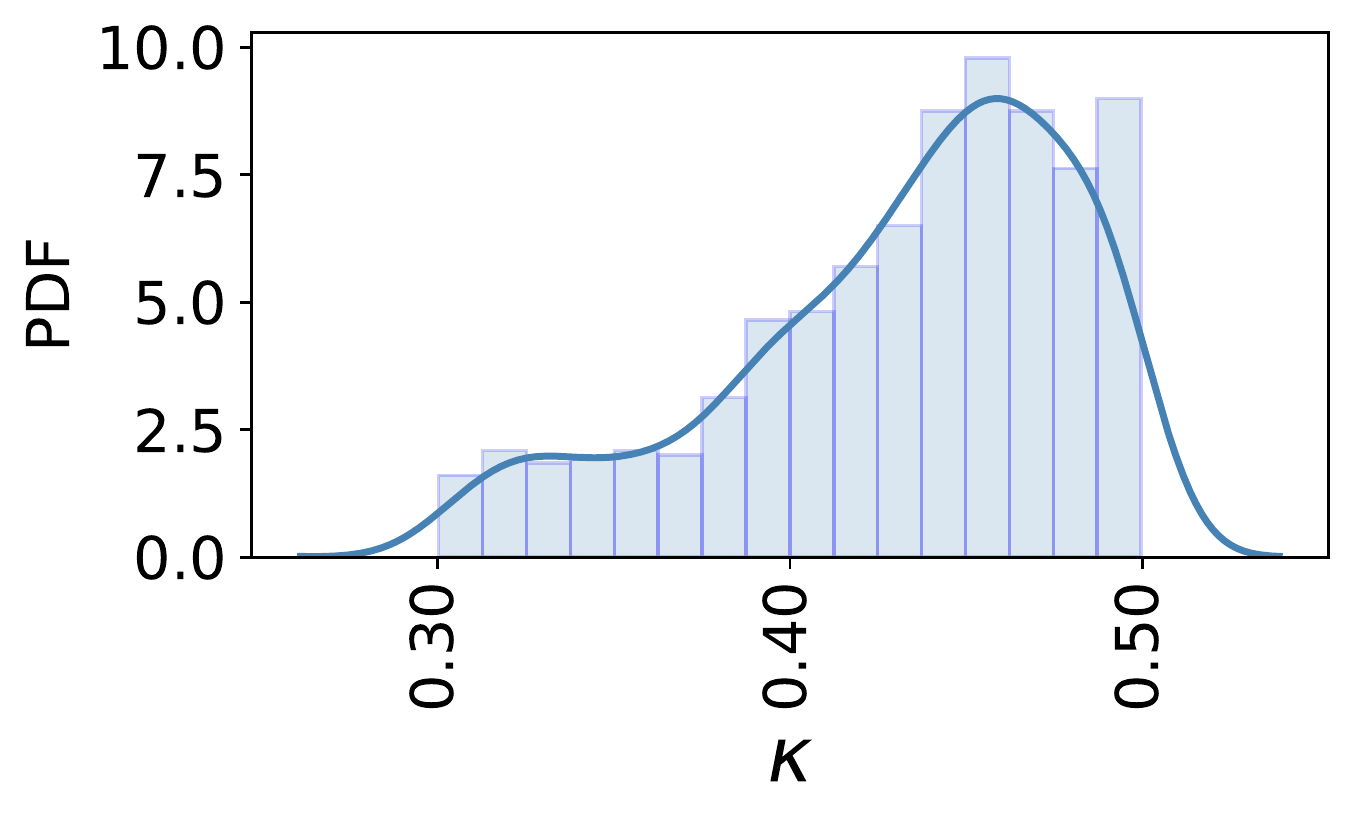} \\
    \end{tabular}
    \caption{(Top) Joint and marginal PDFs, and (bottom) sample posterior distribution of $H_{\rm bubble}/h$ at $x/h=2.5$ for (left) Case-A and (right) Case-B data sets. In the top plot, the contours belong to the joint PDF with associated values specified in the color bar.}
    \label{fig:phill_jpdf_caseAB}
\end{figure}

The response surface and PDF of the QoI obtained from only the HF data of Case-A and Case-B are illustrated in \fig~\ref{fig:phill_pdfiso_HF}.
For case-A with $n_1=4$ HF data, the PDF of the QoI has a plateau which makes it clearly different from the reference PDF in \fig~\ref{fig:phill_ground2.5}. 
By adding only one more DNS data point and obtaining~Case-B, the response surface becomes more similar to the reference, however, the associated PDF is still single mode. 
The improved predictions through the application of the HC-MFM are shown in \fig~\ref{fig:phill_pdfiso_MF}. 
Compared to the HF-data in \fig~\ref{fig:phill_pdfiso_HF}, the PDFs of the QoI clearly exhibit a second peak for the values between~$0.5$ and~$0.6$. 
This peak has been introduced by adding the LF data, see \fig~\ref{fig:phill_pdf_RANS}(f), and this is, in fact, the task for the multifidelity model to adjust the involved hyperparameters such that the fusion of the data at the two fidelities leads to a PDF similar to the reference. 
Clearly, for Case-B with only~5 DNS data samples included, the PDF and response surface of the QoI are very close to the ground truth in \fig~\ref{fig:phill_ground2.5} (9~DNS). 
This can also be confirmed by plotting the associated HC-MFM predictions against the reference data at all test points in the $\alpha-\gamma$ plane, see \fig~\ref{fig:phill_jpdf_caseAB}. 
The joint PDF of these two sets of data for both considered cases is narrow, specifically for Case-B, and hence implies a low point-to-point deviation of the predictions from the reference values. 
It is also interesting to look at the posterior distribution of the RANS calibration parameter~$\betas$. 
It is not surprising that the resulting PDFs from the multifidelity data sets Case-A and Case-B are different in spite of having the same uniform prior distribution.
Two important observations here are the following: First, in contrast to the previous examples in the present study, even with a small number of HF-data, \ie~Case-A, a significantly informative PDF for the LF calibration parameter is obtained. 
Second, for Case-B, the posterior distribution of~$\betas$ is very similar to the reference case where~$9$ DNS data sets are used, see \fig~\ref{fig:phill_refPostBeta}. 
Thus, depending on the case, the HC-MFM is capable of calibrating the model parameters in a fairly accurate way at the same time of constructing an accurate predictive model. 
This somewhat challenges the conclusion which could be drawn from the previous examples in the present study and also in Goh \et~\cite{goh:13}, where the priority of the HC-MFM is found to make accurate predictions for the QoI rather than providing accurate calibration of the fidelity-specific parameters.

To conclude the periodic hill example, we can quantify stochastic moments of the QoI as well as the Sobol sensitivity indices~\cite{sobol:01} due to the uncertainty in~$\alpha$ and~$\gamma$.
These UQ measures are integral quantities over the admissible range of the parameters and can be computed using the reference data (all available DNS), LF, HF, and MF data sets. 
Note that for the LF data, the uncertainty in~$\betas$ is also taken into account.
Noting the parameters~$\alpha$,~$\gamma$, and~$\betas$ are uniformly distributed and independent from each other, the results summarized in \tab~\ref{tab:phill_UQresults} are obtained using the generalized polynomial chaos expansion~\cite{xiu:02} for the reference and LF data sets, and the Monte Carlo method for the multifidelity cases. 
All the UQ analyses have been performed using UQit~\cite{uqit:21}. 
In general, for all cases but the pure LF data, the prediction of the mean and standard deviation of the QoI are close to the reference. 
For the total Sobol indices, the closest estimates to the reference values is obtained from the HC-MFM applied to Case-B, and on the second rank, Case-A. 
Noting the improvement of the Sobol indices accuracy in each of the multifidelity cases compared to the estimates from the associated HF data, the effectiveness of the HC-MFM is once again confirmed. 
This is an important outcome considering the forward UQ problems and global sensitivity analyses are of most relevance in CFD applications.

\begin{table}[h]
\centering
\captionsetup{justification=centering,margin=1cm}
\caption{Estimated mean, standard deviation, and total Sobol indices of the QoI~$R=H_{\rm bubble}/h$ at~$x/h=2.5$ due to the uncertainty in~$\alpha$ and~$\gamma$. For the LF (RANS) data the uncertainty due and sensitivity with respect to parameter~$\betas$ is also included. For the Case-A and Case-B data sets used for multifidelity modeling, see \fig~\ref{fig:phill_caseABSamples}. }\label{tab:phill_UQresults}
\begin{tabular}{*7c}
\toprule\toprule
{} & {} & \multicolumn{2}{c}{Moments due to $\alpha$, $\gamma$} & \multicolumn{3}{c}{Total Sobol Indices of $R$ w.r.t.} \\
{} & Data set & $\BE[R]$ & $\BS[R]$ & $\alpha$ & $\gamma$ & $\betas$ \\
\midrule
{} & Reference & 0.48104 & 0.05027 &  0.94551 & 0.08094 & --\\
{} & Low-Fidelity (LF) & 0.51999 & 0.01666  & 0.44275 & 0.02689 & 0.60623 \\
 \midrule
Case-A & High-Fidelity (HF) & 0.50047 & 0.05045 & 0.81448 & 0.19144 & --\\
{} & Multifidelity (MF) & 0.49209 & 0.05238 & 0.88133 & 0.12680 & --\\
 \midrule
Case-B & High-Fidelity (HF) & 0.46981 & 0.05198 & 0.82327 & 0.18237 & --\\
{} & Multifidelity (MF) & 0.48437 & 0.05215 & 0.91355 & 0.10097 & --\\
\bottomrule
\end{tabular}
\end{table}

\subsection{Keeping the RANS Parameter Fixed}
\begin{figure}
    \centering
    \begin{tabular}{cc}
         \includegraphics[scale=0.4]{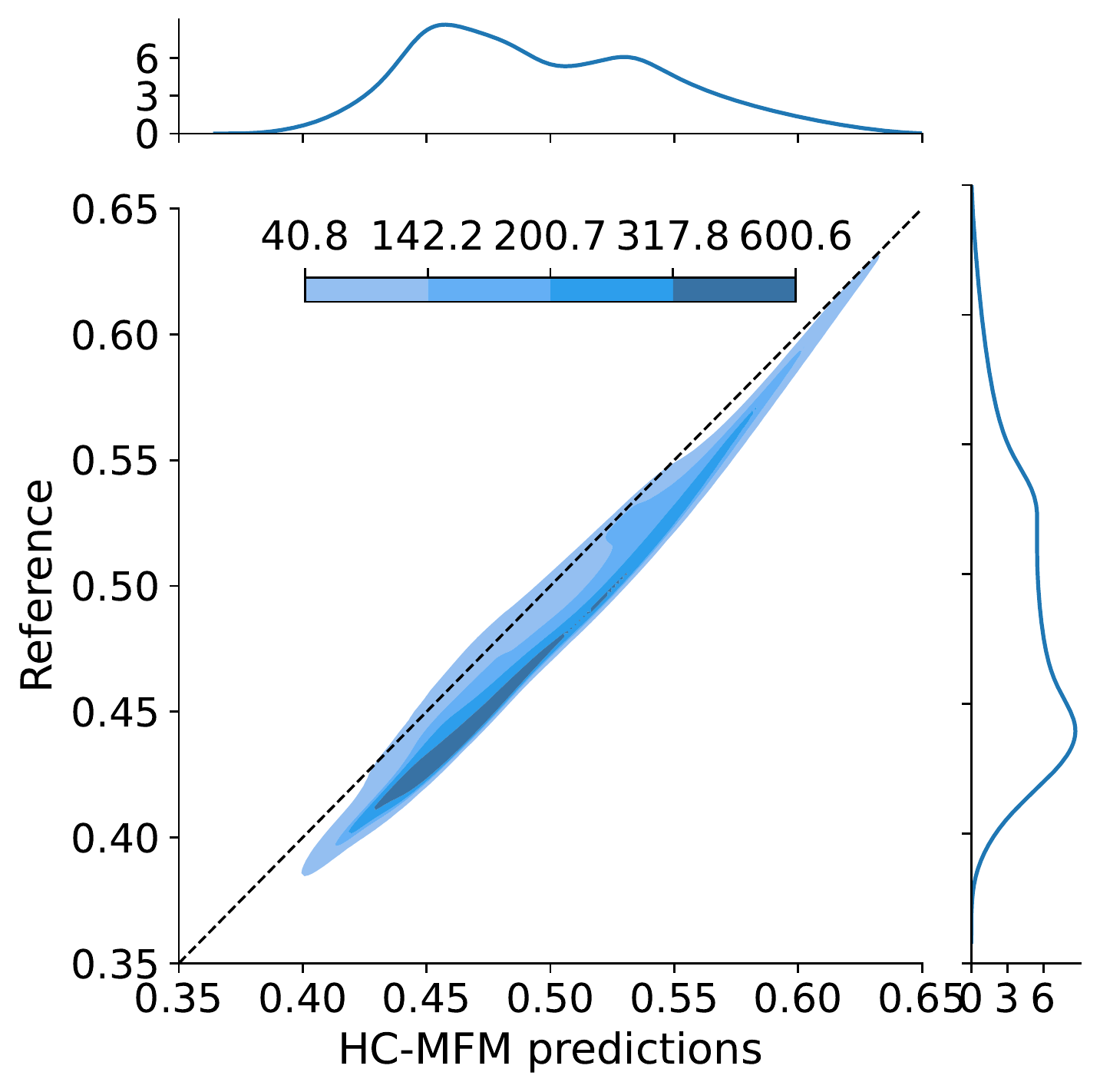} &
         \includegraphics[scale=0.4]{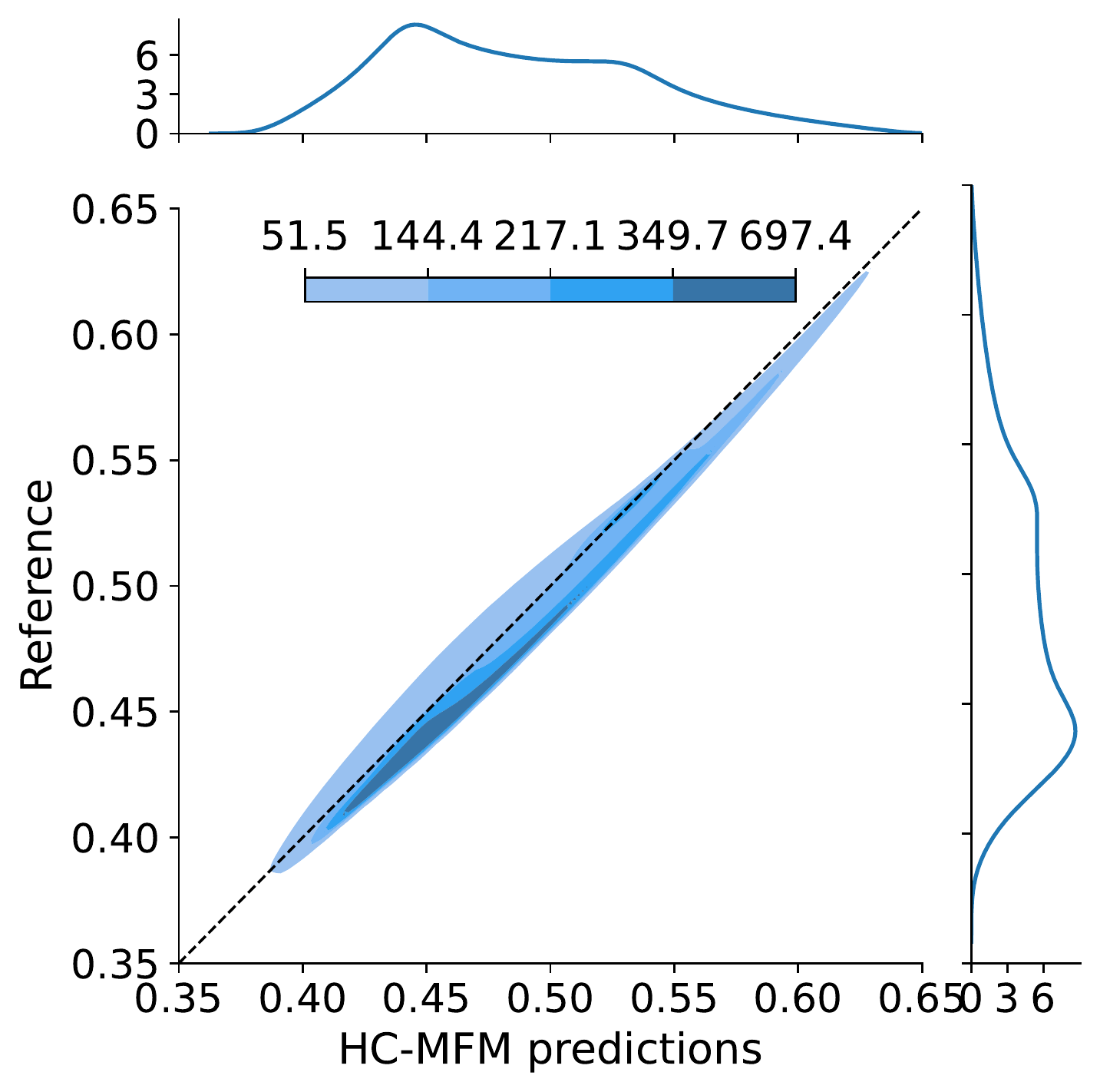} \\
    \end{tabular}
    \caption{Joint and marginal PDFs of $H_{\rm bubble}/h$ at $x/h=2.5$ for Case-B data sets using fixed values of~$\kappa$ equal to (left)~$0.4$ and (right)~$0.433$. Note that the PDF of the QoI due to the variation of~$\alpha$ and~$\beta$ corresponding to theses~$\kappa$ values is plotted in \fig~\ref{fig:phill_pdf_RANS}~(c) and~(d), respectively.}
    \label{fig:phill_noPar_jpdf_caseB}
\end{figure}

In all the examples in the present study, the fidelity-specific calibration parameters are involved in the multifidelity modeling and posterior distributions for them are learned during the construction of the MFM. 
But, the methodology behind the HC-MFM described in \sect~\ref{sec:method} is general and flexible to be directly applicable to the cases where the fidelity-specific parameters are kept fixed.
To demonstrate this, let us apply the HC-MFM to the example of the periodic hill and use the RANS data at constant values of~$\kappa$, the RANS modeling parameter in \eq~(\ref{eq:kappa_beta}). 
We use the Case-B data sets shown in \fig~\ref{fig:phill_caseABSamples} which means having~$5$ and~$25$ samples for the DNS and RANS simulations, respectively, in the $\alpha-\gamma$ space. 
The validation of the PDF of the QoI, $H_{\rm bubble}/h$ at $x/h=2.5$, of the HC-MFM for two values of~$\kappa$ is represented in \fig~\ref{fig:phill_noPar_jpdf_caseB}.
Adopting $\kappa=0.433$ ($\beta_\infty^*=0.084$) shows a clear improvement in the predictions compared to the standard value $\kappa=0.4$ ($\beta_\infty^*=0.09$).
This observation is consistent with the posterior PDF of~$\kappa$ in \fig~\ref{fig:phill_jpdf_caseAB}, where the mode of the distribution is higher than~$0.4$.
From this test, not only the validity of the HC-MFM for fixed values of the fidelity-specific parameters is confirmed, but also the fact that such accuracy-controlling parameters should be actively part of the data generation and hence the construction of HC-MFM is emphasized. 

Another important point is that the good predictive accuracy in \fig~\ref{fig:phill_noPar_jpdf_caseB} is obtained despite the poor correlation between the DNS and RANS data.
This is because of accurate construction of the model discrepancy term in \eq~(\ref{eq:mfModel_phill}) and also accurate estimation of various hyperparameters. 
It is also noteworthy that the present example is comparable to what is performed by Voet \et~\cite{voet:21} using~$7$ DNS and~$30$ RANS data sets using different multifidelity modeling approaches while fixing the value of the coefficients in the RANS closure model.
However, a direct comparison between the two studies is not possible since in \rf~\cite{voet:21}, the plots of the MFM predictions versus reference values of the QoI as in \fig~\ref{fig:phill_noPar_jpdf_caseB} are not provided.

\subsection{Impact of Replacing the MCMC by a Point-Estimator}\label{sec:MCMC_MAP}
In all the examples presented in this study, the HC-MFM is constructed using the MCMC method to draw samples from the posterior distribution of the hyperparameters and calibration parameters, \ie~$\bm{\beta}$ and~$\fthet$, respectively, in the Bayes formula~(\ref{eq:bayes}).
Similarly, to predict the sample distribution of the QoI, direct samples from these posterior distributions are used in the HC-MFM. 
As an alternative to these sample-based methods within the Bayesian framework, point estimators such as maximum a-posteriori probability (MAP) and maximum likelihood estimators (MLE) can be adopted. 
The point-estimated values are considered to be the representatives of the corresponding distribution.
Note that the use of the uniform priors in \eq~(\ref{eq:bayes}) makes the MAP estimations identical to MLE's. 
Our investigations showed that using point estimators instead of the MCMC method, could deteriorate the accuracy of the HC-MFM predictions, independent from how the LF and HF data are combined. 

\begin{figure}[!h]
    \centering
    \begin{tabular}{cc}
    \includegraphics[scale=0.4]{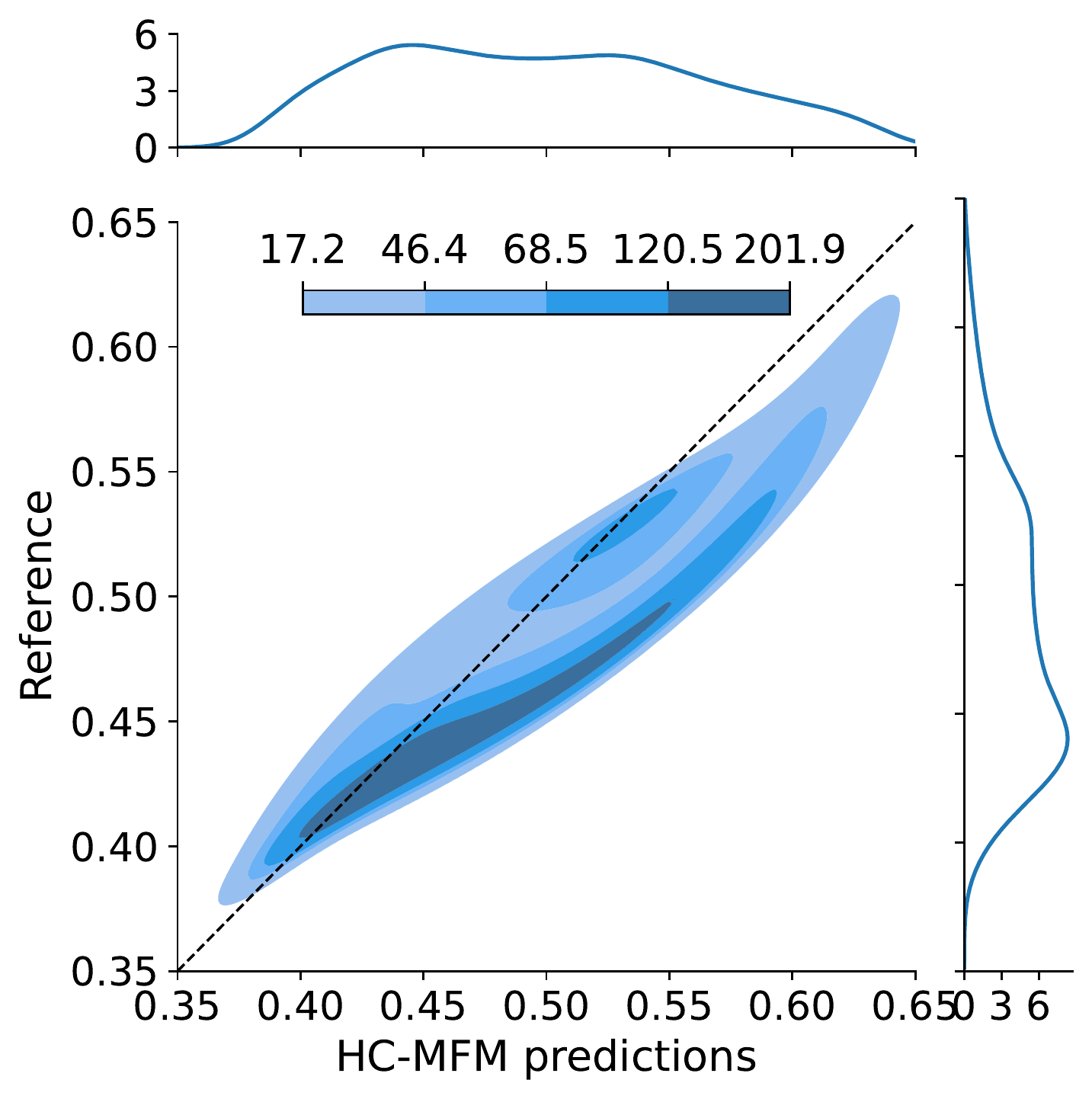} &     
    \includegraphics[scale=0.4]{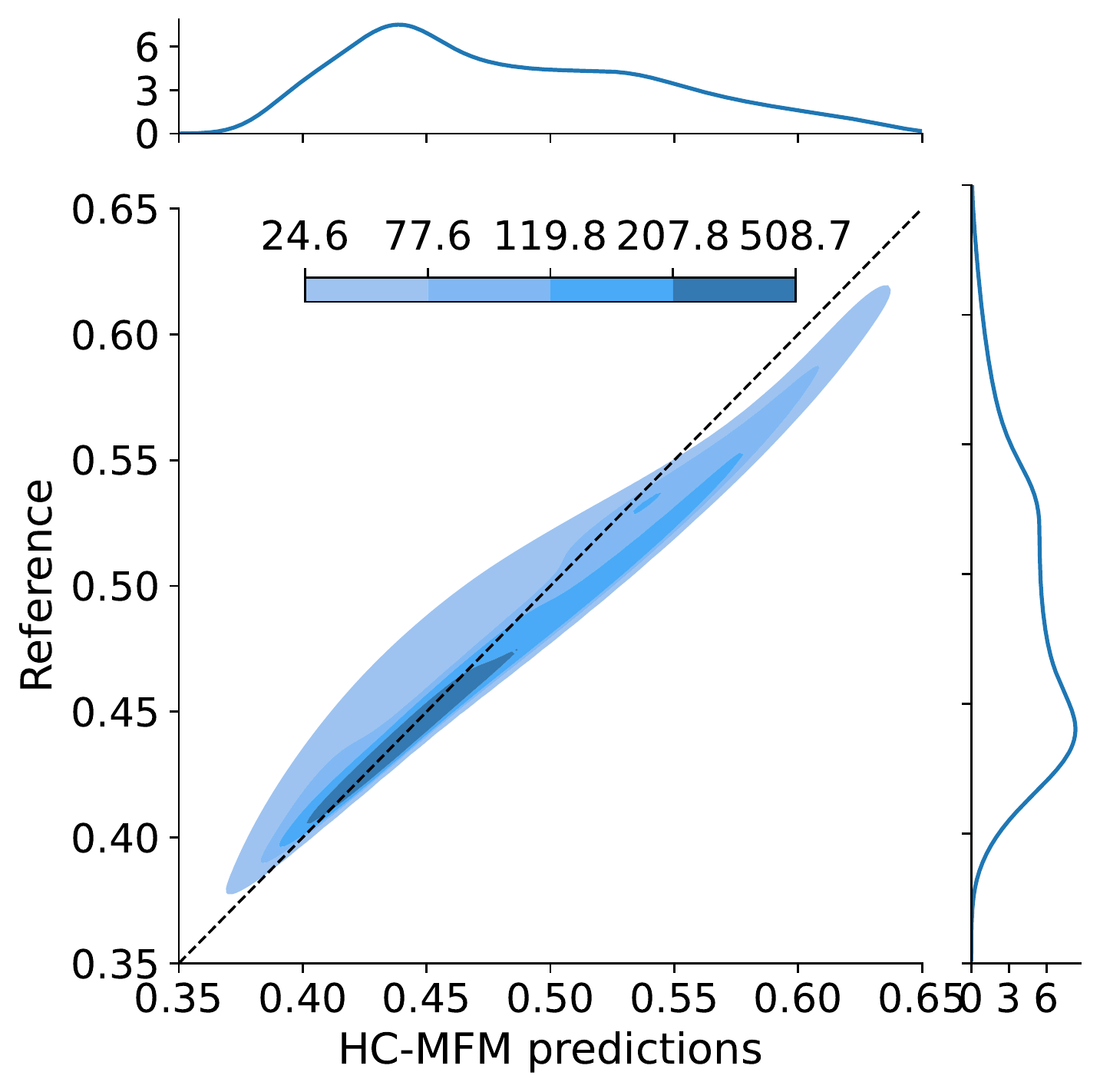} \\
    \end{tabular}
    \caption{(Top) Joint and marginal PDFs, and (bottom) sample posterior distribution of $H_{\rm bubble}/h$ at $x/h=2.5$ for (left) Case-A and (right) Case-B data sets. Here a MAP estimator is used to construct the HC-MFM, in contrast to \fig~\ref{fig:phill_jpdf_caseAB} and the rest of the examples in the present study which are obtained using an MCMC method. }
    \label{fig:phill_jpdf_caseAB_MAP}
\end{figure}

For instance, according to \fig~\ref{fig:phill_jpdf_caseAB_MAP}, the PDF of the QoIs predicted by the HC-MFM using the MAP estimator is significantly worse than what is given by the MCMC method as shown in \fig~\ref{fig:phill_jpdf_caseAB}.
This is an important message of the present study noting that all previous multifidelity studies in the literature relevant to the fluid flows have been based on using the point estimators, see \eg~Voet \et~\cite{voet:21} where the MLE is adopted. 

\section{Summary and Conclusions}\label{sec:conclusion}
The Bayesian hierarchical multifidelity model with automatic calibration (HC-MFM) developed by Goh~\et~\cite{goh:13} is adapted to several examples relevant to wall-bounded turbulent flows.
The HC-MFM is general, accurate, applicable to an arbitrary number of fidelity levels and well-suited to the simulations of turbulent flows since as a part of the MFM construction the fidelity-specific parameters can be automatically calibrated using the training data of higher fidelities.
This is an important feature noting that in all approaches for simulation of turbulence, different numerical and modeling uncertain parameters can influence the accuracy of the QoIs.
Because of using the Gaussian processes, the predictions made by the HC-MFM are accompanied with confidence intervals. 
Moreover, it is possible to incorporate the observational uncertainties in the data at all fidelity levels, and hence perform various UQ analyses for combinations of different types of uncertainties, see \rf~\cite{uqFrame:22}.

Based on the examples, the following main conclusions can be made.
1. For a fixed number of HF training data, the HC-MFM prioritizes the prediction of QoIs so that they become as close as possible to the HF validation data, while the posterior distributions of the calibration parameters are found to be accurate only if sufficiently many LF training data are provided.
A similar conclusion was drawn by Goh~\et~\cite{goh:13} by systematically increasing the amount of both HF and LF data.
For the periodic hill subject to geometrical uncertainties, \sect~\ref{sec:phill_MFM}, fixing the number of the LF data and considering two sets of HF data, the posterior distribution of the RANS (LF) parameter was found to be close to what would be obtained by using all the available HF data.
This, again, confirms the importance of providing sufficient LF training data through well exploring the space of the design and calibration parameters. 
2. When there are more than one QoI, the posterior distribution of the calibration parameters may depend on the QoI, see \sect~\ref{sec:airfoil}.
Therefore, the calibration parameters are more numerical than physical and hence, predictions by the HC-MFM for a QoI can be more accurate than the case of a-priori calibrating the low-fidelity models against high-fidelity data of another QoI (an example is calibrating a RANS closure model by the HF data of the lift coefficient, and then using the calibrated model in a simulation aiming for the drag coefficient with optimal accuracy).
3. As show in \sect~\ref{sec:MCMC_MAP}, the method for estimating the hyperparameters and parameters in the HC-MFM can significantly affect the resulting predictive accuracy. In fact, the MCMC sampling method is shown to result in more accurate predictions compared to a point estimator like MAP. 
This important point is usually overlooked in most of, if not all, the previous studies regarding the multifidelity modeling in CFD. 
4. If the fidelity-specific calibration parameters are kept fixed, the HC-MFM is still applicable without any need to modifying its general formulation. Obviously, the predictive accuracy of the model will depend on the validity of the value chosen for such parameters when generating the training data. 
The success of the HC-MFM relies on the accurate modeling of the discrepancy terms between different fidelities, and also the use of the MCMC methods to find optimal values for underlying hyperparameters through exploration of the parameter space.

The present study may be extended in several directions.
For instance, in addition to the scalar QoIs, spatio--temporal fields can be considered in the HC-MFM, making it possible to predict full flow fields by combining different fidelities. Such an approach may be particularly interesting as an alternative to the more black-box machine-learning tools when it comes to super-resolution and related methods. Another potential extension is in the combination of the HC-MFM with a Bayesian optimization for CFD applications and turbulent flow problems~\cite{morita:22}. In this case, the surrogate for the optimizer is based on the MFM, and is thus potentially cheaper to evaluate during the optimization process. In particular, applications in flow control for turbulence, where the main computational time lies in the evaluation of the objective function (\ie~the CFD solver), may greatly benefit from a well-calibrated multifidelity model.

\section*{Acknowledgments}
This work has been supported by the EXCELLERAT project, which has received funding from the European Union's Horizon 2020 research and innovation programme under grant agreement No 823691.
Additional funding was provided by the Knut and Alice Wallenberg Foundation (KAW). The computations were enabled by resources provided by the Swedish National Infrastructure for Computing (SNIC), partially funded by the Swedish Research Council through grant agreement no. 2018-05973.

\appendix

\section{PDF of A Set of Standard Distributions}
For the prior distribution of the parameters and hyperparameters of the multifidelity models in \sect~\ref{sec:results}, a set of standard distributions was used. 
\tab~\ref{tab:pdfs} summarizes the PDF and associated support of such distributions.

\begin{table}[h]
\centering
\captionsetup{justification=centering,margin=1cm}
\caption{PDF and associated support of the standard distributions used in \sect~\ref{sec:results}.}\label{tab:pdfs}
\begin{tabular}{*3c}
\toprule\toprule
Distribution & PDF & Support in $x$ \\
\midrule
Uniform ($\cU$) & $\rho(x;\alpha,\beta)=(\beta-\alpha)^{-1}$ & $[\alpha,\beta]$ \\
Gaussian ($\cN$) & $\rho(x;\alpha,\beta)=
(2\pi \beta^2)^{-1/2} \exp \left(-\frac{1}{2} (\frac{x-\alpha}{\beta})^2 \right)$ & $(-\infty,\infty)$ \\
Half-Cauchy ($\mathcal{HC}$) &
$\rho(x;\alpha) = 2\left[\pi \alpha \left(1+ \left(\frac{x}{\alpha}\right)^2\right) \right]^{-1}$ &
$[0,\infty)$ \\
Gamma ($\Gamma$) & $\rho(x;\alpha,\beta)=\beta^\alpha x^{\alpha-1} \exp(-\beta x) / \Gamma(\alpha)$  & $(0,\infty)$ \\
\bottomrule
\end{tabular}
\end{table}

\bibliographystyle{abbrv}  
\bibliography{bib_MFM}

\end{document}